\newcommand\aap{A\&A}                
\newcommand\apj{ApJ}                 
\newcommand\apjl{ApJL}                
\newcommand\araa{ARA\&A}             
\newcommand\mnras{MNRAS}             
\newcommand\na{New~Astron.}          
\newcommand\nar{New~Astron.~Rev.}    
\newcommand\nat{Nature}              
\newcommand\prd{Phys. Rev.~D}        
\newcommand\prl{Phys. Rev.~Lett.}    
\newcommand\pasj{PASJ}               
\newcommand\physrep{Phys.~Rep.}      

\documentclass[%
 reprint,
 amsmath,amssymb,
 aps,
]{revtex4-2}

\def\st${\ensuremath\mathrm}

\usepackage{comment}
\usepackage{cancel} 
\usepackage{graphicx}
\usepackage{dcolumn}
\usepackage{bm}
\usepackage{xcolor}
\usepackage{hyperref}
\usepackage[normalem]{ulem}
\usepackage[mathlines]{lineno}

\begin{document}

\preprint{APS/123-QED}

\title{
Optical and UV Flares from Binary Black Hole Mergers in Active Galactic Nuclei
}


\author{J. C. Rodr\'iguez-Ram\'irez$^{1}$}
\email{rod.ram.astro@gmail.com}
\author{R. Nemmen$^{3,4}$}
\author{C. R. Bom$^{1,2}$}
\affiliation{
$^1$CBPF—Centro Brasileiro de Pesquisas F\'isicas, Xavier Sigaud st. 150, 22290-180, Rio de Janeiro, Brazil\\
$^{2}$Centro Federal de Educa\c{c}\~{a}o Tecnol\'{o}gica Celso Suckow da Fonseca, Rodovia M\'{a}rcio Covas, lote J2, quadra J - Itagua\'{i}, Brazil \\
$^{3}$Instituto de Astronomia, Geof\'{\i}sica e Ci\^encias Atmosf\'ericas, Universidade de S\~ao Paulo, S\~ao Paulo, SP, 05508-090, Brazil\\
$^4$Kavli Institute for Particle Astrophysics and Cosmology, Stanford University, Stanford, CA 94305, USA 
}


%


\date{\today}

\begin{abstract}
Stellar mass, binary black hole (BBH) mergers dominates the sources of gravitational wave (GW) events so far detected by the LIGO/Virgo/KAGRA (LVK) experiment. 
The origin of these BBHs is unknown, and no electromagnetic (EM) counterpart has been undoubtedly associated to any of such GW events.
The thin discs of active galactic nuclei (AGNs) might be viable environments where BBHs can form and at the same time produce observable radiation feedback.
This paper presents new physically motivated light-curve (LC) solutions for thermal flares driven by the remnant of a BBH merger within the disc of an AGN. 
Following previous analyses, we consider that the BBH likely creates an under-density cavity in the disc prior to its coalescence.
Depending on the merger conditions, 
the black hole (BH) remnant can leave the cavity, 
interact with the unperturbed disc, and drive a transient BH wind.
The wind expels disc material that expands above and below the disc plane and we consider the emission of these plasma ejections as the EM counterpart to the merger GW.
We model the LC of such eruptions as mini supernovae explosions finding that
stellar mass merger remnants can drive distinguishable flares in optical and UV bands, with time lags of 
$\sim 10-100$ days after the GW event, when occurring in AGN with central engines of $\sim 10^{6-8}$ M$_\odot$.
The Vera C. Rubin Observatory can potentially detect the optical component of these EM counterparts up to redshifts of $z \sim 0.7 - 1.2$, accordingly.
The present model provide constraints on multi-messenger observables,
such as the binary effective spin, mass ratio, remnant mass, and the time delay of the EM signature. These constraints are useful to 
localise flaring AGNs as sources of multi-messenger emission following GW alerts.

\end{abstract}

\maketitle


\section{\label{sec:level1}Introduction
}

The measurement of gravitational waves (GWs) by the LIGO/Virgo/KAGRA (LVK) experiment opens a new window to explore our Universe in an unprecedented way. Among the types of GWs the experiment can detect, those from stellar mass binary black hole (BBH) mergers dominate the measurements during the LVK O1-O4 observational runs~\cite{bbh_o3,Sadiq:2023zee,petrov_2022,abbott2020prospects}. Currently, the astrophysical environments where such BBHs form are not known, and multiple channels have been proposed as plausible origins~\cite{bbh_chanel01,bbh_chanel02,bbh_chanel03}.

Electromagnetic flares accompanying BBH mergers could unveil their astrophysical hosts. The search for such multi-messenger (MM) associations is the subject of current active research involving follow-up campaigns~\cite{connaughton_2016, Bagoly_2016, becerra_2021, Graham_2020, ohgami_2023,Santos_2024}, statistical analyses of observed data~\cite{Graham_2020, Graham_2023,Palmese_2021,Santini_2023,Morton_2023,Veronesi_2025},
as well as theoretical efforts to understand the production of detectable electromagnetic counterparts~\cite{Perna_2016, Janiuk_2017, McKernan_2019, Wang_2021b, Kimura_2021, Tagawa_2023, Rodriguez-Ramirez_2024, Chen_2024}. GWs and radiation from a common origin provide the means of measuring the distance and redshift of the source independently. In this regard, MM events triggered by BBH mergers can also find cosmological applications~\cite{Gayathri_2020, Haster_2020, Bom_2023, Alfradique_2024}. Contrary to mergers of compact objects involving neutron stars, BBH mergers do not produce radiation by themselves and require occurring in gas-rich environments to produce significant EM signals.

The central regions of active galactic nuclei (AGNs) possess suitable conditions to cluster stellar mass black holes~\cite{Morris_1993,Miralda_2000,Hailey_2018} and to induce BBH pairing and merging~\cite{McKernan_2012, Bartos_2017, Tagawa_2020, Rowan_2023, Whitehead_2024, Dittmann_2024, Calcino_2024}. The scenario of mergers in AGN discs appears to naturally explain puzzling features revealed by LVK measurements, such as the anti-correlation of mass ratio and effective spin ($q-\chi_\mathrm{eff}$; ~\cite{Callister_2021,Santini_2023}), as well as the production of intermediate mass black holes as merger remnants~\cite{Kimball_2020,McKernan_2012}. Interestingly, AGN thin discs are environments dense enough where BBH mergers can produce significant radiation feedback~\cite{Bartos_2017,McKernan_2019,Kimura_2021}.

A number of optical AGN flares measured by the Zwicky Transient Facility (ZTF)~\cite{ZTF_2018} have been investigated as possible EM counterparts to certain BBH GW events of the LVK experiment~\cite{Graham_2020,Graham_2023}. These MM associations are currently debated, mostly because the spatial correlation among the different messengers is challenging~\cite{DePaolis_2020, Ashton_2021, Palmese_2021, Morton_2023, Veronesi_2025}. Given the current LVK localization volume in the sky, if AGNs indeed host the observed BBH mergers, the possible EM counterpart of a single GW event should be found among hundreds or thousands of known AGNs~\cite{Bom_2023}. An additional challenge in spotting such counterparts is that AGNs typically exhibit intrinsic variability on timescales from days to years~\cite{VandenBerk_2004, Yu_2022}, and different astrophysical phenomena can lead to flaring and emission enhancement in the AGN emission~\cite{Ross_2018, deGouveia_2010, Scepi_2021, Grishin_2021, Chan_2019}. Physical MM models for BBH mergers are then crucial to favor or disfavor candidate AGN flares as GW counterparts.

A number of astrophysical scenarios involving BBH mergers in AGNs leading to EM signals have been analysed in previous works, most of them considering gas feedback launched by the merger remnant. EM signals powered by radiation-driven outflows from the merger remnant have been analysed by \cite{Wang_2021a,Wang_2021b, Kimura_2021}. The merger remnant could also produce collimated, relativistic jets, and the interaction of such jets with the disc under different conditions has been analysed by \cite{Tagawa_2023, Tagawa_2023b, Rodriguez-Ramirez_2024} predicting different EM signatures.

In this paper, we analytically derive new optical light-curve profiles of thermal flares energised by the remnant of a BBH merger that occurs within the thin disc of an AGN. We follow the scenario proposed by \cite{Kimura_2021}, and consider that the merger likely occurs within an under-density cavity created by the binary in the thin disc prior to coalescence \cite{Kimura_2021, Chen_2023, Tagawa_2023, Tagawa_2023b}. The radiation counterpart is then energised when the kicked remnant penetrates the unperturbed disc and produces hyper-Eddington outflows. Unlike \cite{Kimura_2021}, who focuses on the outflow breakout emission, here we consider the long-term emission produced when the outflow drives disc ejection and the latter expands and cools outside the disc, resulting in the production of UV and optical flares (the breakout emission predicted in \cite{Kimura_2021} mainly falls in the soft X-ray bands).

\begin{figure}
   \centering
   \includegraphics[width=\hsize]{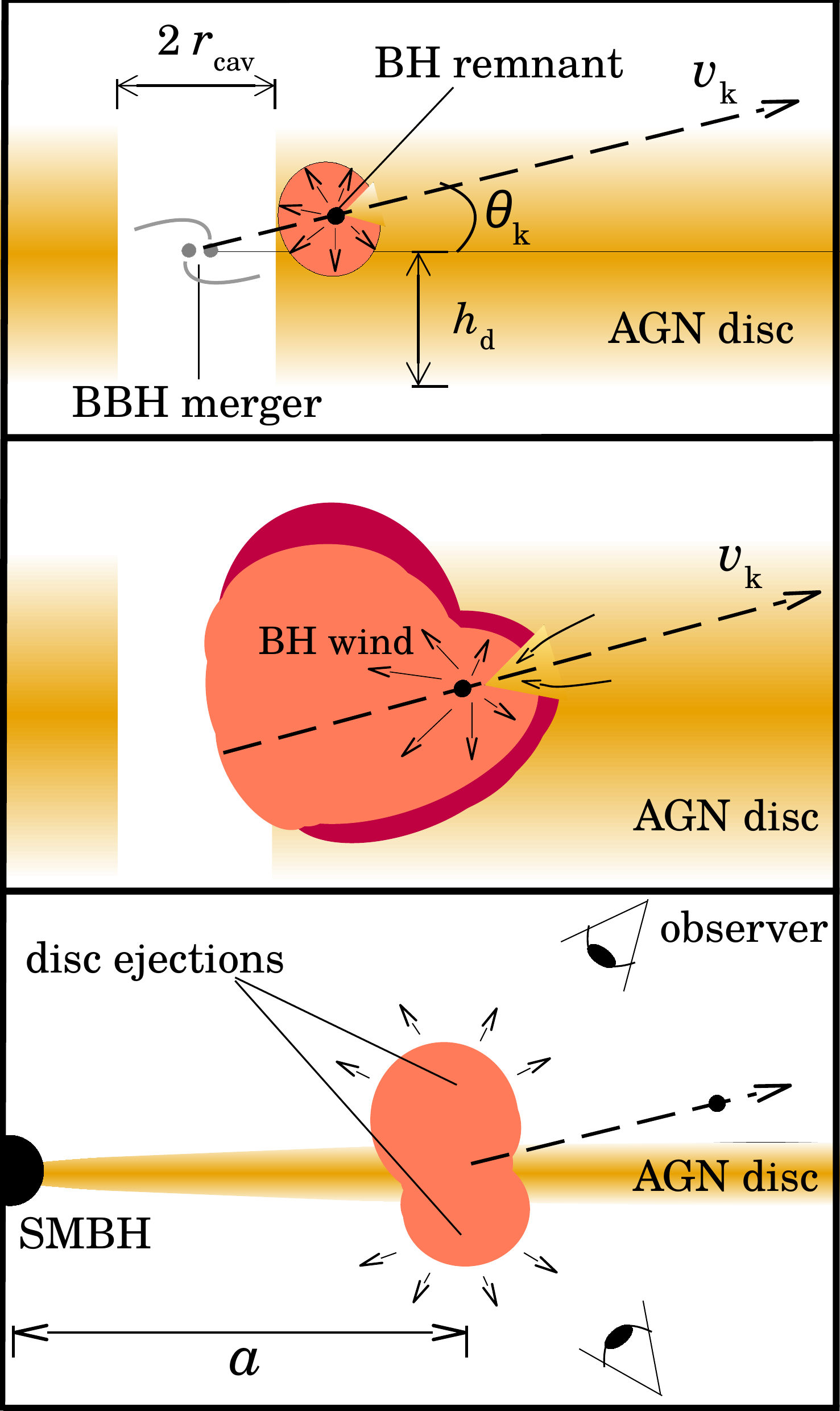}
      \caption{
Sketch of the EM counterpart scenario discussed in this work.
Upper: Radial cut of an AGN thin disc with vertical density stratification, where the BBH coalescence occurs within a pre-merger cavity. The merger remnant is kicked after coalescence and penetrates the unperturbed disc region.
Middle: While travelling within the AGN disc, the remnant experiences a highly super-Eddington inflow of gas, leading to the production of a quasi-isotropic wind that expels the disc material above and below the disc plane.
Lower: The plasma ejections produce an electromagnetic flare after expanding enough to allow the escape of internal photons.
              }
         \label{fig:sketch}
\end{figure}

This paper is organised as follows. In the next section, we derive an analytical model for the merger EM counterpart. In Section~\ref{sec:emission}, we present the spectra and optical light curves of the predicted emission and compare them with the sensitivity of optical instruments.
We analyse constraints on the binary $q-\chi_\mathrm{eff}$ correlation and the flare time delay in Section~\ref{sec:q_Xeff}.
Finally, we provide a summary and conclusions in Section~\ref{sec:conclusions}.

\section{Electromagnetic counterpart: analytical considerations}
\label{sec:analytic_model}

We consider a BBH of total mass $M_\bullet$, embedded in, and co-rotating with an AGN thin disc, at a distance $a$ from the central SMBH of mass $M_\mathrm{S}$. According to the analysis of \cite{Kimura_2021}, in this situation the BBH coalescence would likely occur within an under-density cavity of size scale equivalent to the Hill radius of the BBH system:
\begin{equation}
r_\mathrm{cav} \approx a \left(\frac{1}{3}\frac{M_\bullet}{M_\mathrm{S}}\right)^{1/3}.
\end{equation} 
Following such prediction and for analytical convenience, here we consider a BBH coalescence at the mid-plane of a thin disc, and within a cylindrical cavity of radius $r_\mathrm{cav}$, as depicted in Figure~\ref{fig:sketch}.

A burst of GWs is released at coalescence, and the merger remnant is born with a mass $\sim M_\bullet$\footnote{The mass of the remnant is typically slightly smaller than the mass of the BBH due to the release of gravitational waves at coalescence. Nevertheless, for practical purposes proposed in this work, we do not make a distinction between the masses of the BBH and the remnant.} with a likely gravitational kick velocity of magnitude $v_\mathrm{k}$ in the frame of the coalescence.
The magnitude and direction of this gravitational recoil is sensitive to the mass ratio $q$ and spins of the binary components previous to the merger \cite{Campanelli_2007,Lousto_2010,Lousto_2012,Varma_2022},
(we discuss the distribution of the kick magnitude and orientation in more detail in Section~\ref{sec:q_Xeff}), and is generally
highly supersonic with respect to the disc material.

\begin{figure}
   \centering
   \includegraphics[width=\hsize]{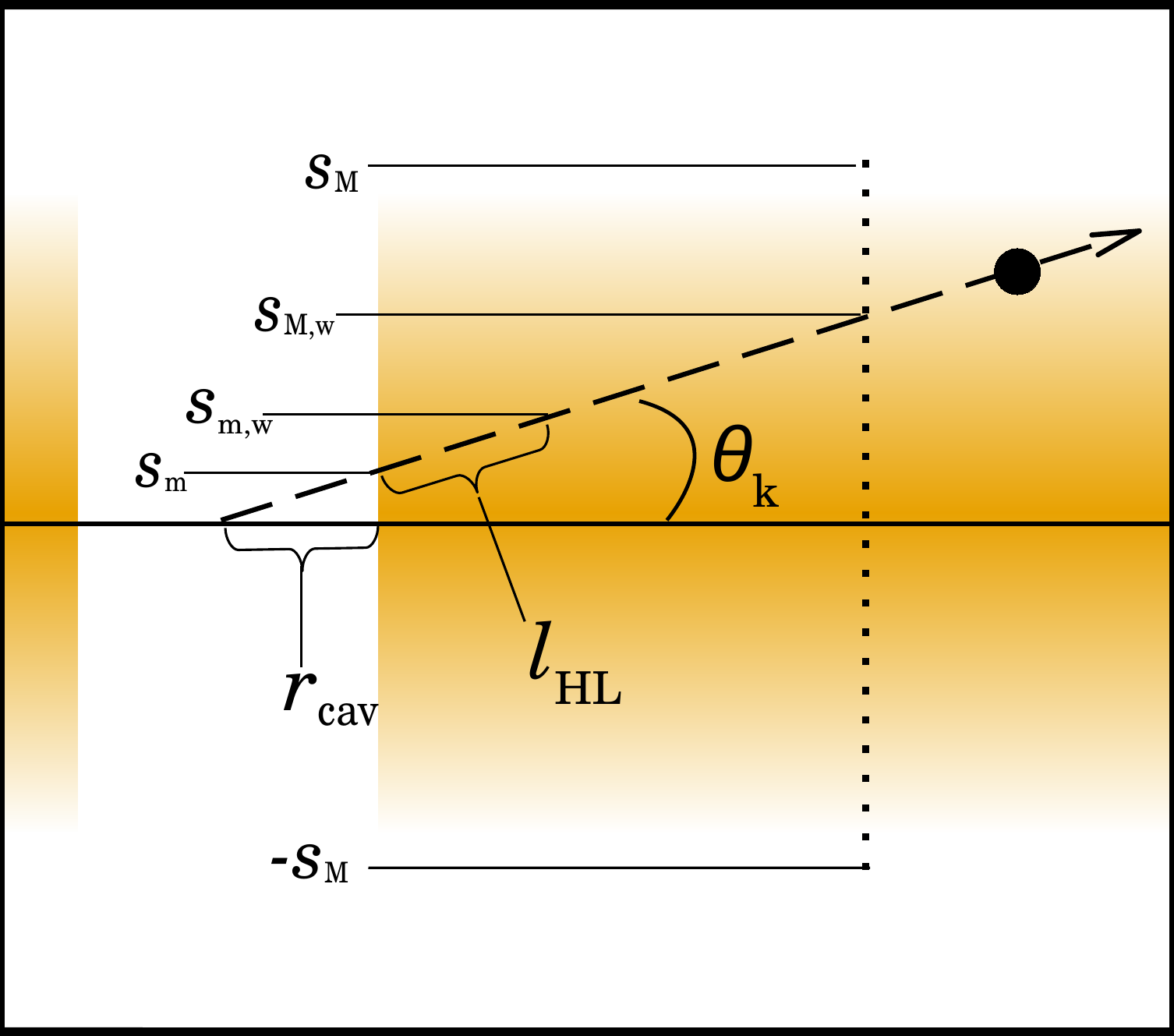}
      \caption{
Sketch illustrating the characteristic lengths and vertical heights $s_i$ defined in the text for calculating the disc mass expelled by the BBH merger remnant. Similar to the upper panel in Figure~\ref{fig:sketch}, the dashed thick line represents the trajectory of the merger remnant with angle $\theta_\mathrm{k}$ relative to the disc mid-plane.  
In the proposed scenario, the BH wind is ignited after the remnant travels the length $l_\mathrm{HL} = 2 G M_\bullet / c^2$ within the disc and is quenched upon reaching the height $s_\mathrm{M,w} = h_\mathrm{d}$. This transient wind expels the disc material located above and below the BH trajectory.  
The vertical dotted line indicates the wind quenching location along the BH trajectory and defines the volume of the ejected material in the proposed model (see the text).
              }
         \label{fig:sketch_l}
\end{figure}
On the other hand, the radial drift velocity of the disc matter is highly subsonic \cite{SS_1973}, and thus the cavity crossing time of the remnant $r_\mathrm{cav}/v_\mathrm{k}$ is much shorter than the timescale of cavity replenishment \cite{Kimura_2021}.
Therefore, the interaction of the remnant with the unperturbed disc (and thus the possible production of electromagnetic feedback) crucially depends on the kick direction $\theta_\mathrm{k}$ with respect to the disc plane. We expect no interaction and hence no radiation counterpart for remnants with $\theta_\mathrm{k} \rightarrow \pi/2$.
On the other hand, a significant interaction between the remnant and the unperturbed disc is expected for small $\theta_\mathrm{k}$ (see Figure~\ref{fig:sketch}), which is the case we focus on in this work.

A kicked remnant of mass $M_\bullet$ penetrating the disc volume perturbs the disc material due to its gravitational influence. Such interaction can be initially assessed with Bondi-Hoyle-Lyttleton (BHL) theory. In a medium of gas density $\rho_\mathrm{d}$, the classical BHL inflow onto the kicked remnant is approximately $\dot{M}_\mathrm{B} \sim 4\pi\rho_\mathrm{d} (G M_\bullet)^2 / v_\mathrm{k}^3$. Normalising this inflow rate by the Eddington accretion limit of the remnant, gives for typical parameters of the problem:
\begin{equation}
\frac{\dot{M}_\mathrm{B}}{\dot{M}_\mathrm{Edd}}
\sim \; 2 \times 10^4  
\left(
\frac{n_\mathrm{d}}{10^{14}\mathrm{cm}^{-3}}
\right) 
\left(
\frac{M_\bullet}{50\mathrm{M}_\odot}
\right)
\left(
\frac{400\mathrm{km/s}}{v_\mathrm{k}}
\right)^{3},
\
\end{equation}
where $n_\mathrm{d}$ is the disc number density.
Under such highly super-Eddington inflow, the remnant is expected to develop radiation-driven winds and/or bipolar collimated jets. Recent general relativistic magnetohydrodynamic (GRMHD) simulations of BHL accretion performed by \cite{Kaaz_2023} show the production of Blandford-Znajek jets \cite{Blandford_1977} by spinning black holes (BHs) with a fixed spin direction, when traveling through a medium with plasma beta $\beta \gtrsim 10$. It is not clear, however, what the AGN disc's $\beta$ is where BBHs likely coalesce. Furthermore, it is not clear whether the BBH coalescence produces a remnant with a fixed spin direction. BHs with precessing spin may launch precessing jets that could globally resemble wide-angle bipolar outflows.
Given the aforementioned uncertainties for the production of powerful collimated jets, in the present work we conservatively consider that the remnant drives a 
wide-angle, bipolar, transient wind (i.e. a quasi-spherical outflow) during its trajectory within the disc. We assume that the remnant captures matter from the environment while simultaneously driving outflows. These inflow-outflow processes have been analysed in the context of dynamical friction of travelling black holes (BHs) with feedback \cite{Gruzinov_2020, Li_2020}, and to explain the emission of ultra-luminous X-ray sources as supercritical accretors \cite{Fabrika_2021, Abaroa_2023}.

The outcome of the remnant-disc interaction depends on the properties of the disc surrounding the merger, such as gas density, temperature, and disc thickness, which vary with the radial distance from the AGN central engine. One of the most employed models for describing AGN thin discs are those of Shakura \& Sunyaev \cite{SS_1973}, Sirko \& Goodman \cite{SG_2003}, and Thompson, Murray, \& Quataert \cite{TQM_2005} (hereafter SS73, SG03, and TQM05, respectively).
The SG03 model extends the pioneering work of SS73 by incorporating  heating mechanisms to prevent material collapse due to gravitational instabilities at the outer disc regions. The TQM05 model address this by implementing mass loss due to star formation feedback (see \cite{Gangardt_2024} for a detailed comparison).

Considering the characteristics of the aforementioned disc models, in the present study we focus on AGN discs with relatively low accretion rates and following the SG03 description, for the following reasons.
The SG03 model allows for a broader exploration of merger locations compared to SS73, whose validity is limited to regions where the disc remains gravitationally stable.
While the TQM05 model provides a more accurate description for AGNs with high accretion rates, such systems are expected to produce substancial disc feedback. This feedback introduces additional sources of emission enhancement and flaring, complicating the detection of the EM counterparts discussed here.
Finally, for AGNs with SMBHs ranging within $10^{6-8}$ M$_\odot$ (including most SMBHs in the local Universe), the TQM05 model predicts significantly lower gas densities and smaller aspect ratios compared to the SG03 model, as shown in Appendix~\ref{app:dmods}.
This suggests that the emission mechanism discussed here would produce weaker electromagnetic signatures in TQM05 discs than in SG03 discs.

At a given radial location $a$ in the disc, we consider the matter density as vertically stratified following a Gaussian profile
(which is a suitable hydrostatic solution for the vertical disc structure):

\begin{equation}
\label{rho_vertical}
\rho_\mathrm{d}(a,s) = \rho_\mathrm{c}(a)
\exp
\left\{ 
-\frac{1}{2} \left[\frac{s}{h_\mathrm{d}(a)}\right]^2
\right\},
\end{equation}
being $\rho_c$ the density at the disc mid-plane and $h_\mathrm{d}$ the disc scale height, which depend on the radial location $a$.

In the present analysis, we asses the inflow onto the BH remnant based on the classical BHL theory. 
Therefore we assume this approach as valid when 
\begin{equation}
R_\mathrm{HL}=\frac{2GM_\bullet}{v_\mathrm{k}^2}<h_\mathrm{d}.
\label{RHL_hd_cond}
\end{equation}
Once the remnant enters the disc, it takes a period to capture enough material to ignite the wind. We estimate this period as the crossing time of the BHL radius in the source frame:
\[
\Delta t'_\mathrm{HL} = \frac{R_\mathrm{HL}}{v_\mathrm{k}} = \frac{2GM_\bullet}{v_\mathrm{k}^3}.
\]
Therefore, given the kick angle 
$\theta_\mathrm{k}$ with respect to the disc mid-plane,
the BH wind is ignited when the remnant is located at the height: 
\begin{equation}
\label{smw}
s_\mathrm{m,w} = \tan\theta_\mathrm{k} 
r_\mathrm{cav} + \sin\theta_\mathrm{k}R_\mathrm{HL}.
\end{equation}
above the mid-plane (see Figure~\ref{fig:sketch_l}). The wind is then quenched when the BH remnant captures insufficient inflow material to fuel the outflow.
We approximate this quenching point when the remnant is located at the height:
\begin{equation}
\label{sMw}
s_\mathrm{M,w}=h_\mathrm{d},
\end{equation}
above the mid-plane.
At this height, the disc density has diminished to $60\%$ of its mid-plane value (see equation~\ref{rho_vertical}).
At vertical locations $s> h_\mathrm{d}$, the BH can still experience inflow of disc material and produce outflows (although this feedback weakens as the disc gas density drops). Nevertheless, at $s> h_\mathrm{d}$, the disc atmosphere is strongly stratified (see equation~\ref{rho_vertical}), and the classical BHL theory (formulated for a massive body travelling within an external, uniform medium) does not apply.
Given these conditions, we restrict our analysis to an outflow injected 
within the heights $s_\mathrm{m,w}$ and $s_\mathrm{M,w}$
(see Figure~\ref{fig:sketch_l}). Our analysis is then valid when $s_\mathrm{m,w}<s_\mathrm{M,w}$, which, using equations (\ref{smw}) and (\ref{sMw}), results in the following condition:
\begin{equation}
\tan\theta_\mathrm{k} \frac{r_\mathrm{cav}}{h_\mathrm{d}}+
\sin\theta_\mathrm{k}\frac{R_\mathrm{HL}}{h_\mathrm{d}}
<1.
\label{smw_sMw_condition}
\end{equation}

The energy injected by the BH produces
two expansive ejections of plasma above and below the accretion disc as depicted in Figure~\ref{fig:sketch} (bottom panel).
In this work, we consider the emission produced by such disc eruptions as the electromagnetic counterpart to the preceding merger gravitational waves (GWs), and we approximate the associated light curves (LCs) as those of explosive blast waves, similar to approaches in \cite{Pihajoki_2016, Rodriguez-Ramirez_2020, Rodriguez-Ramirez_2024}.
In Subsection~\ref{subsec:ejections}, we estimate the amount of matter and energy in the disc ejections, and in Subsection~\ref{subsec:emission_of_ejection} we assess their emission.

\subsection{The energy and mass of the disc ejections}
\label{subsec:ejections}

During its passage through the disc, the remnant drives a transient outflow which we refer to as the "BH wind". This BH wind expels disc matter that emerges from the disc on the same and opposite sides relative to the direction where the remnant leaves the disc (see Figure~\ref{fig:sketch}, bottom). These ejections are referred to as the "upper" and "lower" eruptions, respectively.

Following \cite{Wang_2021a, Wang_2021b}, we parameterise the kinetic power of the BH wind as
\begin{equation}
\label{LwLE}
L_\mathrm{w} = \eta_\mathrm{w} \dot{M}_\mathrm{in} c^2,
\end{equation} 
where $\eta_\mathrm{w} \ll 1$ is taken as an efficiency parameter and $\dot{M}_\mathrm{in}$ is taken as a fraction $\eta_\mathrm{in} < 1$ of the classical BHL accretion rate: 
\begin{equation}
\dot{M}_\mathrm{in} = \eta_\mathrm{in} 4\pi\rho_\mathrm{d} \frac{(GM_\bullet)^2}{(c_\mathrm{s}^2 + v_\mathrm{k}^2)^{3/2}},
\end{equation}
where $\rho_\mathrm{d}$ and $c_\mathrm{s}$ are the local density and speed of sound in the disc, which we obtain using the SG03 disc model.

The wind kinetic power can also be expressed as
\begin{equation}
\label{LwGw}
L_\mathrm{w} = (\Gamma_\mathrm{w} - 1 ) \dot{M}_\mathrm{w} c^2,
\end{equation}
where $\dot{M}_\mathrm{w}$ and $\Gamma_\mathrm{w}$ are the mass loss rate and the terminal Lorentz factor of the wind flow, respectively. Following \cite{Fukue_2004}, we consider that the remnant grows at the critical accretion rate $\dot{M}_\mathrm{cr} \equiv \eta\dot{M}_\mathrm{Edd} = L_\mathrm{Edd}/c^2$ with $\eta=0.1$, and the excess mass inflow is expelled to form the wind. Since the rate of mass inflow is typically highly super-Eddington in the present problem, the wind mass loss rate can then be approximated as $\dot{M}_\mathrm{w} \approx \dot{M}_\mathrm{in}$. Under this consideration, the wind terminal velocity $v_\mathrm{w}$ is obtained by combining Equations (\ref{LwLE}) and (\ref{LwGw}), giving:
\begin{equation}
\Gamma_\mathrm{w} = 1 + \eta_\mathrm{w},
\end{equation}
and hence
\begin{equation}
v_\mathrm{w} = c \left(1 - \Gamma_\mathrm{w}^{-2}\right)^{1/2} \approx c \sqrt{2 \eta_\mathrm{w}},
\end{equation}
where the last approximation is valid when $\eta_\mathrm{w} \ll 1$.

We assume equipartition of the total wind power for energising the upper and lower ejections of matter, which is a reasonable approximation for small kick angles $\theta_\mathrm{k}$. 
Thus, we calculate the energy stored by the BH wind in each ejection of plasma as:
\begin{equation}
\label{E0}
E_0 = \frac{1}{2}\int dt L_\mathrm{w} = 
\frac{\eta_\mathrm{w}\eta_\mathrm{in}2\pi(GM_\bullet c)^2}{\sin\theta_\mathrm{k} v_\mathrm{k}(c_\mathrm{s}^2 +v_\mathrm{k}^2)^{3/2}} \Sigma_\mathrm{E},
\end{equation}
with
\begin{equation}
\Sigma_\mathrm{E} \equiv \int_{s_\mathrm{m,w}}^{s_\mathrm{M,w}} ds
\rho_\mathrm{d}(a,s),
\end{equation}
where the vertical stratification follows equation \ref{rho_vertical} and the local speed of sound is constant along the vertical direction.
An order of magnitude estimation for the energy $E_0$ calculated above can be evaluated considering some mean constant density 
$\bar{\rho}_\mathrm{d}$ and approximating 
$(s_\mathrm{M,w}-s_\mathrm{m,w})/\sin\theta_\mathrm{k}\approx h_\mathrm{d}$, giving:
\begin{align}
\nonumber
E_\mathrm{0} & \approx 
2\pi\eta_\mathrm{w}\eta_\mathrm{in} 
\bar{\rho}_\mathrm{d} h_\mathrm{d}
(G M_\bullet c)^2/v_\mathrm{k}^4
\sim 1.561 \times10^{49}\, \mathrm{erg}
\\
\nonumber
&\times\left(
\frac{\eta_\mathrm{w}}{0.05}
\right)
\left(
\frac{\eta_\mathrm{in}}{0.1}
\right)
\left(
\frac{M_\bullet}{20 \mathrm{M}_\odot}
\right)^{2}
\left(
\frac{v_\mathrm{k}}{200\,\mathrm{km s}^{-1}}
\right)^{-4}
\\
&\times 
\left(
\frac{M_\mathrm{S}}{10^7\,\mathrm{M}_\odot}
\right)
\left(
\frac{a/R_\mathrm{g}}{10000}
\right)
\left(
\frac{h_\mathrm{d}/a}{0.005}
\right)
\left(
\frac{\bar{n}_\mathrm{d}}{10^{14}\,\mathrm{cm}^{-3}}
\right).
\label{E_0ll}
\end{align}


The disc ejections are composed of material from the BH wind and
the matter it sweeps up from the disc environment.
Ultimately, the wind material is taken from the disc itself through the BHL inflow (see Figure~\ref{fig:sketch}).
Therefore, we consider the mass of the upper and lower ejections
to originate in the regions of the disc above and below the BH trajectory
(dashed line in Figure~\ref{fig:sketch_l}),
up to the point where we assume the wind is quenched (dotted line in Figure~\ref{fig:sketch_l}; see also equation~\ref{sMw}).
As previously stated, we model the quasi-spherical BH wind as bipolar outflows with a wide opening half-angle $\theta_\mathrm{w}$. 
We then assess the masses of the upper and lower ejections as the material enclosed within the volume generated by the upper and lower triangles of half-opening angles $\theta_\mathrm{w}$, translated along the BH trajectory from the disc entry point to the wind quenching point (horizontally bounded by the cavity-disc interface and the vertical dotted line sketched in Figure~\ref{fig:sketch_l}).
It can be straightforwardly shown that these upper and lower volumes $V_0^\pm$ are given by:
\begin{align}
\label{V0p}
V_0^{+} &=
\int_{s_\mathrm{m}}^{s_\mathrm{M,w}} ds A_\triangle^+(s)
+ \int_{s_\mathrm{M,w}}^{s_\mathrm{M}}ds A^+(s)\\
\label{V0m}
V_0^{-} &= 
\int_{-s_\mathrm{M}}^{s_\mathrm{m}}ds A^-(s) 
+ \int_{s_\mathrm{m}}^{s_\mathrm{M,w}} ds A_\triangle^-(s)
\end{align}
and the corresponding enclosed masses are:
\begin{align}
\label{M0p}
M_0^{+} &=
\int_{s_\mathrm{m}}^{s_\mathrm{M,w}} ds A_\triangle^+(s)\rho_\mathrm{d}(s)
+ \int_{s_\mathrm{M,w}}^{s_\mathrm{M}}ds A^+(s)\rho_\mathrm{d}(s)\\
\label{M0m}
M_0^{-} &= 
\int_{-s_\mathrm{M}}^{s_\mathrm{m}}ds A^-(s)\rho_\mathrm{d}(s)  
+ \int_{s_\mathrm{m}}^{s_\mathrm{M,w}} ds A_\triangle^-(s)\rho_\mathrm{d}(s)
\end{align}
where
\begin{equation}
\begin{bmatrix}
A^+ \\
A_\triangle^+ \\
A_\triangle^- \\
A^-
\end{bmatrix}
=\frac{\tan\theta_\mathrm{w}}{\tan\theta_\mathrm{k}}
\begin{bmatrix}
(s_\mathrm{M,w}-s_\mathrm{m})(2s-s_\mathrm{M,w}-s_\mathrm{m}) \\
(s-s_\mathrm{m})^2 \\
(s_\mathrm{M,w}-s)^2 \\
(s_\mathrm{M,w}-s_\mathrm{m})(s_\mathrm{m}+s_\mathrm{M,w}-2s)
\end{bmatrix}.
\end{equation}
We limit the upper and lower volumes vertically at $\pm s_\mathrm{M}=\pm2h_\mathrm{d}$.
This choice for the upper and lower limits is motivated by the fact that
at such heights, the disc density has dropped substantially (to $\sim10\%$ of the mid-plane density), and at the same time
such vertical heights are attained by the wind shock front in 
shorter time scales compared to the total crossing time of the remnant
(this is further detailed by equation \ref{At_bo} and Figure~\ref{fig:AtComps}).
Summarising, the characterictic heights employing to estimate mass and energy stored in the disc eruption (see equations \ref{E0}, \ref{V0p}-\ref{M0m}) are
$s_\mathrm{m}=\tan\theta_\mathrm{k}r_\mathrm{cav}$, $s_\mathrm{m,w}=\tan\theta_w r_\mathrm{cav}+\sin\theta_\mathrm{k}R_\mathrm{HL}$, $s_\mathrm{M,w}=h_\mathrm{d}$, $s_\mathrm{M}=2 h_\mathrm{d}$.

An order of magnitude estimation for the volumes $V_0^\pm$ can be obtained approximating them as half of the  cylinder's volume of 
diameter $\frac{h_\mathrm{d}}{\tan\theta_\mathrm{k}} -r_\mathrm{cav}$
and height $s_\mathrm{M,w}=2h_\mathrm{d}$ (see Figure~\ref{fig:sketch_l}):
\begin{equation}
V^\pm_0\approx \frac{\pi}{2}
\left(
\frac{h_\mathrm{d}}{\tan\theta_\mathrm{k}} - r_\mathrm{cav}
\right)^2 h_\mathrm{d}\approx
\frac{\pi h_\mathrm{d}^3}{2\theta_\mathrm{k}^2},
\label{V0_approx}
\end{equation}
where the second approximation 
holds for small angles $\theta_\mathrm{k}$, which is the case of interest.
The last expression in equation~\ref{V0_approx}, clearly over-estimates the
volumes given by equations \ref{V0p}-\ref{V0m}, but can be useful for 
a quick order or magnitude evaluation. Normalising the simplified estimation \ref{V0_approx}
by $R_g^3$ 
($R_\mathrm{g}=GM_\mathrm{S}/c^2$, the gravitational radius of the central SMBH)
one obtains:
\begin{equation}
\frac{V_0}{R_\mathrm{g}^3} \approx 6.446\times10^6
\left(\frac{10^\circ}{\theta_\mathrm{k}} \right)^2 
\left(\frac{h_\mathrm{d}/a}{0.005} \right)^3 
\left(\frac{a/R_\mathrm{g}}{10^4} \right)^3.
\label{V0_Rg3}
\end{equation}

\subsection{The emission of the disc ejections}

Following \cite{Ivanov_1998, Pihajoki_2016, Rodriguez-Ramirez_2024}, we approximate  the disc ejections as spherical blast waves in homologous expansion \cite{Arnett_1980, Arnett_1996, Chatzopoulos_2012}. 
The initial radii of such spheres $R_0^\pm$ are set so that they contain the volumes of the ejected material (see quations~\ref{V0p}-\ref{V0m}):
\begin{equation}
R^\pm_0 = \left(\frac{3V_0^\pm}{4\pi}\right)^{1/3},
\end{equation}
where the superscript ``$\pm$'' labels quantities corresponding to the upper (+) and lower (-) ejections.
Since the gas density drastically drops outside the disc, we assume the external radii of the spheres to undergo free expansions:
\begin{equation}
\label{R}
R^\pm(t') = R^\pm_0 + u_0^\pm (t' - t'^\pm_0),
\end{equation}
where $u_0^\pm$ is the expansion velocity and $t'$ is the elapsed time in the AGN's frame, being $t' = 0$ the moment of the coalescence. The radii given by equation~\ref{R} is defined for $t' \geq t'^\pm_0$, and we estimate the initial expansion time as:
\begin{equation}
t'^\pm_0 = \Delta t'_\mathrm{cav} + \Delta t'_\mathrm{HL} + \Delta t'^\pm_\mathrm{bo},
\label{t_0}
\end{equation}
where $\Delta t'_\mathrm{cav} = r_\mathrm{cav} / (v_\mathrm{k} \cos{\theta_\mathrm{k}})$ is the period in which the remnant enters the unperturbed disc, $\Delta t'_\mathrm{HL} = 2GM_\bullet / v_\mathrm{k}^3$ is the elapsed time before the BH wind is ignited, and $\Delta t'^\pm_\mathrm{bo}$
is the time in which the disc material is expelled above en bellow the disc.
We estimate the $\Delta t'^\pm_\mathrm{bo}$
period as the time the BH wind takes to reach the upper and lower boundaries $\pm s_\mathrm{M}$.
Employing the solution of Waever et al. \cite{Weaver_1977}
for the evolution of a stellar wind bubble 
$R_w = (\frac{250}{308\pi})^{1/5}L_\mathrm{w}^{1/5}\rho_\mathrm{ext}^{-1/5} t^{3/5}$ and
taking the density of the external medium $\rho_\mathrm{ext}$ as  the  maximum density within the  upper and lower break out distances, this results in:
\begin{align}
\label{At_bo}
\nonumber \Delta t'^\pm_\mathrm{bo}= & 
\, v_\mathrm{k}
\left[\frac{250}{77} \eta_\mathrm{w} \eta_\mathrm{in} (G M_\bullet c)^2 \right]^{-1/3}
\\
& \times \begin{cases}
(s_\mathrm{M} - s_\mathrm{m,w})^{5/3}, 
 \\ \\
(\rho_\mathrm{c}/\rho_\mathrm{m,w})^{1/3}(s_\mathrm{M} + s_\mathrm{m,w})^{5/3}. 
  \end{cases}
\end{align}

When the BH wind is quenched, two ejections of disc material expand above and bellow the disc.
To obtain the expansion velocity of such expelled material, we assume equipartition of their total energy (see equation \ref{E0}) among thermal and kinetic energy, i.e., $E^\pm_\mathrm{0,th} = E^\pm_\mathrm{0,kin} = E_0 / 2$. Thus, the expansion velocity of the ejected material is taken as:
\begin{equation}
u_0^\pm = \left(\frac{E_0}{M_0^\pm}\right)^{1/2},
\label{u0}
\end{equation}
where $E_0$ and $M_0^\pm$ are given by equations (\ref{E0}) and (\ref{M0p}-\ref{M0m}), respectively.

As we show in Appendix \ref{app:radp}, the gas of the plasma ejections is initially radiation pressure dominated. In this case, the equation of state is approximated as $P_0 = \frac{1}{3}a_\mathrm{r}T_0^4$, with $a_\mathrm{r}$ being the radiation constant and $T_0$ the initial temperature. On the other hand, assuming the idealised situation in which the plasma ejections have a homogeneous distribution of matter, their thermal energy and gas pressure can be related as $P_0V_0 = E_\mathrm{0,th}(\gamma_a - 1)$, with $\gamma_a = 4/3$ being the adiabatic index appropriate for a radiation-dominated gas. Thus, we calculate the initial temperature
of the plasma ejections as:
\begin{equation}
T_0^\pm = \left( \frac{E_\mathrm{0}}{2 a_\mathrm{r}V_0^\pm} \right)^{1/4}.
\end{equation}

To illustrate an order of magnitude estimation for typical values of the temperature $T_0^\pm$, we the use energy and initial volume of the plasma given by equations \ref{E_0ll} and \ref{V0_approx}, giving:
\begin{align}
\nonumber
&T_0^{+}\approx 
\frac{1}{v_\mathrm{k}}
\left(
\frac{\theta_\mathrm{k}GM_\bullet c}{h_\mathrm{d}}
\right)^{1/2}
\left(
\frac{2\eta_\mathrm{w}\eta_\mathrm{in}\rho_\mathrm{d}}{a_\mathrm{r}}
\right)^{1/4} =
1.184 \times 10^{5} \,\mathrm{K}\\
\nonumber
& \times \left(
\frac{200\,\mathrm{km s}^{-1}}{v_\mathrm{k}}
\right)
\left(
\frac{0.005}{h_\mathrm{d}/a}
\right)^{1/2}
\left(
\frac{5000}{a/R_\mathrm{g}}
\right)^{1/2}
\left(
\frac{10^7\, \mathrm{M}_\odot}{M_\mathrm{S}}
\right)^{1/2} \\
&\times\left(
 \frac{\theta_k}{10^\circ}
\right)^{1/2}
\left(
\frac{M_\bullet}{20 \,\mathrm{M}_\odot}
\right)^{1/2}
\left(
\frac{\eta_\mathrm{w}}{0.05}
\right)^{1/4}
\left(
\frac{\eta_\mathrm{in}}{0.1}
\right)^{1/4}
\left(
\frac{ n_\mathrm{i}}{10^{14}\, \mathrm{cm}^{-3}}
\right)^{1/4}.
\label{T0approx}
\end{align}

For simplicity, we do not calculate the rising emission produced by the disc eruptions and assume that photons remain trapped until their diffusion timescale matches the expansion dynamical time, which occurs over the period \cite{Arnett_1996,Chatzopoulos_2012}:
\begin{equation}
\label{Atph}
\Delta t'^\pm_\mathrm{ph}= \sqrt{\frac{9\kappa_\mathrm{T} M_0^\pm }{4\pi^3 c u_0^\pm}},
\end{equation}
after the expansion is initiated, being
$\kappa_\mathrm{T}$ the Thompson's cross section for electron  scattering.
The bolometric luminosity of the plasma ejection is then calculated as \cite{Arnett_1980, Arnett_1996, Chatzopoulos_2012}:
\begin{equation}
\label{Lt}
L^\pm(t') = 
\begin{cases}
  L_0^\pm \exp\left\{- \frac{t'^2}{{t^\pm_g}^2} - \frac{2R^\pm_0 t'}{u^\pm_0 {t^\pm_g}^2} \right\} & \;\mathrm{for}\; t'\geq t'_0 + \Delta t'^\pm_\mathrm{ph}, \\
  \\
0, & \;\mathrm{for}\; t'< t'_0 + \Delta t'^\pm_\mathrm{ph},
  \end{cases}
\end{equation}
being
\begin{equation}
\label{L0}
L_0^\pm =  \frac{a_\mathrm{r}c(2\pi T^\pm_0 R^\pm_0)^4}{27\kappa_\mathrm{T} M_0^\pm}, 
\end{equation}
the maximum bolometric luminosity, 
$ t^\pm_\mathrm{g} = \sqrt{2} \Delta t'^\pm_\mathrm{ph}$, and  $a_\mathrm{r}$ the radiation constant.

The observed spectrum is calculated as black body emission of a spherical volume with the effective temperature
\begin{equation}
T^\pm_\mathrm{eff}(t') =
\left[
\frac{L^\pm(t')}{4\pi \sigma_\mathrm{SB} R^\pm(t')^2}
\right]^{1/4},
\label{Teff}
\end{equation}
where $\sigma_\mathrm{SB}$ is the Stefan-Boltzmann constant. Thus, the flux density of the plasma ejections at the time $t$ and frequency $\nu$ in the observer's frame (at Earth) is:
\begin{equation}
\label{EMflux}
\nu F^\pm_\nu(t) = \nu'
\pi \left[\frac{R^\pm(t')}{D_\mathrm{L}}\right]^2
B_{\nu'}\left[T^\pm_\mathrm{eff}(t')\right],
\end{equation}
where $z$ and $D_\mathrm{L}$ are the redshift and distance of the source, $t' = t / (1+z)$ and $\nu' = (1+z) \nu$ are the time and radiation frequency in the AGN's frame, and $B_{\nu'}$ is the black body spectral radiance.

\label{subsec:emission_of_ejection}

\begin{figure}
   \centering
   \includegraphics[width=\hsize]{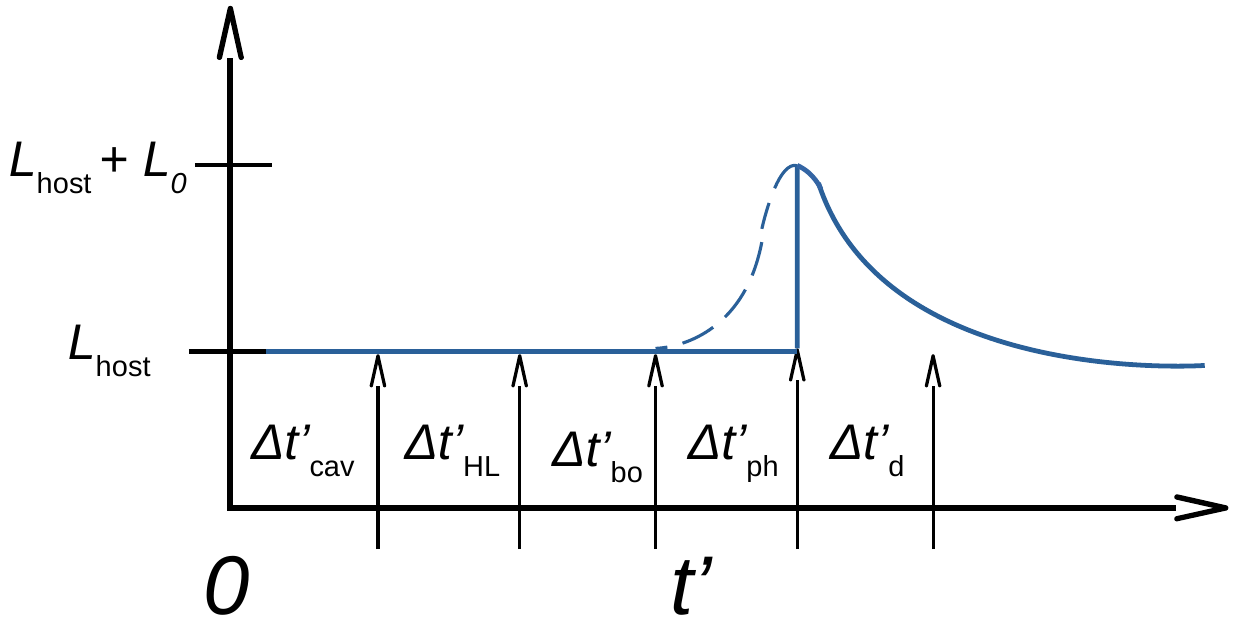}
      \caption{
Schematic representation of the sequence of periods (not to scale) that encompasses the flare counterpart to a BBH merger in the LC model discussed here (solid curve). The $y$-axis is the bolometric luminosity of the LC, with $L_0$ being the maximum bolometric luminosity of the BBH-driven flare (see equation \ref{L0}). The $x$-axis is the elapsed time after the GW event in the source frame. The dashed curve represents the rise of the flare, not modelled in the present approach (see the text).
}
         \label{fig:LC_sketch}
\end{figure}

\label{subsec:tdelay}
In this approach, the time delay in the observer's frame between the GW event and the appearance of the flare is then:
\begin{equation}
\label{Atl}
\Delta t^\pm_\ell = (1+z)\left(t'_0 +\Delta t'^\pm_\mathrm{ph}\right),
\end{equation}
where $t_0$ is the starting time of the ejection's expansion in the AGN frame estimated by equation (\ref{t_0}), and the period $\Delta t'^\pm_\mathrm{ph}$ is given by equation (\ref{Atph}). 
In the observer's frame, the bolometric luminosity decays in the characteristic period (see equation~\ref{Lt}):
\begin{equation}
\Delta t^\pm_\mathrm{d} = (1+z)\sqrt{\frac{9\kappa_\mathrm{T} M_0^\pm}{2\pi^3 c u_0^\pm}}.
\label{Atdure}
\end{equation}
which we consider as the duration of the flare.
Figure~\ref{fig:LC_sketch} presents a pictorial illustration for the period components of  counterpart flare discussed in this section.

\subsection{The model parameter space}
\label{sec:pspace}

The EM counterparts discussed in this paper are attributed to ejections of plasma driven by a kicked remnant of a BBH merger within an AGN disc. We approximate the thermal emission of such disc eruptions as that of spherical blast waves in homologous expansion \cite{Arnett_1980, Chatzopoulos_2012}. The observed emission corresponds to the photosphere flux given by equation \ref{EMflux}, which in the formulation presented here is evaluated by specifying:
    the distance of the source $D_\mathrm{L}$,
    the mass $M_\mathrm{S}$ and accretion rate $\dot{M}_\mathrm{S}$ of the AGN's central engine,
    the radial distance of the coalescence $a$,
    the mass of the merger remnant $M_\bullet$,
    the angle $\theta_\mathrm{k}$ and velocity $v_\mathrm{k}$ of the post-merger gravitational kick,
    the BH wind opening angle $\theta_\mathrm{w}$,
    and the inflow and outflow efficiencies $\eta_\mathrm{in}$ and $\eta_\mathrm{w}$, respectively, of the BH remnant.
In what follows, we suggest ranges for the aforementioned parameters, suitable for the assumptions on which the present emission model is built.

The source distance of interest ranges within $D_\mathrm{L} \sim [300, 3000]$ Mpc, which is compatible with the distances of the BBH coalescences observed during the first three runs of Advanced LIGO \cite{Mandel_2022} and with the distances of AGNs potentially observed with current and forthcoming telescopes. 

We consider SMBHs with masses within the range of $10^{6-8}$ M$_\odot$ as such SMBHs dominate the mass function for AGNs in the local Universe \cite{Greene_2007, Li_2011}.
We are interested in EM counterparts with luminosities comparable to or larger than that of the hosting AGN. Such emissions could occur more effectively in thin discs with relatively low accretion rates, and thus we consider systems with $\dot{m}_\mathrm{S} \equiv \dot{M}_\mathrm{S}/\dot{M}_\mathrm{S,Edd} \sim [0.005, 0.1]$. Below the lower limit of this range the disc experiences a transition to a geometrically thick accretion flow  \cite{Narayan_2008}, limiting the application of the present analysis.

The location in the disc where mergers are likely to occur is not well constrained; thus, we consider merger locations ranging within $a \sim 5 \times 10^{2-4}  R_\mathrm{g}$, consistent with results from previous analyses \cite{Tagawa_2020, Grishin_2024}. The remnant masses of interest fall within $M_\bullet \sim [10, 150]$ M$_\odot$, which encompasses 
the inferred remnant masses of previous LVK detections \cite{Mandel_2022}. 

According to the present analysis, the post-merger kick angle should be small enough to allow sufficient interaction of the remnant with the unperturbed disc region. Thus, we suggest kick angles within $\theta_\mathrm{k} \sim [5^\circ, 15^\circ]$. 
The kick velocities in the frame of the coalescence 
could be as large as $v_\mathrm{k} \sim 2000$ km s$^{-1}$ \cite{Gonzalez_2007, Campanelli_2007, Lousto_2012, Varma_2022}.
Nevertheless, kicks within the previous $\theta_\mathrm{k}$ range
could be limited to $v_\mathrm{k} \sim [100, 400]$ km s$^{-1}$ (assuming a merger with orbital plane parallel to the disc plane, see Section~\ref{sec:q_Xeff}), which is the range we consider here.

Recent hydrodynamical simulations of BBHs embedded within AGN discs conducted by
\cite{Dittmann_2024} and \cite{Calcino_2024},
exhibit inflows onto the binaries of about 8\% of the classical Bondi rate.
However, the nature of the remnant's inflow analysed here differs from these previous studies 
as we examine accretion onto a single BH moving at supersonic speeds (then undergoing ``wind accretion'')
and producing outflows.
Thus, for the inflow efficiency parameter, we propose the range of $\eta_\mathrm{in}\sim[0.05,0.5]$, which encompasses the results of \cite{Dittmann_2024} and \cite{Calcino_2024}.

The outcome of hyper-Eddington inflows onto black holes is poorly studied. Conservatively, here we consider a wind efficiency ranging within $\eta_\mathrm{w} \sim [0.01, 0.1]$, consistent with previous studies on BH super winds \cite{Wang_2021a, Abaroa_2023}. In addition, a given  parameter configuration within the intervals discussed above must fulfil the conditions (\ref{RHL_hd_cond}) and (\ref{smw_sMw_condition}), i.e.,
\begin{equation}
\boldsymbol{\max} \left\{ R_\mathrm{HL}\,,
\,\tan\theta_\mathrm{k} r_\mathrm{cav} + 
\sin\theta_\mathrm{k}R_\mathrm{HL}
\right\} < h_\mathrm{d} .
\label{condition}
\end{equation}
to produce valid emission solutions within the present approach.

When employing the present emission model to interpret flares of a particular AGN considering a given GW event, the quantities $D_\mathrm{L}$, $M_\mathrm{S}$, $\dot{M}_\mathrm{S}$, and $M_\bullet$ can be constrained by observations. In this case, the possible predicted flares have six free parameters, namely, $a$, $v_\mathrm{k}$, $\theta_\mathrm{k}$, $\theta_\mathrm{w}$, $\eta_\mathrm{in}$, and $\eta_\mathrm{w}$.

In the next section, we illustrate spectral energy distributions and LC profiles using parameter configurations within the aforementioned ranges. Such calculations are obtained employing the full mathematical expressions presented in this section; we do not use the approximated expressions, which were aimed at illustrating order of magnitude estimations.

\section{Emission profiles}
\label{sec:emission}

The flares discussed here can potentially be identified as EM counterparts provided their emission is comparable to or brighter than that of the hosting AGN. This comparison can be done at a given observed frequency $\nu$ through the ratio:
\begin{equation}
r_\nu \equiv \frac{F_\nu}{F_{\nu,\mathrm{host}}} = \frac{4 \pi D_\mathrm{L}^2 \nu' F_\nu'}{\nu L_{\nu,\mathrm{host}}},
\label{rnu}
\end{equation}
where $F_\nu$ and $F_{\nu,\mathrm{host}}$ are the emission fluxes of the disc eruption (equation \ref{EMflux}) and of the hosting AGN, respectively. For simplicity, we consider the host emission as stationary and assess its differential luminosity as a function of the mass $M_\mathrm{S}$ and accretion rate $\dot{M}_\mathrm{S}$ onto the central SMBH, as done in \cite{Tagawa_2023,Rodriguez-Ramirez_2024}:
\begin{equation}
\label{hostEM}
\nu L_{\nu, \mathrm{host}} = \left( \frac{f_\mathrm{n}}{l_\mathrm{ref}} \right) \nu L_{\nu, \mathrm{ref}}.
\end{equation}
In this approach, the reference differential luminosity $\nu L_{\nu,\mathrm{ref}}$ is taken
from the average AGN spectrum derived in \citep{Ho_2008} (their Figure 7),
$l_\mathrm{ref}$ is the reference luminosity of this spectrum at $\lambda_\mathrm{ref} = 4400 $ \AA,
and $f_\mathrm{n}$ is a modulating factor
\begin{equation}
f_\mathrm{n} = 10^{44}\mathrm{erg}\,\mathrm{s}^{-1}
\left(
\frac{M_\mathrm{s}}{10^8 \mathrm{M}_\odot}
\right) 
\left(
\frac{\dot{M}_\mathrm{s} c^2}{L_\mathrm{Edd}(M_\mathrm{s})}\right)
\left(
\frac{10}{f_\mathrm{c}}
\right).
\label{fn}
\end{equation}
Setting $f_\mathrm{c}=3$, we obtain an AGN spectrum consistent with the emission of the corresponding thin disc (see Figures~\ref{fig:SEDs1e6}-\ref{fig:SEDs1e7}).

We derive LC profiles of the EM counterparts as the sum of the host (equation \ref{hostEM}) and the disc eruption (equation \ref{EMflux}):
\begin{equation} 
F^\pm_{\nu,\mathrm{flare}} = (1+z)
\left(
\frac{\nu L_{\nu,\mathrm{host}}}{\nu' 4\pi D_\mathrm{L}^2 } + F^\pm_{\nu'}
\right),
\label{Fnu_flare}
\end{equation}
where $\nu'$ and $\nu = \nu'/ (1+z)$ are the emission frequencies at the source and observer's frames, respectively, and $z$ is the source redshift which we relate to the luminosity distance $D_\mathrm{L}$ through a $\Lambda$CDM cosmological model constrained with Planck 2018 results \cite{PlanckCollab_2020}.

Non-blazar AGNs typically exhibit stochastic variability in optical and X-ray bands representing about 10\% of their average emission over time scales ranging from days to years \cite{VandenBerk_2004, Yu_2022}. The origin of such variability is not established, as it could be driven by multiple processes in the AGN disc. Therefore, here we consider $r_\mathrm{th}=0.25$ as a reasonable threshold to distinguish the flares predicted in this work from their host emission, and define as detectable, those solutions leading to $r_\nu \geq r_\mathrm{th}$ (see equation \ref{rnu}).

The emission scenario presented in this paper suggests, in principle, that the largest ratios $r_\nu$ would be produced by high-mass BH remnants with low kick velocities, as the flare energy scales as a function of $v_\mathrm{k}^{-4}$ and $M_\bullet^{2}$ (see equation~\ref{E_0ll}). Nevertheless, towards this regime (low $v_\mathrm{k}$ and high $M_\bullet$), the condition \ref{condition} also becomes increasingly difficult to satisfy, making the emission mechanism discussed here uncertain.

On the other hand, a large $M_\bullet$ and low $v_\mathrm{k}$ could satisfy the condition \ref{condition} if the merger occurs at sufficiently large radial distances $a$ within the disc, as the disc height increases with the distance to the SMBH. However, in such cases, the flare time delay also increases, making the multi-messenger association more challenging. A more detailed analysis of flare delays for detectable signatures is presented in Sections~\ref{subsec:tdelay} and \ref{sec:q_Xeff}.

In the following subsections, we illustrate spectral energy distributions (SEDs) and LC profiles predicted by the present model using a fiducial BH remnant of 20 M$_\odot$ with inflow/outflow efficiencies of 
$\eta_\mathrm{in}=0.1$ and $\eta_\mathrm{w}=$0.05, and wind opening angle of $\theta_\mathrm{w}=45^\circ$. The other model parameters are varied as indicated.

\subsection{Spectral energy distributions}
\begin{figure}
   \centering
   \includegraphics[width=\hsize]{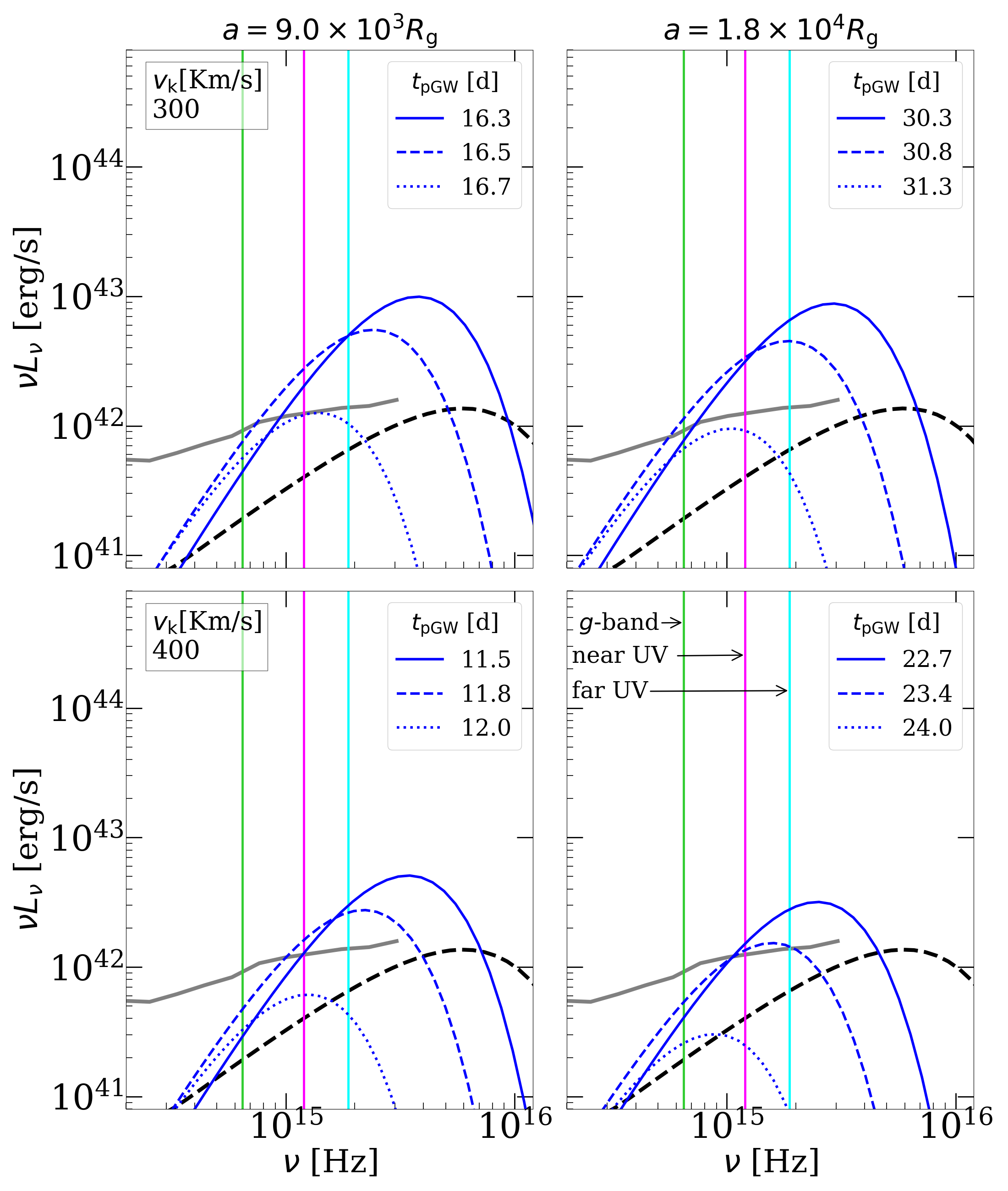}
      \caption{
SED snapshots of the disc eruption (the merger EM counterpart) in an AGN with a central engine of $10^6$ M$_\odot$. 
The SEDs are calculated at different times, 
as indicated by the legends in each frame. The first curve of this family corresponds to the maximum bolometric luminosity of the flare. The emission of the associated thin disc and host galaxy are over-plotted with the thick dashed and solid curves, respectively. The vertical lines indicate the optical g-band, near UV, and far UV frequencies. Curves on the left and right panels are calculated assuming a merger at $9000$ and 18000 $R_\mathrm{g}$ from the central SMBH, whereas upper and lower panels consider kick velocities of 300 and 400 km s$^{-1}$. Other model parameters shared by all the panels are specified in the text.
              }
         \label{fig:SEDs1e6}
\end{figure}

\begin{figure}
   \centering
   \includegraphics[width=\hsize]{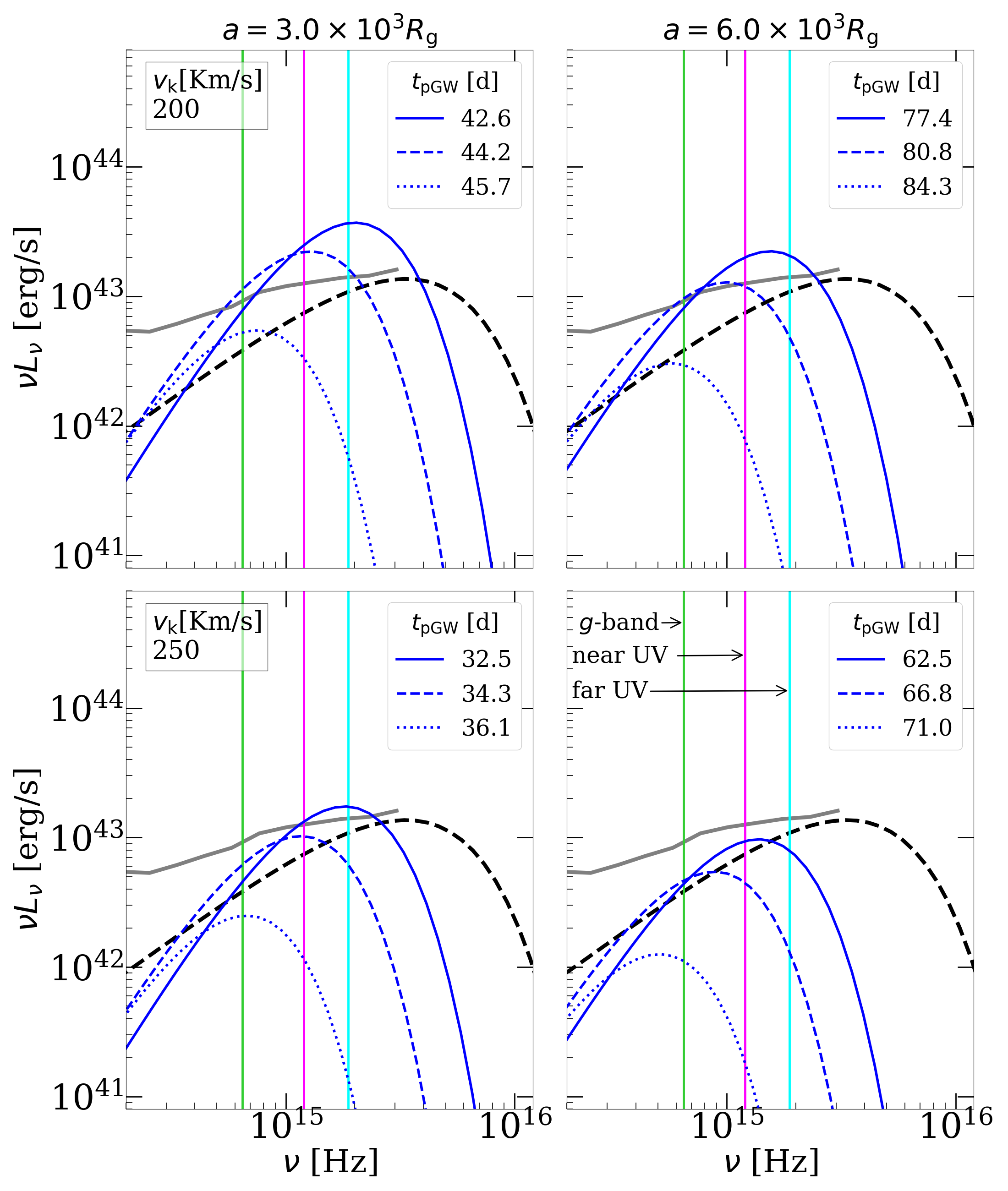}
      \caption{
Same as in Figure~\ref{fig:SEDs1e6}, but considering an SMBH mass of $10^7$ M$_\odot$, and different merger distances $a$ and kick velocities $v_\mathrm{k}$ as indicated.
              }
         \label{fig:SEDs1e7}
\end{figure}

The EM counterpart discussed here is approached as blackbody emission from the photosphere of the disc eruption (see equation \ref{EMflux}).
As such, the appropriate frequency for observing these flares is
$\nu_\mathrm{pk}\sim k_\mathrm{B} T_\mathrm{eff}/[(1+z)h]$ and somewhat lower frequencies.
We note that, at maximum bolometric luminosity, the effective temperature $T_\mathrm{eff}$ scales inversely with the
disc height $h_\mathrm{d}$ 
(see equations \ref{T0approx}-\ref{Teff}). 
Therefore, we expect the flares to shine at lower frequencies as the hosting SMBH is larger.
This is illustrated in Figures~\ref{fig:SEDs1e6}-\ref{fig:SEDs1e7}, where we present SEDs of the disc eruptions considering AGNs with central engines of $10^6$ and $10^7$ M$_\odot$.

The SEDs displayed in Figure~\ref{fig:SEDs1e6} correspond to an ``upper eruption'' of a merger with a kick angle of
$\theta_\mathrm{k}=10^\circ$, in an AGN with $M_\mathrm{S}=10^6$ M$_\odot$, $\dot{m}$=0.025,  located at $D_\mathrm{L}=1600$ Mpc. In each frame, we display the SEDs of the disc eruption at different observed times after the detection of the GW event, as labelled, starting with the time when photons can escape from the plasma ejection (given by equation \ref{Atl}). We over-plot the spectra of the associated thin disc emission (derived from the SG03 model, thick dashed curve) and of the host galaxy given by equation \ref{hostEM} (thick solid curve). The vertical lines indicate optical ($g$-band), near UV, and far UV frequencies, as labelled. Left/right panels of Figure~\ref{fig:SEDs1e6} consider different merger locations $a$ and upper/lower panels different kick velocities $v_\mathrm{k}$, as indicated. These values for $a$ and $v_\mathrm{k}$ are chosen so that conditions \ref{RHL_hd_cond} and \ref{smw_sMw_condition} are fulfilled.

Figure~\ref{fig:SEDs1e6} suggests that merger flares in discs with SMBHs of $\sim 10^6$ M$_\odot$ have time lags (after the GW event) of week scales and radiate most of their luminosity at far ultraviolet and somewhat higher frequencies. These flares can clearly surpass the host emission at far ultraviolet. Certain solutions, especially those with relatively low kick velocities (near the threshold where condition \ref{condition} is satisfied), produce emission comparable to the host at optical bands.

Figure~\ref{fig:SEDs1e7} shows SEDs of the merger flare with the same parameter set as in Figure~\ref{fig:SEDs1e6}, but considering a central SMBH of $10^7$ M$_\odot$, and merger locations and kick velocities consistent with condition \ref{condition}. As expected, these flares peak at lower frequencies than in the $M_\mathrm{S}=10^6$ M$_\odot$ case, and can be suitably detected at optical and UV frequencies. On the other hand, the time scale for the flare lags increases by nearly three times with respect to the $M_\mathrm{S}=10^6$ M$_\odot$ case.

Following the trend exhibited by the SEDs of Figures \ref{fig:SEDs1e6} and \ref{fig:SEDs1e7}, one might expect that mergers in AGNs with SMBHs of $10^8$ M$_\odot$ and heavier can be detectable at optical and infrared frequencies. 
Nevertheless, the flare time lags in these larger AGNs would likely increase to several tens to hundred day scales, which would challenge potential MM associations. We analyse the flare time lag and duration in more detail in Section \ref{subsec:Atld}.

\subsection{Optical light-curves}

In the previous subsection, we show that the EM counterpart can be comparable to or brighter than the emission of the host galaxy at optical bands in AGNs with central engines of $10^{6-7}$ M$_\odot$. In this subsection, we derive LC profiles of the counterpart in the optical g-band, illustrating the ``upper'' and ``lower'' disc eruptions, the effect of the AGN's accretion rate, and the remnant's kick velocity. Furthermore, we consider different redshifts for the source to compare the emission with the limiting magnitude of different optical instruments.

\begin{figure*}
   \centering
   \includegraphics[width=\hsize]{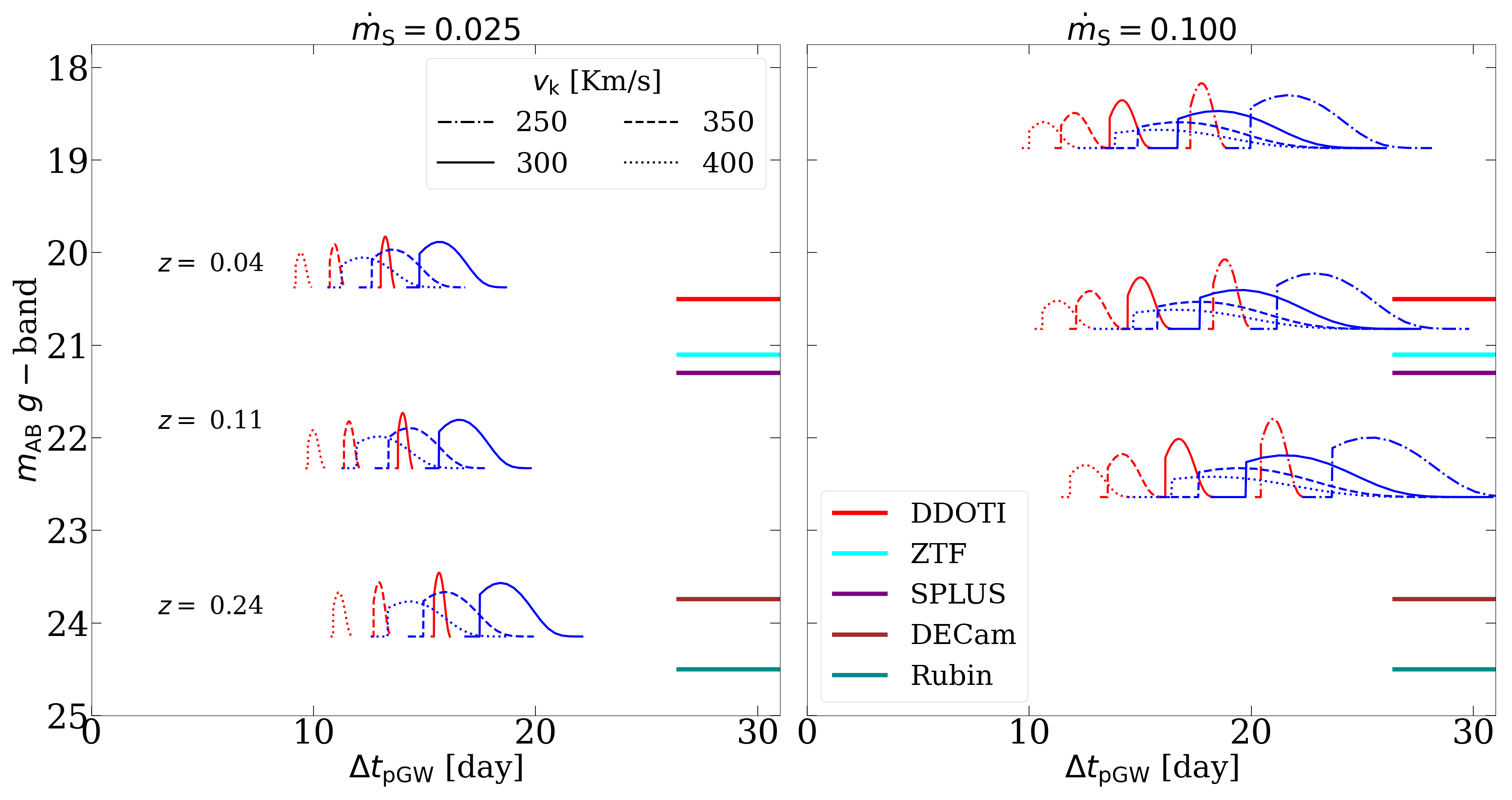}
      \caption{
LCs in the optical $g$-band of the EM counterpart model for a BBH merger in an AGN disc. The figure illustrates possible LC profiles driven by a merger with the following model parameters: $M_\mathrm{S} = 10^6$ M$_\odot$, $a=9000 R_\mathrm{g}$, $M_\bullet = 20$ M$_\odot$, $\theta_\mathrm{k}=10^\circ$, $\eta_\mathrm{in}=0.1$  and $\eta_\mathrm{w}=0.05$ (see the text for details). We explore different redshifts for the source which results in the upper, middle, and lower curve families as well as different kick velocities, as labelled. Red and blue curves correspond to the ``upper'' and ``lower'' eruptions of the model. Left and right panels correspond to AGNs accreting at different rates $\dot{m}_\mathrm{S}$, as indicated. The horizontal lines indicate limiting magnitudes of different optical instruments as labelled (see details in the text).
              }
         \label{fig:profilesLVK1e6}
\end{figure*}

Figure~\ref{fig:profilesLVK1e6} displays the profiles of possible LCs driven by a BBH merger at $9000$ $R_\mathrm{g}$ from an SMBH of $10^6$ M$_\odot$ with a kick angle of $\theta_\mathrm{k}=10^{\circ}$ of the merger remnant. Curves in the left and right panels of Figure~\ref{fig:profilesLVK1e6} are obtained assuming an AGN accreting at 0.025 and 0.1 Eddington accretion rates, respectively. In each frame, upper, middle, and lower curve families are obtained assuming different redshifts of the source, as indicated. Curves with different line styles are obtained using different kick velocities as labelled, whereas red and blue curves correspond to the upper and lower disc eruptions, respectively. The LCs are calculated as given by equation \ref{Fnu_flare} and converted to AB magnitudes employing the \texttt{Python} package \textit{Speclite}\footnote{\href{https://speclite.readthedocs.io/en/latest/index.html}{https://speclite.readthedocs.io/en/latest/index.html}} with the $g$ filter of the Sloan Digital Sky Survey (SDSS, \cite{blanton2017sloan}). For reference, the horizontal lines on the right in each panel indicate the limiting magnitude of different optical instruments, namely DDOTI (1000s exposure, 3-$\sigma$, white filter \cite{becerra_2021, Dichiara_2021}), ZTF (30s exposure, 5-$\sigma$ detection, g filter \cite{ZTF_2018}), SPLUS (3$\times$30s, 3-$\sigma$, g filter \cite{MendesdeOliveira_2019,Almeida-Fernandes_2022}), DECam (60s exposure, 5-$\sigma$ detection, g filter \cite{Bom_2023}), and the forthcoming Vera C. Rubin Observatory ($2\times15$s exposure, 5-$\sigma$ detection, g filter \cite{LSST_2019}), as labelled.

Figure~\ref{fig:profilesLVK1e6} shows how the flare profiles are impacted by the remnant's kick velocity $v_\mathrm{k}$ and by the SMBH accretion rate $\dot{m}_\mathrm{S}$. The emission time lag is shorter for larger values of $v_\mathrm{k}$ since the disc crossing time (and hence the time for the production of the disc eruptions) is reduced (see equation \ref{t_0}). Furthermore, the amplitude of the flare is reduced as $v_\mathrm{k}$ increases since the energy stored in the disc ejections is approximately proportional to $v_\mathrm{k}^{-4}$ (see equation \ref{E0}). We observe that for larger $\dot{m}_\mathrm{S}$ the overall flux (i.e., AGN plus merger counterpart) increases, hence increasing the distance up to which the source can be detected.
These LCs show, for instance, that in AGNs of $10^6$ M$_\odot$ accreting at $\dot{m}_\mathrm{S}=$0.025 - 0.1, the flares could be detected by ZTF up to $z\sim$ 0.04- 0.1, and up to $z\sim$ 0.25-0.5 by forthcoming Rubin. The curves show that upper eruptions appear always somewhat before and have shorter durations than the lower counterparts. This is because upper ejections are always less massive than the lower ones, and therefore their photon diffusion time (see equations \ref{Atph} and \ref{u0}) and flare characteristic period (equation \ref{Atdure}) are always shorter in the upper flares.

\begin{figure*}
   \centering
   \includegraphics[width=\hsize]{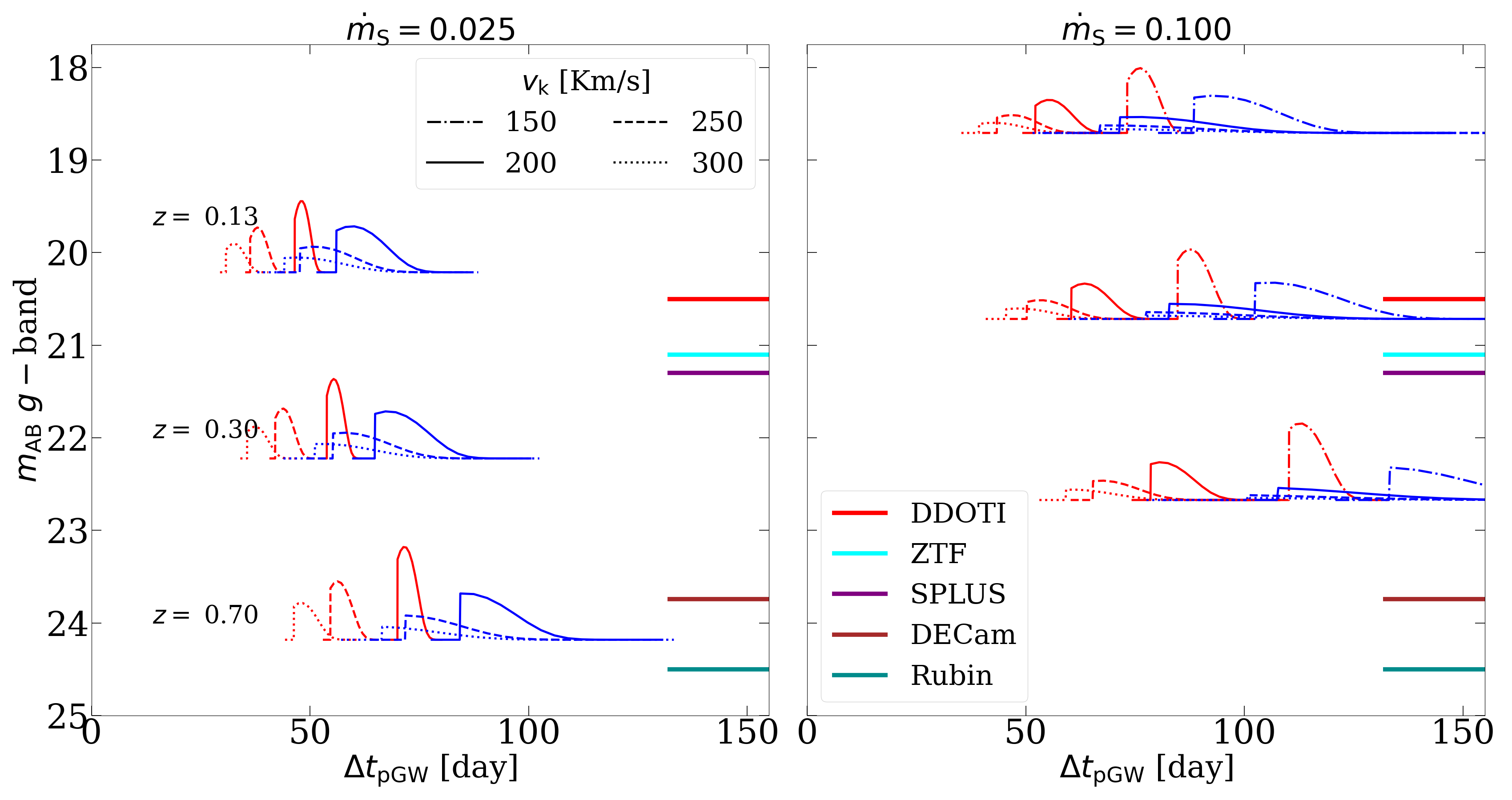}
      \caption{
Same as in Figure~\ref{fig:profilesLVK1e7}, but considering an SMBH of $M_\mathrm{S}=10^{7}$ M$_\odot$, $a=4000 R_\mathrm{g}$ and different kick velocities $v_\mathrm{k}$ and source redshifts $z$, as indicated.
              }
         \label{fig:profilesLVK1e7}
\end{figure*}

In Figure~\ref{fig:profilesLVK1e7}, we present possible LCs of a flare triggered by a merger with the same parameters as in Figure~\ref{fig:profilesLVK1e6}, but considering an SMBH of $10^7$ M$_\odot$, and exploring different values for the kick velocity and the redshift of the source, as indicated. The flare time lags falls within 40-20 days, which represents an increase of about 3-5 times compared to such period in Figure~\ref{fig:profilesLVK1e6}. However, the effect on the LC profiles due to variations of $v_\mathrm{k}$ and $\dot{m}_\mathrm{S}$ is qualitatively similar. We also note that in this $M_\mathrm{S}=10^7$ M$_\odot$ case, lower values for $v_\mathrm{k}$ are allowed by condition \ref{condition}, and that the overall emission can be detected at larger redshifts compared to the $M_\mathrm{S}=10^6$ M$_\odot$ case. For instance, curves in Figure~\ref{fig:profilesLVK1e7} show that in discs accreting at $\dot{m}_\mathrm{S}=0.025 - 0.1$, the flares could be detected by ZTF up to $z\sim 0.1- 0.3$, and up to $z\sim 0.7 - 1.2$ by forthcoming Rubin.

The SEDs and LCs presented in Figures \ref{fig:SEDs1e6}-\ref{fig:profilesLVK1e7} suggest that, in general, the flares are more distinguishable from their host as the merger location $a$ and the kick velocity $v_\mathrm{k}$ are smaller.

\subsection{The flare delay and duration.}
\label{subsec:Atld}

The SEDs and LCs examples displayed in Figures~\ref{fig:SEDs1e6}-\ref{fig:profilesLVK1e7}, show that the time lag $\Delta t_\ell$ and the duration $\Delta t_\mathrm{d}$ of the flares are notably impacted by the mass of the hosting SMBH $M_\mathrm{S}$, the location of the merger $a$, and the kick velocity $v_\mathrm{k}$. Here we describe how the present emission model predicts $\Delta t_\ell$ and $\Delta t_\mathrm{d}$ in a more systematic way.

\begin{figure}
   \centering
   \includegraphics[width=\hsize]{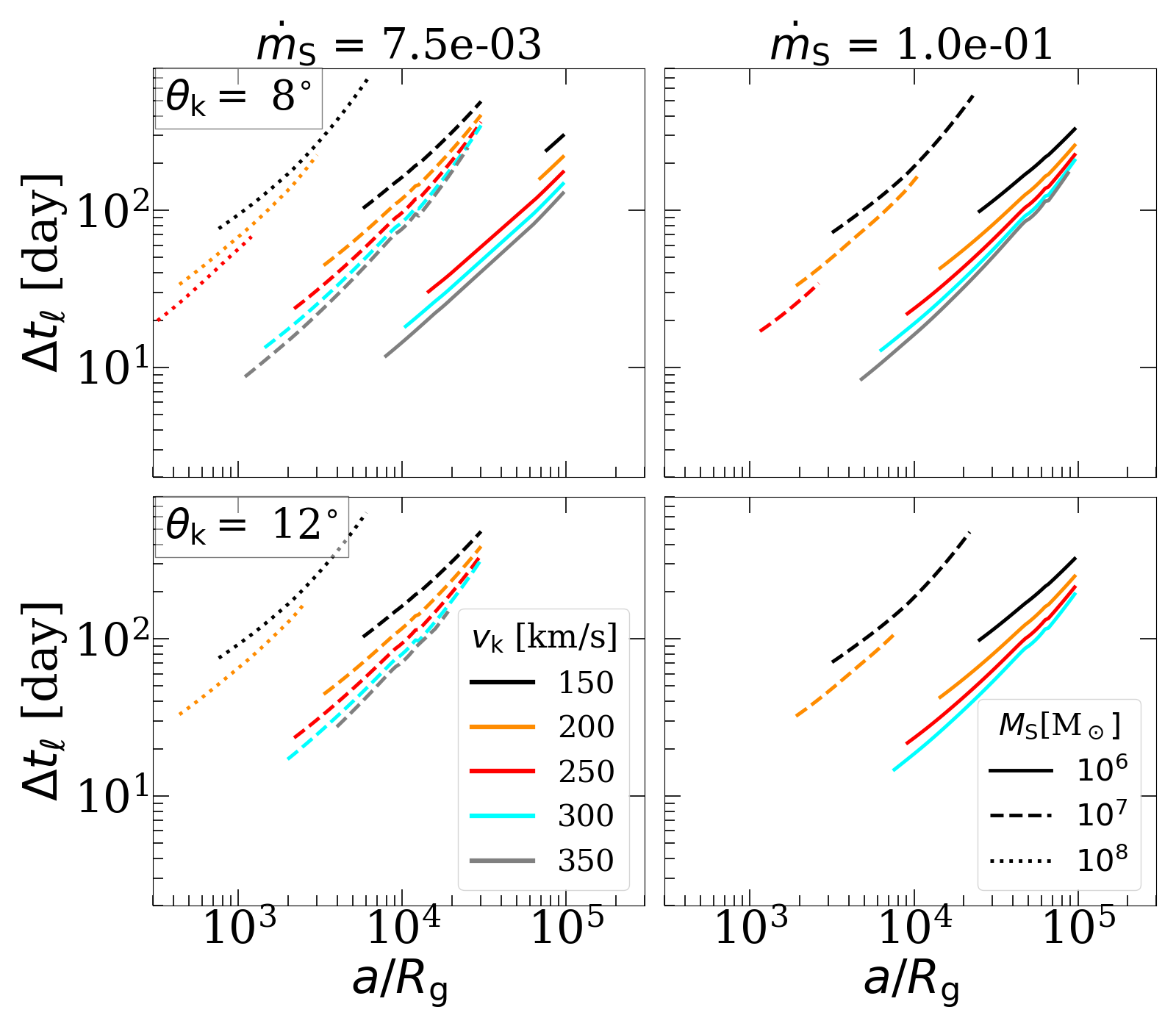}
      \caption{      
Time lag of the counterpart flare after the GW event as a function of the radial location $a$ of the merger in the AGN disc. Curves in the left and right panels are obtained considering different SMBH accretion rates, whereas upper and lower panels depict solutions for different kick angles, as labelled. In each panel, the different curves correspond to flares in AGNs with different SMBH masses, as well as different magnitudes of remnant kick velocity, as indicated. The curves terminate/initiate when conditions \ref{condition} and $F_\nu/F_{\nu,\mathrm{host}} \leq 0.25$ are no longer satisfied (see the text for details).
}
         \label{fig:Atl}
\end{figure}

In Figure~\ref{fig:Atl}, we display $\Delta t_\ell$ given by equation \ref{Atl} as a function of the merger location $a$ in the AGN disc. In this example, we consider a source located at $D_\mathrm{L}=$1600 Mpc and derive the $\Delta t_\ell$ vs $a/R_\mathrm{g}$ curves exploring different values for $v_\mathrm{k}$, $\theta_\mathrm{k}$, $M_\mathrm{S}$, and $\dot{m}_\mathrm{S}$. We restrict these curves to fluxes with ratios $r_\nu \geq 0.25$ in the optical $g$-band (see equation \ref{Fnu_flare}) together with condition \ref{condition}. Thus, the curves in Figure~\ref{fig:Atl} initiate/terminate when the aforementioned conditions are no longer satisfied. In each panel, the curves corresponding to different SMBH masses are labelled with different line styles, whereas curves of different kick velocities are labelled with different colours. Computations in the upper and lower panels are obtained assuming different kick angles $\theta_\mathrm{k}$, whereas left and right panels correspond to different accretion rates $\dot{m}_\mathrm{S}$, as labelled.

The curves in Figure~\ref{fig:Atl} show that the time lag of the EM counterpart increases monotonically with the merger location  as $\propto (a/R_\mathrm{g})^{\alpha_\ell}$ with $\alpha_\ell \approx 1$. As already exhibited by the SEDs and LCs of previous subsections, Figure~\ref{fig:Atl} shows that $\Delta t_\ell$ increases as the SMBH mass is larger (at a given $a/R_\mathrm{g}$) and diminishes for larger kick velocities. Interestingly, the condition $r_\nu \geq r_\mathrm{th}$ together with condition \ref{condition} associates the mass of the hosting SMBH to a  
characteristic range of $a/R_\mathrm{g}$ within which 
visible flares could be produced.
For instance, we see that in AGNs with SMBHs of $10^6$ M$_\odot$, visible flares would be more likely produced
at $a\sim10^{4-5}R_\mathrm{g}$, whereas they would be better produced at $a\sim 5\times10^{2-3}R_\mathrm{g}$ for SMBHs of
$10^8$ M$_\odot$.

\begin{figure}
   \centering
   \includegraphics[width=\hsize]{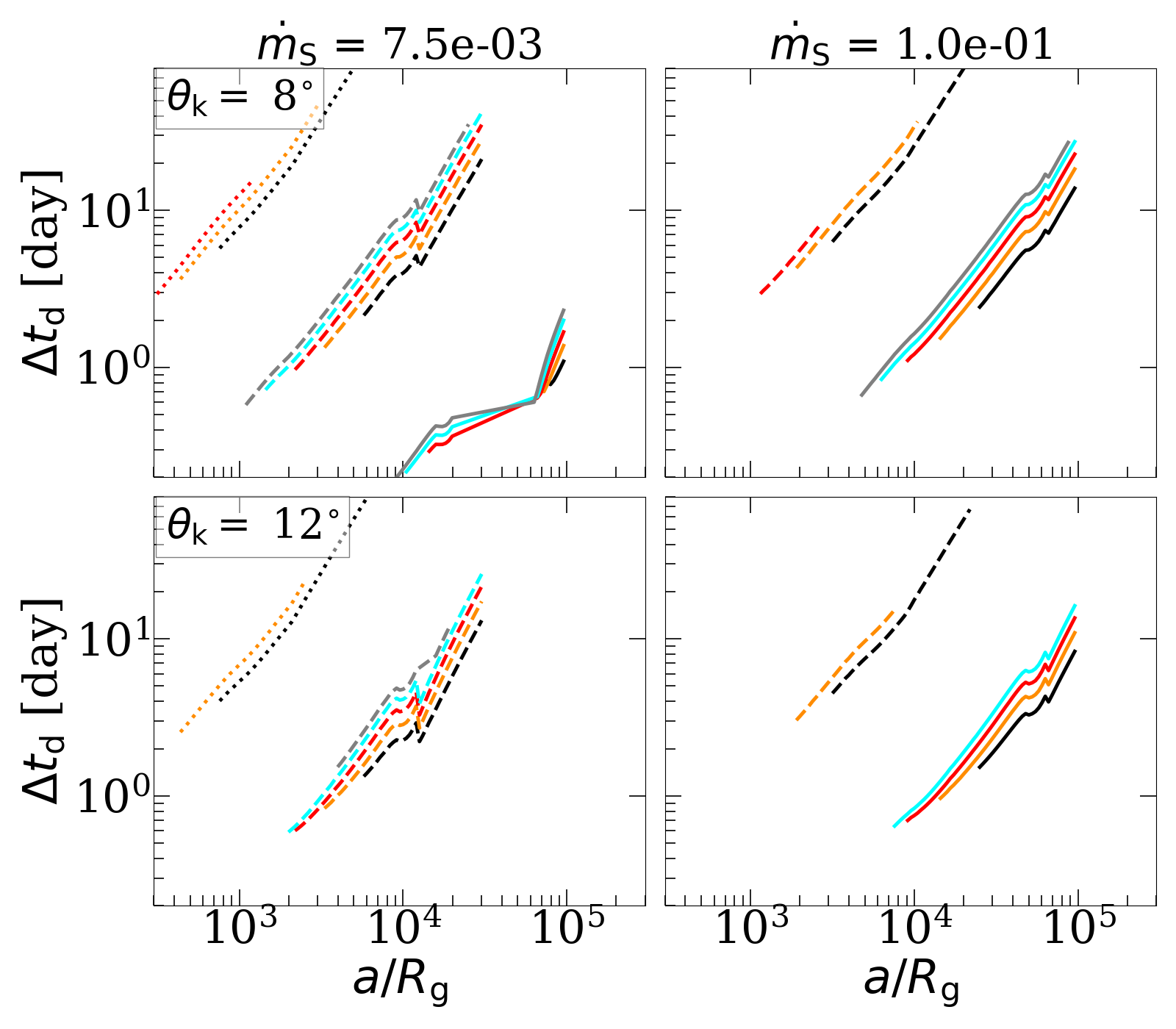}
      \caption{
Same as in Figure~\ref{fig:Atl}, but displaying the flare duration $\Delta t_\mathrm{d}$.
              }
         \label{fig:Atd}
\end{figure}

In Figure~\ref{fig:Atd}, we display the flare duration given by equation \ref{Atdure}  and associated to the curves of Figure~\ref{fig:Atl}. This figure shows that $\Delta t_\mathrm{d}$ is generally one order of magnitude smaller than the flare time lag.
We observe that flare duration
$\Delta t_\mathrm{d}$ increases with the magnitude of the kick velocity $v_\mathrm{k}$, different to $\Delta t_\ell$.

The present EM counterpart model is built such that the flare time lag is composed of a sequence of periods, namely the cavity crossing time $\Delta t_\mathrm{cav}$, the time for the wind ignition $\Delta t_\mathrm{HL}$,
the time for the wind to expel the disc matter $\Delta t_\mathrm{bo}$,
and the time for photons to escape the matter ejections $\Delta t_\mathrm{ph}$
(see Section~\ref{subsec:tdelay} for details).
The contribution of these periods to the flare time lag vary according to the model parameter configuration.

\begin{figure}
   \centering
   \includegraphics[width=\hsize]{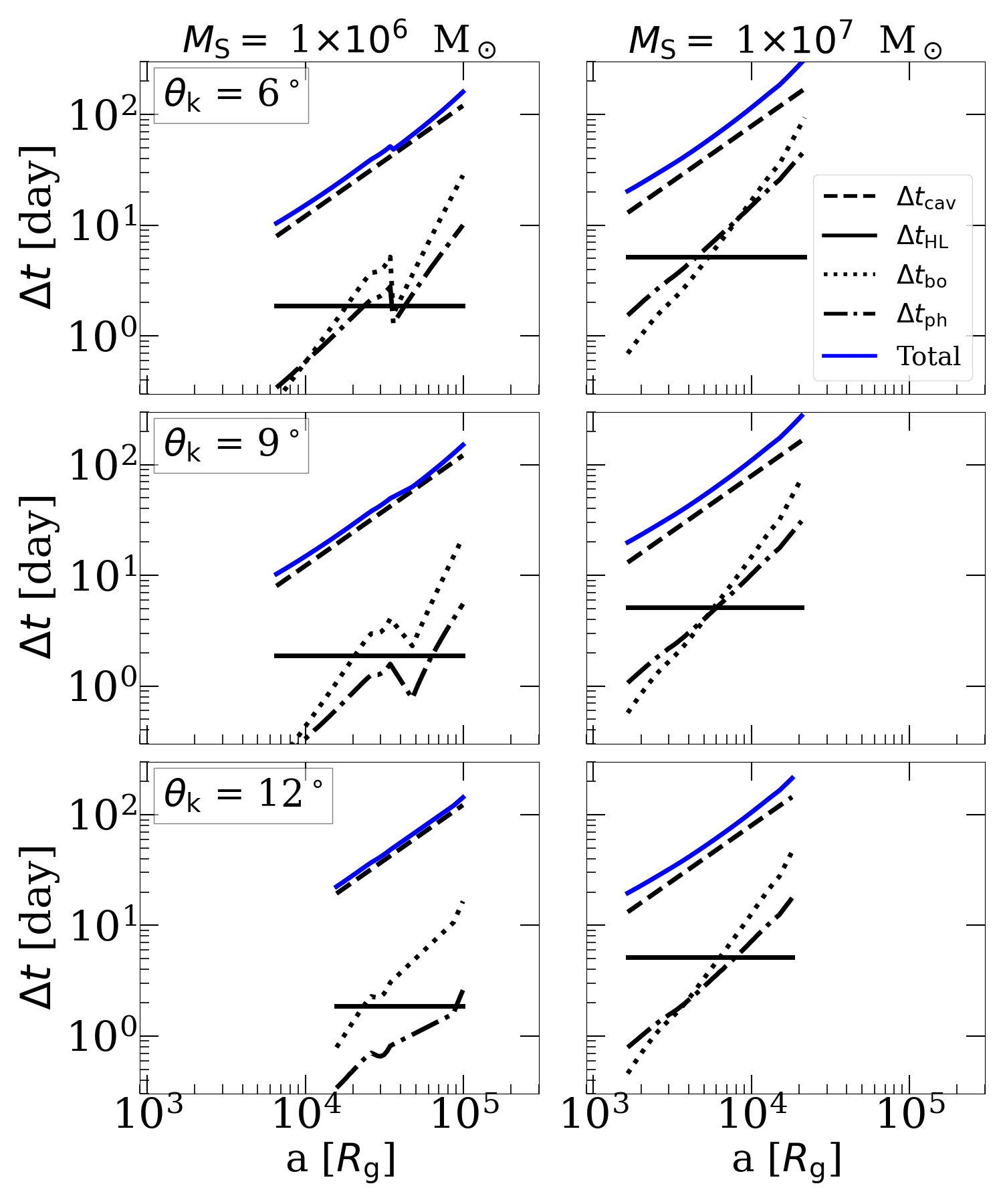}
      \caption{
Examples of the flare time lag and its components as functions of the merger location $a$. These periods are calculated in the observer frame. The curves begin and end when the condition for the flux ratio $r_\nu > 0.25$ (see equation \ref{rnu}) and the condition \ref{condition} are no longer satisfied. The parameter configurations for each panel are given in the plot labels and the text.              }
         \label{fig:AtComps}
\end{figure}

In Figure~\ref{fig:AtComps}, we illustrate the components of the time lag as a function of merger location, considering an AGN at a distance of 1600 Mpc, accreting at a rate of $\dot{m}_\mathrm{S}=0.025$, and within the conditions \ref{condition} and $r_\nu \geq r_\mathrm{th}$ (similarly as in Figures~\ref{fig:Atl}-\ref{fig:Atd}). Curves in the upper, middle, and lower panels are calculated with different kick angles as indicated. Based on the $M_\mathrm{S} - v_\mathrm{k}$ relation discussed in the previous subsection, the curves in the left panels are obtained assuming an SMBH of $M_\mathrm{S}=10^6$ M$_\odot$ with a remnant of $v_\mathrm{k}=350$ km s$^{-1}$, whereas curves in the right panels assume $M_\mathrm{S}=10^7$ M$_\odot$ and  $v_\mathrm{k}=250$ km s$^{-1}$.
We observe that, generally, $\Delta t_\mathrm{cav}$ is the dominant component, whereas the photon escape time $\Delta t_\mathrm{ph}$ is subdominant.

\section{Flare delay, kick velocity, and the $q-\chi_\mathrm{eff}$ correlation}
\label{sec:q_Xeff}

The present MM emission model can be used to investigate constraints on MM observables, such as the binary effective spin $\chi_\mathrm{eff}$, the mass ratio $q = m_2/m_1$, the merger remnant mass $M_\bullet$, and the time delay $\Delta t_\mathrm{\ell}$ of the EM signature.
In this formulation, the flare delay $\Delta t_\ell$ can be linked to the BBH effective spin by parametrising the remnant's kick velocity $v_\mathrm{k}$ and direction $\theta_\mathrm{k}$ (relative to the disc plane; see Figure~\ref{fig:sketch}) in terms of the dimensionless spins $\bar{\chi}_1$, $\bar{\chi}_2$, and the masses $m_1$, $m_2$ of the merging BHs.
To achieve this, we use the empirical formulae of \cite{Lousto_2010, Lousto_2012} (see also \cite{Fragione_2021}), which provide analytic fits to numerical relativity simulations of BBH merger recoils.

In terms of the kick velocity components, the kick angle relative to the disc plane is given by:
\begin{equation}
\theta_\mathrm{k} = \arcsin\left\{\frac{v_{||} }{ |\vec{v}_\mathrm{k}| }\right\},
\label{thk_reparam}
\end{equation}
where $|\vec{v}_{\mathrm{k}}|$ is the kick velocity magnitude, and $v_{||}$ is its component normal to the disc plane.
Assuming the BBH orbital plane is aligned with the AGN disc plane, the Cartesian components of the recoil velocity  
$(v_x, v_y, v_\parallel)$  
can be expressed in terms of the spin vectors $\vec{\chi}_1, \vec{\chi}_2$ and the mass ratio $q = m_2/m_1 \leq 1$ as given by \cite{Lousto_2010, Lousto_2012}:  
\begin{align}
\label{v_paralel}
&v_\mathrm{||} = \frac{16\eta^2}{1+q}
\left[
v_0 + v_A S_{||} + v_B S_{||}^2 + v_C S_{||}^3
\right]\\
\nonumber&\,\,\,\,\,\,\,\,\,\,\,\,\,\,\,\,\,\,\,\,\,\,\,\,\,\,\,\,\,\times|\vec{\chi}_{\perp,2} - q \vec{\chi}_{\perp,1}|\cos\left(\delta_\triangle-\delta_1\right),
\\ 
\label{vx}
&v_\mathrm{x} = v_m +\cos\xi v_\perp,\\
\label{vy}
&v_\mathrm{y} = v_\perp\sin\xi,
\end{align}
with
\begin{align}
&v_m = A\eta^2\sqrt{1-4\eta}(1+B\eta),\\
&v_\perp = \frac{H\eta^2}{1+q}
\left(
\chi_{||,2} - q\chi_{||,1}
\right),\\
&\eta = \frac{q}{(1+q)^2},
\end{align}
where we adopt the following coefficient values:  
$A=1.2\times10^4$ km/s,  
$B=-0.93$,  
$H=6.9\times10^3$ km/s,  
$v_0= 3678$ km/s,  
$v_A=2481$ km/s,  
$v_B=1793$ km/s,  
$v_C=1507$ km/s,  
and $\xi=145^\circ$,  
as given in \cite{Lousto_2012} and references therein.
In the equations above,  
$\vec{\chi}_{\perp,1}$ and $\vec{\chi}_{\perp,2}$ are the projections of the spin vectors onto the orbital plane (i.e., in-plane components), and  
$S_\parallel$ is the component parallel to the binary's orbital angular momentum 
of the vector
\begin{equation}
\vec{S} = 2\frac{\vec{\chi}_2 + q^2\vec{\chi}_1}{(1+q)^2}.
\end{equation}
In equation \ref{v_paralel},  
$\delta_\triangle$ is the angle between the in-plane component of  
\begin{equation}
\vec{\Delta} = \frac{(m_1+m_2)^2}{(1+q)} 
(\vec{\chi}_2 - q\vec{\chi}_1),
\end{equation}
and the in-fall direction at merger, while $\delta_1$ is an angular phase dependent on the binary's initial separation.

We define the $x$-direction of the spin vectors and kick velocity as parallel to the in-fall direction at merger and set $\delta_1 = 0$.
The choice of in-plane coordinates does not affect the emission outcome discussed in Section~\ref{sec:analytic_model}, as we assume the local environment around the merger to be azimuthally symmetrical.
Finally, we parameterise the spin vector components as:
\begin{align}
\label{X_z}
&\chi_{\parallel,i} = \chi_i\cos\varphi_i, \\
\label{X_x}
&\chi_{x,i} =  \chi_i\sin\varphi_i \cos\phi_i,\\
\label{X_y}
&\chi_{y,i} =  \chi_i\sin\varphi_i \sin\phi_i,
\end{align}
where $\varphi_i$ and $\phi_i$ are the polar and azimuthal angles,  
$\chi_i\equiv|\vec{\chi}_i|$ are the dimensionless spin magnitudes,  
and $i=1,2$ label the heavier and lighter BHs, respectively.  
Thus, the mass ratio $q$ and effective spin $\chi_\mathrm{eff}$ of a BBH with spins $\vec{\chi}_i$ and masses $m_i$ are given by:
\begin{equation}
q=\frac{m_2}{m_1}\leq 1,\,\,\,\,\,
\chi_\mathrm{eff} =
\frac{\chi_1\cos\varphi_1+q\chi_2\cos\varphi_2}{1+q} \in[-1,1].
\label{defXeff}
\end{equation}

Summarising, the observables ($\chi_\mathrm{eff}, q, M_\bullet, v_\mathrm{k}, \theta_\mathrm{k}, \Delta t_\ell$) can be predicted using the MM emission model from Section~\ref{sec:analytic_model} along with equations (\ref{thk_reparam})-(\ref{defXeff}), for a given configuration $(m_i, \varphi_i, \phi_i, \chi_i)$ of the BBH components.

BBH mergers detected by the LVK experiment show a trend where the effective spin 
$\chi_\mathrm{eff}$ increases towards positive values as the $q$ ratio decreases  
\cite{Callister_2021, Abbott_2023, Adamcewicz_2022}.  
This suggests a bias in the AGN GW channel towards alignment of the heavier BH's spin with the orbital angular momentum (i.e., $\varphi_1 \sim 0$, see equation \ref{defXeff}), possibly due to torques exerted by gas accretion.

\begin{figure*}
   \centering
   \includegraphics[width=\hsize]{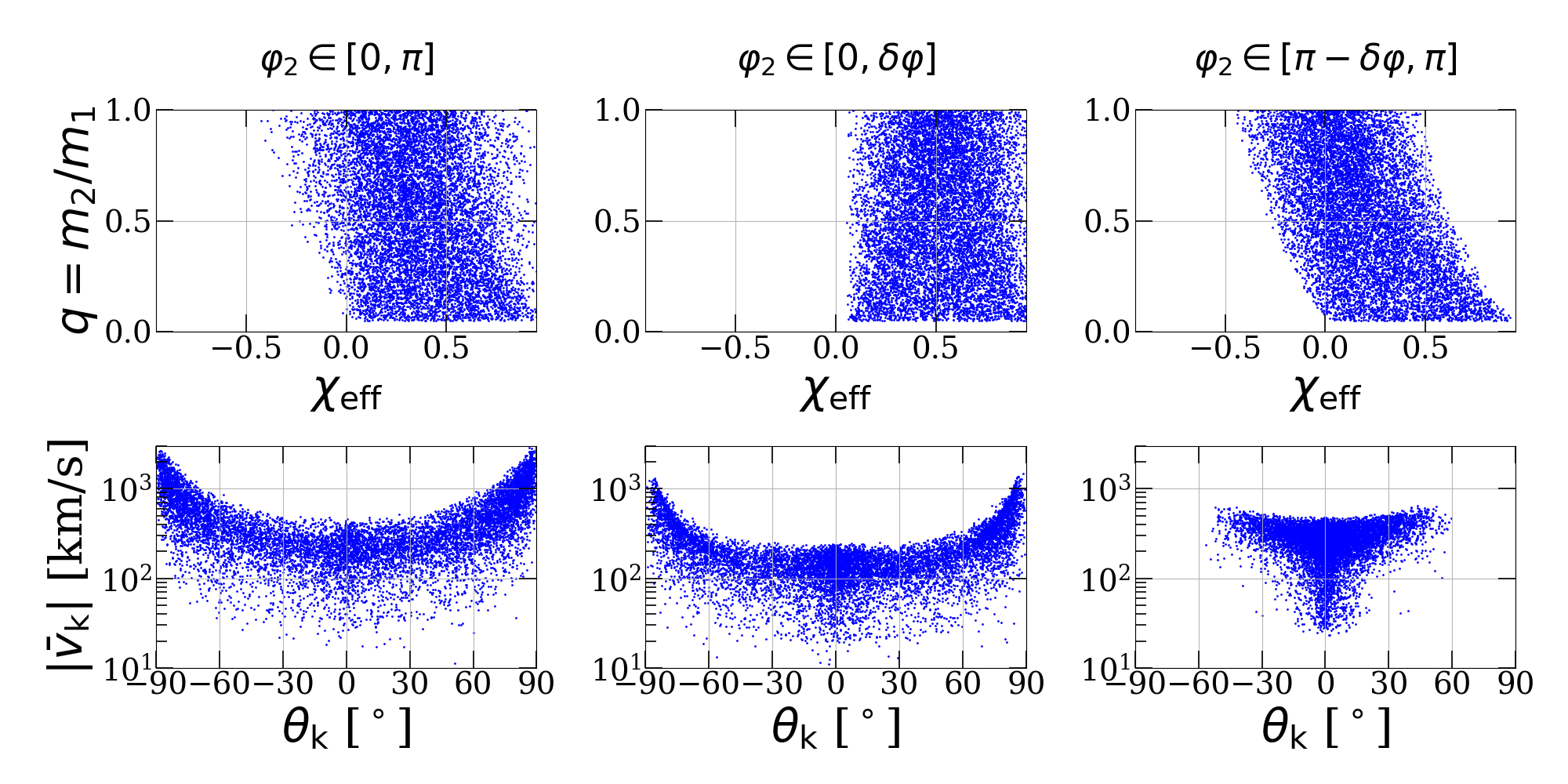}
      \caption{
Upper: Merging BBHs in the $q-\chi_\mathrm{eff}$
space, that results from sampling 
$10^4$ configurations of spin and masses of the binary components (see the text).
Lower: Correlation of the kick magnitude and its direction relative to the AGN disc plane, associated to the sampled configurations of the upper panels. The kick velocity and direction are obtained adopting an analytical parametrisation presented in \cite{Lousto_2010,Lousto_2012}, and detailed in the text.
All panels were obtained assuming the spin of the heavier BH aligned with the orbital spin, whereas in the left, centre, and right panels, the polar angle of the lightest BH spin, $\varphi_2$ is constrained as indicated.
              }
         \label{fig:Xeff_q}
\end{figure*}

\begin{figure}
   \centering
   \includegraphics[width=\hsize]{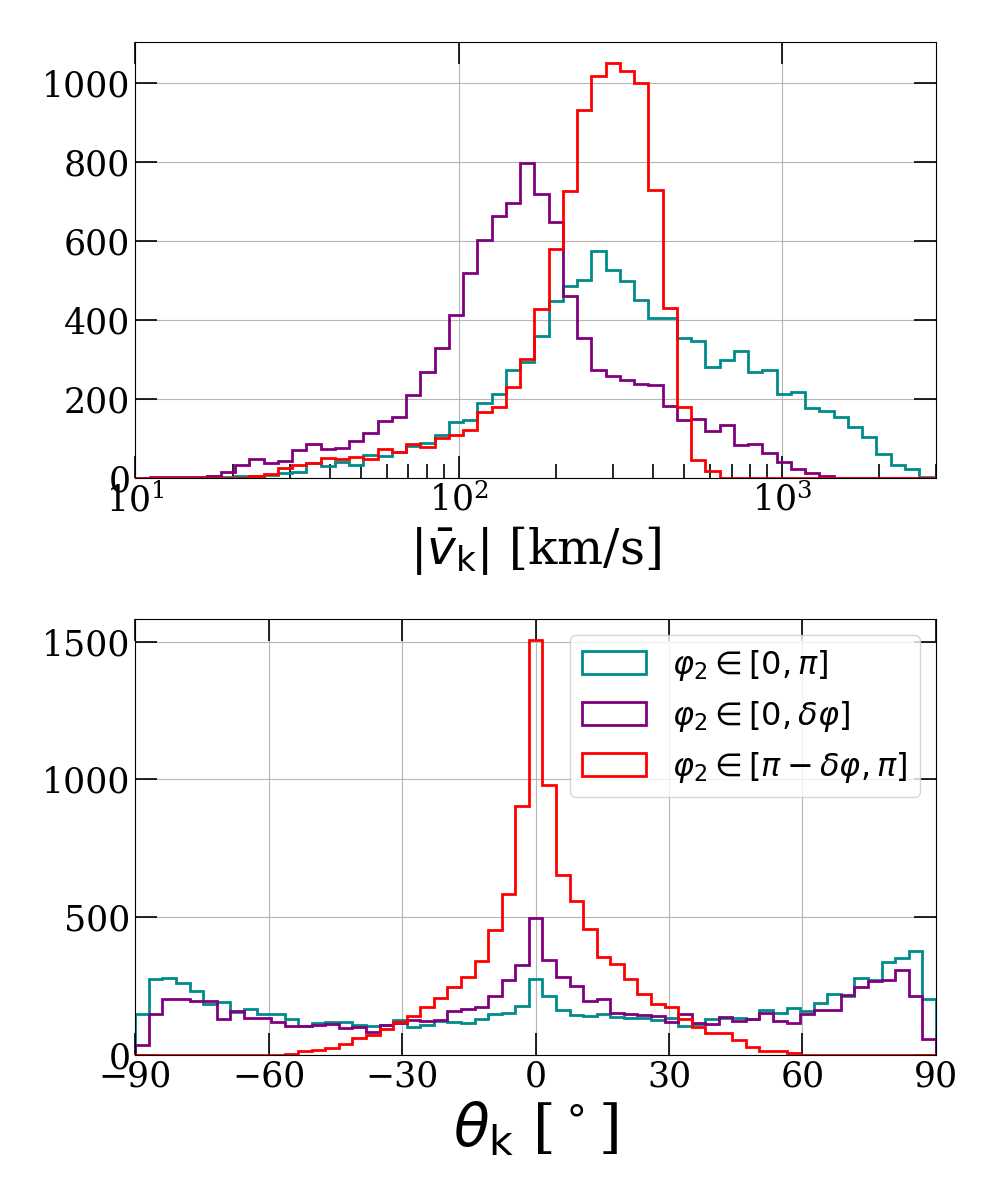}
      \caption{
Distribution of the kick velocity magnitudes (upper) and directions relative to the disc plane (lower) associated to the lower panels of Figure~\ref{fig:Xeff_q}.
              }
         \label{fig:vk_dist}
\end{figure}

In Figures~\ref{fig:Xeff_q}-\ref{fig:vk_dist}, we show the distributions of kick velocities and directions,  
and their associated configurations in the $q-\chi_\mathrm{eff}$ space,  
resulting from imposing the $\varphi_1 = 0$ constraint on the recoil velocity model of equations (\ref{v_paralel})-(\ref{vy}).  
We consider three scenarios for the spin of the lighter BH:  
(i) a uniform angular distribution, (ii) a quasi-aligned spin, and (iii) a quasi-anti-aligned spin relative to the orbital angular momentum.  
To emulate these cases, we keep $\varphi_1 = 0$ and sample $10^4$ parameter configurations within the intervals  
$q \in [0.05,1]$,  
$\chi_{1,2} \in [0.05,1]$,  
$\phi_{1,2} \in [0,2\pi]$,  
with the spin of the lighter BH sampled as:  
\begin{enumerate}
\renewcommand{\labelenumi}{(\roman{enumi})}
\item
$\varphi_2 \in [0,2\pi]$ (uniform distribution),
\item 
$\varphi_2 \in [0, \delta\varphi]$ (quasi-aligned),
\item
$\varphi_2 \in [\pi - \delta\varphi, \pi]$ (quasi-anti-aligned).
\end{enumerate}
where $\delta\varphi = \pi/10$.

The $v_\mathrm{k}-\theta_\mathrm{k}$ and $q-\chi_\mathrm{eff}$ correlations resulting from the three scenarios above are presented in the plots in different columns of Figure~\ref{fig:Xeff_q}, as indicated. 
Among these, we identify that binaries with quasi-anti-aligned spins (right column in Figure~\ref{fig:Xeff_q}) are the most likely to produce post-merger EM counterparts under the emission scenario discussed in this paper.  
This spin configuration concentrates most of the kick orientations quasi-parallel to the AGN disc plane (as shown in Figure~\ref{fig:vk_dist}), while also being consistent with the observed $q-\chi_\mathrm{eff}$ correlation.  
The $\varphi_1 \in [0,\pi]$ case produces relatively few kicked remnants quasi-parallel to the disc, whereas the quasi-aligned case does not produce null $\chi_\mathrm{eff}$ for $q \rightarrow 1$, as observed.

The $q-\chi_\mathrm{eff}$ relation provides an additional constraint that can be used to favour or rule out candidate EM counterparts under the MM emission scenario discussed in this paper. We illustrate this by deriving the observables ($\chi_\mathrm{eff}$, $q$, $M_\bullet$, $v_\mathrm{k}$, $\Delta t_\ell$) for EM signatures in the optical $g$-band, restricted to flux ratios $r_\nu > 0.25$ (see Section~\ref{sec:emission}), produced by BBH mergers within a SG03 disc, where the BBH merger follows configuration (iii) above (the case of BHs with anti-aligned spins). Given the values for the SMBH mass $M_\mathrm{S}$ and its normalised accretion rate $\dot{m}_\mathrm{S}$, we sample $10^4$ parameter configurations of BBH mergers within the following ranges: \begin{itemize} \item $a \in [300 R_\mathrm{g}, a_\mathrm{max}]$, \item $m_1 \in [5, 80]$, \item $q \in [0.05, 1]$, \item $\chi_1, \chi_2 \in [0.05, 1]$, \item $\phi_1, \phi_2 \in [0, 2\pi]$, \item $\varphi_2 \in [\pi-\delta\varphi, \pi]$. \end{itemize} with $\varphi_1$ fixed at 0. The first range listed corresponds to the radial location of the merger within the AGN disc, with the upper limit determined by the outer boundary of the disc (as specified by the criterion in equation \ref{amax_SG03}). Since $a$ spans several orders of magnitude, we sample it uniformly on a logarithmic scale. As in the example leading to Figures~\ref{fig:Atl} and \ref{fig:Atd}, here we consider the AGN situated at $D_\mathrm{L} = 1600$ Mpc and fix the efficiencies related to the remnant's wind to $\eta_\mathrm{in} = 0.1$ and $\eta_\mathrm{w} = 0.05$.

\begin{figure*}
   \centering
   \includegraphics[width=\hsize]{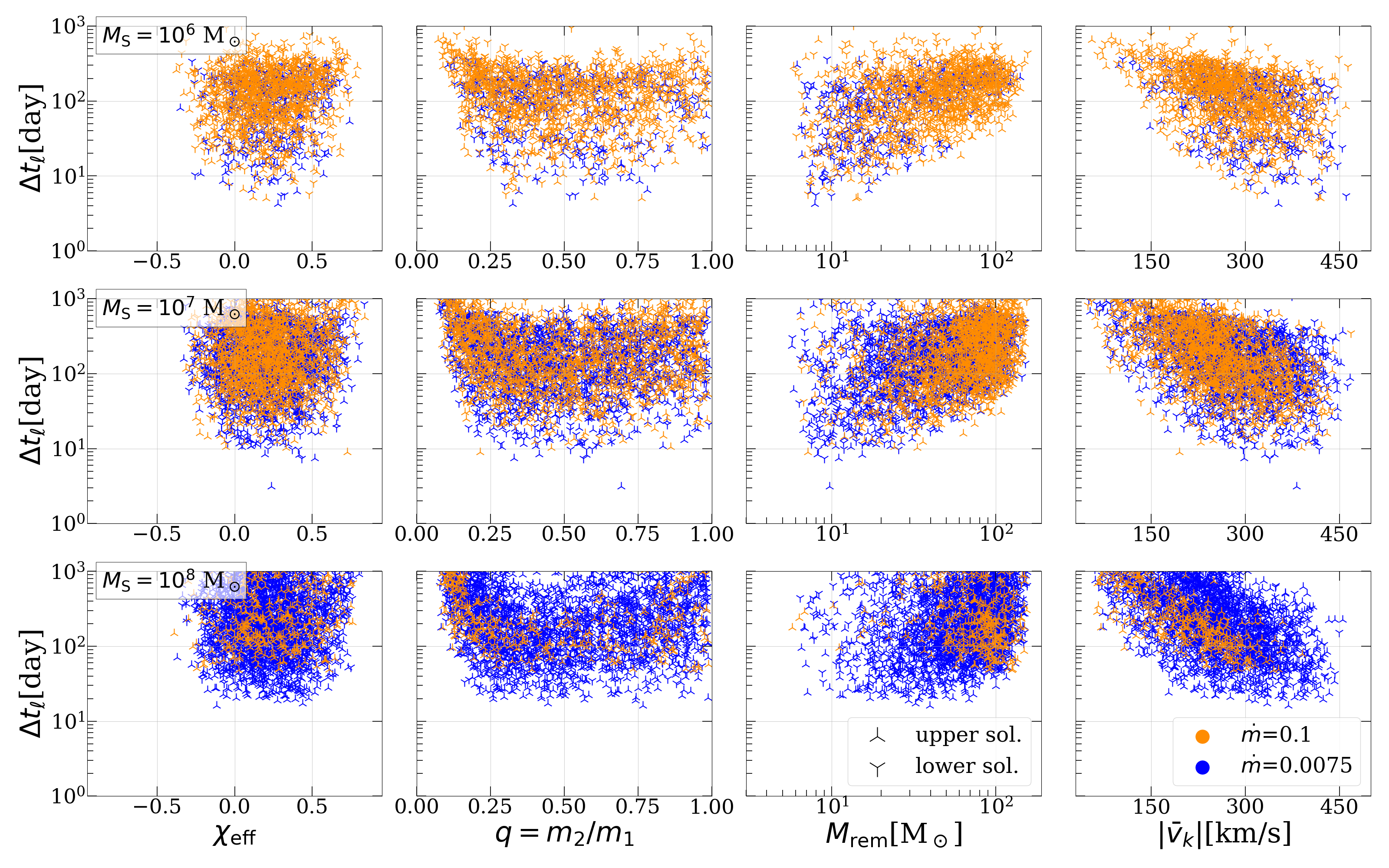}
      \caption{
Correlation of the optical flare delay $\Delta t_\ell$   
with the properties of the BBH merger as plotted in the different columns.
The associations represented with the markers are derived from the $10^4$ parameter configurations sampled as described in the text.
The plots in different rows are obtained considering different masses for the SMBH host, as indicated.
Different colours and orientation of the markers corresponds to different AGN accretion rates and flare directions, (see section~\ref{sec:analytic_model}), as indicated.
              }
         \label{fig:MM_corrs}
\end{figure*}

The predictios for
($\chi_\mathrm{eff}$, $q$, $M_\bullet$, $v_\mathrm{k}$, $\Delta t_\ell$)
 are displayed in Figure~\ref{fig:MM_corrs}. From these outcomes, we observe that, for instance, visible EM counterparts with time lags of $\Delta t_\ell \sim 10$ days can only be produced in hosts with SMBHs of $\sim 10^7$ M$\odot$ or smaller. Such EM signatures are unlikely to be produced by mergers with $q \lesssim 0.2$, $\chi_\mathrm{eff} \gtrsim 0.6$, or remnants of $M_\bullet \gtrsim 40$ M$\odot$. Mergers with these values of $q$, $\chi_\mathrm{eff}$, and $M_\bullet$ in hosts with $M_\mathrm{S} \lesssim 10^7$ M$_\odot$ would produce flares with time delays of $\Delta t_\ell \gtrsim 50$ days. In low-mass AGNs, the EM signature is more likely to be produced when the AGN has a relatively high accretion rate, whereas low accretion rates favour EM emission in high-mass AGNs, according to Figure~\ref{fig:MM_corrs}. We also observe that, in general, the kick velocity must be $\gtrsim 150$ km s$^{-1}$ to produce flares with time delays of $\Delta t_\ell < 50$ days.

\section{Optical counterpart candidates to BBH mergers}
\label{sec:opLCs}

Graham et al. \cite{Graham_2020, Graham_2023} have proposed a number of optical flares measured by the ZTF from known AGNs as possible EM counterparts associated with LVK BBH merger events. 
Such MM associations are currently under debate \cite{Palmese_2021, Morton_2023, Veronesi_2025}, mostly because of the large spatial error of the LVK events.

Among the proposed counterparts, the flare beginning at MJD $\sim$58650 from the AGN J124942.3+344929, located at $z=0.438$,  with a central SMBH mass estimated within $10^{8-9}$ M$_\odot$ \cite{Graham_2020} is perhaps the most extensively discussed candidate. This flare has been associated with the event GW190521, with inferred $\chi_\mathrm{eff} = 0.08$ and $q = 0.8$, producing a remnant BH of $\sim 150$ M$_\odot$ placing it in the category of an intermediate-mass black hole. 
The claimed EM counterpart increased the source's emission by 0.3 mag, peaking approximately 50 days after GW190521.

This flare marginally falls within the parameter space of visible flares derived in Figure~\ref{fig:MM_corrs}. As shown in the figure, remnants of $\sim 150$ M$\odot$ in AGNs with $M_\mathrm{S} \sim 10^8$ M$\odot$ produce flares with time lags of $\Delta t_\ell \gtrsim 50$ days. Furthermore, since J124942.3+344929 is more distant than $D_\mathrm{L} = 1600$ Mpc (the distance assumed in Figure~\ref{fig:MM_corrs}), and its SMBH may be more massive than $10^8$ M$_\odot$ \cite{Graham_2020}, the predicted time delays from our model could be even longer unless more optimistic parameters are considered.

On the other hand, the EM counterparts proposed in \cite{Graham_2023} originate from AGNs at $z > 0.3$ with central SMBHs of $M_\mathrm{S} > 10^8$ M$_\odot$ and exhibit time lags of up to $\sim 200$ days after their potential GW associations. These flares align more comfortably with the parameter space of visible flares presented in Figure~\ref{fig:MM_corrs}. 

A detailed investigation of these EM candidates requires an appropriate model for the baseline emission (i.e., the source’s emission before and after the flare) for each individual AGN. Such considerations can clearly alter the predictions obtained here, where a generic SED template was used to represent the host’s background emission (see equation \ref{hostEM}). Therefore, we leave a comprehensive analysis of the flares proposed in \cite{Graham_2020, Graham_2023} under the MM emission scenario presented here for future work.

\section{Summary and conclusions}
\label{sec:conclusions}

In this paper, we present an analytic model that predicts the time delay, duration, spectrum, and optical LC of EM feedback from stellar mass BBH mergers within thin AGN discs and the associated mass ratio and effective spin of the merger. AGNs have been proposed as promising hosts for the observed GW events associated to BBH mergers measured by the LIGO/Virgo/Kagra experiment. The EM processes considered come from the thermal emission of plasma eruptions expanding outside the disc, expelled from the disc by the BH merger remnant. In our  analytic approach (see section \ref{sec:analytic_model} for details), we assume that the BBH is embedded in the disc and co-rotating. After the merger, the remnant is gravitationally kicked out, accretes from the unperturbed disc and produces a transient quasi-spherical outflow which expels disc material. These ejections expand and cool outside the AGN disc, producing thermal flares which we model here.

In our formulation, the EM counterpart is determined by nine parameters characterising the hosting AGN ($D_\mathrm{L}$, $M_\mathrm{S}$, $\dot{M}_\mathrm{S}$), the BBH merger ($a$, $M_\bullet$, $v_\mathrm{k}$, $\theta_\mathrm{k}$), and the remnant's outflow ($\eta_\mathrm{in}$, $\eta_\mathrm{w}$; see Section~\ref{sec:analytic_model} for details). When the model is applied to interpret a particular LC dataset, MM observations can constrain the parameters $D_\mathrm{L}$, $M_\mathrm{S}$, $\dot{M}_\mathrm{S}$, $M_\bullet$, and $v_\mathrm{k}$ (through $q$ and $\chi_\mathrm{eff}$), and then the possible LCs predicted by the model would depend on the other four parameters.

Under the present scenario, BBH mergers produce EM feedback with the following features.

\begin{enumerate}
\item 
The flare emission can be comparable to or outshine that of the hosting AGN at UV-optical frequencies, for mergers occurring in AGNs with SMBHs of $10^{6-8}$ M$_\odot$.

\item 
The EM counterparts are produced by kicked remnants that leave the AGN disc in the same or opposite direction to the observer (see Figure~\ref{fig:sketch}), which we denominate as ``upper'' and ``lower'' flares, respectively. The upper flares are somewhat brighter and have slightly shorter time delays than their lower flare counterparts.

\item 
The kick angle relative to the disc plane $\theta_\mathrm{k}$ is a crucial parameter that modulates the flare profile and determines whether EM feedback is possible (see e.g., equations \ref{E0} and \ref{condition}). Within the parameter space discussed here, our results suggest that significant EM counterparts are possible when $\theta_\mathrm{k} \lesssim 10^\circ$.
Our results suggest that binaries with quasi-anti-aligned spins are the most likely to produce EM counterparts. This spin configuration concentrates most of the kick orientations at small angles relative to the AGN disc plane while also aligning with the observed $q-\chi_\mathrm{eff}$ correlation trend.

\item 
In AGNs with SMBHs of masses $\lesssim 10^7$ M$\odot$, mergers can produce distinguishable flares with time delays as short as $\sim$10 days, requiring kick velocities of $\gtrsim 200$ km s$^{-1}$. For hosts with SMBHs of $\sim 10^8$ M$\odot$, flares with time lags of a few tens of days are possible.

\item 
EM counterparts from AGNs with SMBHs of $10^{6-8}$ M$_\odot$ are suitable targets for time-domain UV space telescopes like the forthcoming ULTRASAT \cite{Shvartzvald_2024} and UVEX \cite{Kulkarni_2021}, as well as for optical time domain surveys (see Figures~\ref{fig:SEDs1e6}-\ref{fig:SEDs1e7}). 

\item We estimate that ZTF can detect the flares predicted here up to $z \sim 0.04-0.1$ from AGNs with $M_\mathrm{S}=10^{6}$ M$_\odot$, and up to $z \sim 0.1-0.3$ when $M_\mathrm{S} \sim 10^7$ M$_\odot$. Rubin could observe the flares up to $z \sim 0.25-0.5$ and $z \sim 0.7 - 1.2$ when $M_\mathrm{S} \sim 10^6$ and $10^7$ M$_\odot$, respectively.
\end{enumerate}

The MM scenario discussed here was previously proposed by \cite{Kimura_2021}. These authors focused on the emission produced when the remnant's outflow breaks out from the disc, predicting X-ray emissions. In contrast to \cite{Kimura_2021}, here we focus on the long-term emission of the plasma ejection as it expands outside the disc plane, predicting flares at optical-UV frequencies as described above.

The emission of the disc eruptions discussed here relies on complex physics that we have simplified for analytical purposes. For instance, the morphology of the disc ejections (which are the sources of the EM counterparts) can be far from spherical, and their evolution could be more accurately described by including continuous injection of matter during the early expansion stages when the remnant's outflow is active.
Such corrections could be suitably addressed through magneto-hydrodynamical simulations of the present problem combined with numerical radiative transfer. Additionally, the outcome of highly super-Eddington inflow/outflows in travelling black holes is currently not well studied, and here we modelled such processes by introducing the efficiency parameters $\eta_\mathrm{in}$ and $\eta_\mathrm{w}$. Thus, if the emission scenario analysed here is indeed correct, MM observations from BBH mergers in AGNs could also serve to probe the physics of hyper-Eddington accretion onto black holes.

\begin{acknowledgments}
We are thankful to the anonymous reviewer for providing valuable
suggestions that improved the quality of this work. 
JCRR acknowledges support from Rio de Janeiro State Funding Agency FAPERJ, grant E-26/205.635/2022.
CdB acknowledges the financial support from CNPq (316072/2021-4) and from FAPERJ (grants 201.456/2022 and 210.330/2022) and the FINEP contract 01.22.0505.00 (ref. 1891/22). RN acknowledges support from CNPq through a Bolsa de Produtividade and NASA through the {\it Fermi} Guest Investigator Program (Cycle 16).
The authors made use of Sci-Mind servers machines developed by the CBPF AI LAB team and would like to thank P. Russano and M. Portes de Albuquerque for all the support in infrastructure matters.
\end{acknowledgments}

\appendix

\section{AGN thin disc models}
\label{app:dmods}
Here, we compare the radial profiles of AGN thin discs as derived from the models of Shakura \& Sunyaev (\cite{SS_1973}; SS73), Sirko \& Goodman (\cite{SG_2003}; SG03), and Thompson et al. (\cite{TQM_2005}; TQM05), focusing on the physical properties relevant to the multi-messenger emission scenario discussed in this paper.

To calculate the SS73 radial profiles, we employ the prescription presented in \cite{Kato_2008} for the inner, middle,
and outer radial regions (according to the gas/radiation pressure and opacity dominance). To obtain the radial profiles of the SG03 and TQM05 models, we employ the 
\texttt{pAGN}, Python based code
\cite{Gangardt_2024}\footnote{\href{https://github.com/DariaGangardt/pAGN}{https://github.com/DariaGangardt/pAGN}} with some additional specifications motivated as follows.
 
The SG03 disc model predicts an additional SED bump feature at near-infrared compared to the disc spectrum
of the standard SS73 disc.
This additional emission originates at the disc outer regions,
due to the implemented heating mechanism generating radiation pressure
against gravitational collapse.
This emission excess can be employed to limit the size of the disc, 
so that the disc emission matches observations.
To generate emission from SG03 discs consistent with the host galaxy emission template employed in the present study (see equation~\ref{hostEM} of the main text),
we set the outer boundary of the SG03 disc as 
\begin{equation}
a_\mathrm{max}=
\left(
\frac{M_\mathrm{SMBH} }{10^8 \mathrm{M}_\odot} 
\right)^{-1/2}
10^4  R_\mathrm{g}.
\label{amax_SG03}
\end{equation}
with $R_\mathrm{g} = GM_\mathrm{S}/c^2$.

The TQM05 model addresses gas support against self-gravity in the outer region of the disc through star formation by-products. This formulation requires specifying the relation of the velocity dispersion $\sigma$ with the central mass provided by observations, for which we employ the one obtained by \cite{Kormendy_2013} (their equation 7).
A distinctive characteristic of the TQM05 disc solution is that the accretion rate changes as a function of radial distance to the central engine due to mass loss from star formation. To produce TQM05 solutions that can be compared to the SS73 and SG03 counterparts, we modulate the accretion rate $\dot{M}\mathrm{out}$ of the TQM05 disc at a distance $r\mathrm{out}=10^7 R_\mathrm{S}$, where $R_\mathrm{S}=2GM_\mathrm{SMBH}/c^2$, to obtain disc solutions with the desired accretion rate $\dot{M}\mathrm{S}$ at $3R\mathrm{S}$ (the innermost stable circular orbit of a Schwarzschild SMBH).
Specifically, to produce TQM05 disc solutions with accretion rates of $\dot{M}\mathrm{S} = 0.0075$, $0.0100$, $0.0250$, $0.0500$, and $0.1000$ $\dot{M}\mathrm{Edd}$, we set the outer accretion rates to
$\dot{M}\mathrm{out} = 0.450$, $0.920$, $1.972$, $2.925$, and $3.700$ $\dot{M}\mathrm{Edd}$
for an SMBH of $10^8$ M$\odot$, and
$\dot{M}\mathrm{out} = 0.0075$, $0.0100$, $0.0290$, $0.1100$, $0.2970$ $\dot{M}\mathrm{Edd}$
for an SMBH of $10^7$ M$\odot$.
For SMBHs of $10^6$ M$\odot$, our calculations require $\dot{M}\mathrm{out} = \dot{M}_\mathrm{S}$.

\begin{figure*}
   \centering
   \includegraphics[width=\hsize]{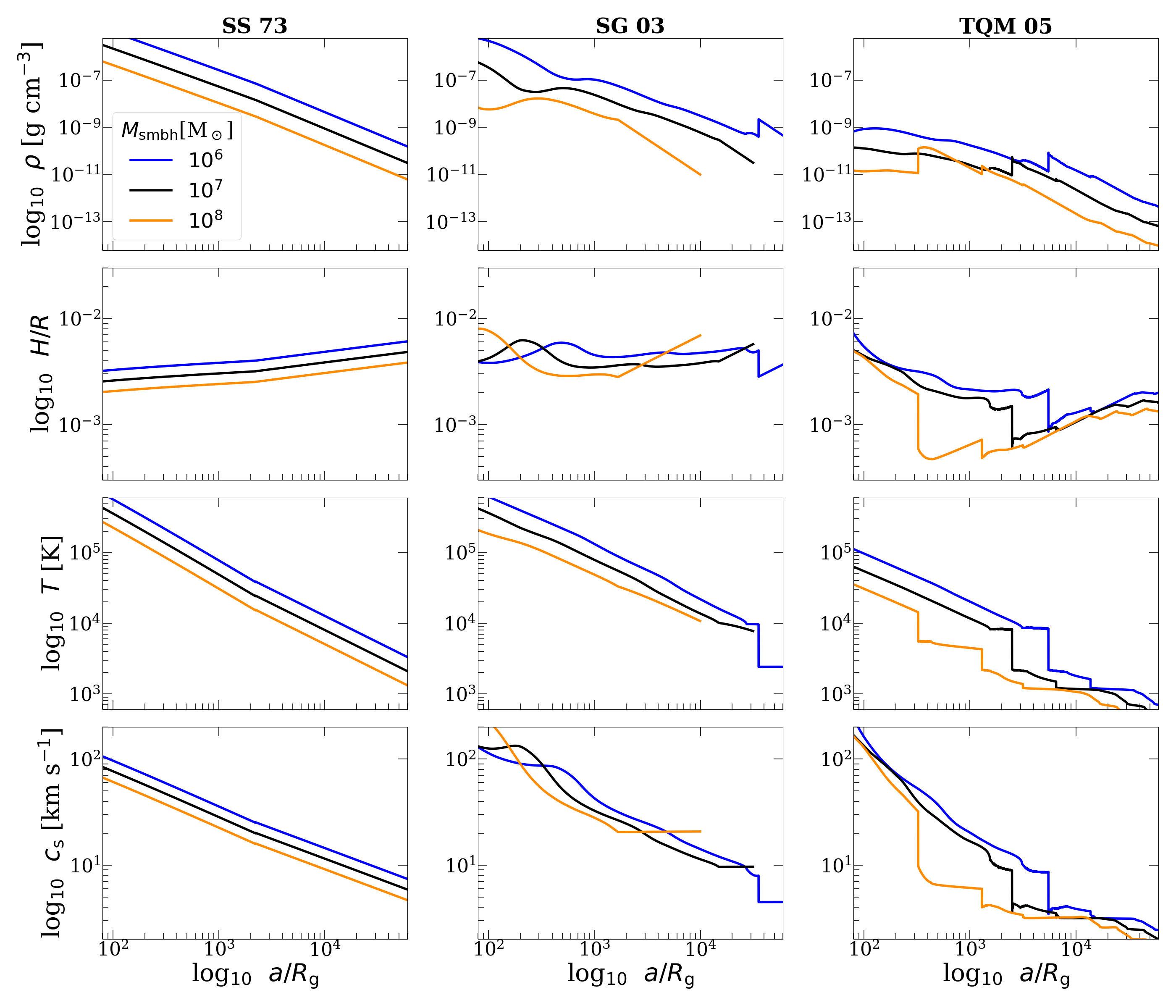}
      \caption{
Radial profiles for the properties of an AGN thin disc, as predicted by the SS73~\cite{SS_1973}, SG03~\cite{SG_2003}, and TQM05~\cite{TQM_2005} models, as indicated (see the text for further details). Solutions for the gas density, aspect ratio, mid-disc temperature, and speed of sound are shown in the different rows of this plot array, from the upper to the lower panels, respectively. All solutions correspond to an accretion rate of $\dot{m}=0.025$ at the central engine, in Eddington units, and curves in different colours correspond to different assumed SMBH masses.
              }
         \label{fig:discs_props}
\end{figure*}

\begin{figure*}
   \centering
   \includegraphics[width=\hsize]{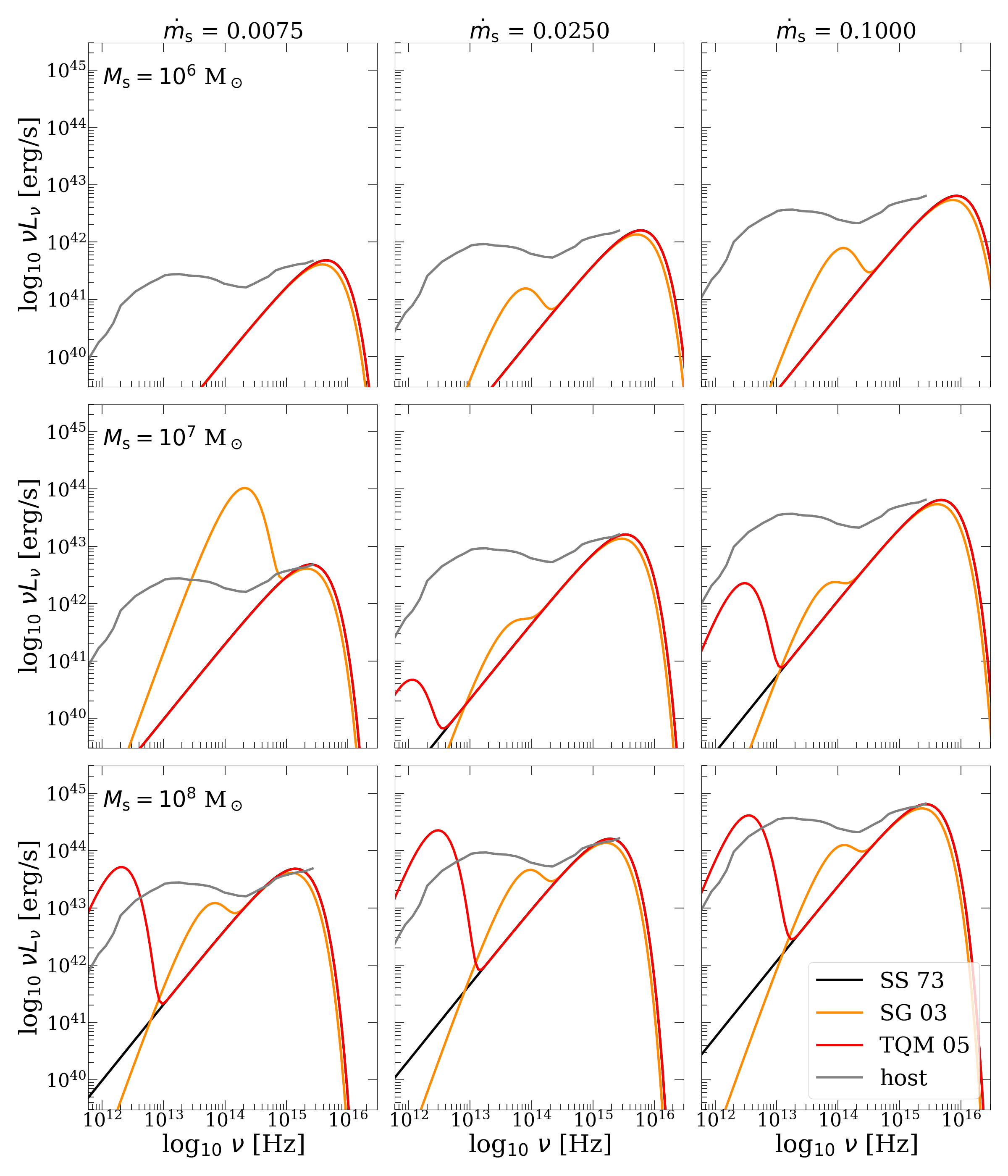}
      \caption{
Spectral energy distributions predicted by the SS73, SG03, and TQM05 AGN disc models described in the text, as indicated. The grey curve represents the spectrum template for the host galaxy adopted in this paper as the quiescent AGN emission. The panels in different columns are obtained by assuming different accretion rates onto the central engine, whereas the panels in different rows correspond to different SMBH masses, as indicated.
              }
         \label{fig:discsSEDs}
\end{figure*}

In Figure~\ref{fig:discs_props}, we illustrate selected disc properties obtained from the SS73, SG03, and TQM05 models as specified above, and in Figure~\ref{fig:discsSEDs}, we show their predicted spectra for different configurations of SMBH masses and accretion rates, as indicated.
We observe that, within the radial domain shown in Figure~\ref{fig:discs_props}, the SG03 model predicts properties of the same order of magnitude as those of the SS73 model. The curves produced by the SG03 model terminate at the outer disc boundary given by equation~\ref{amax_SG03}.
The TQM05 model, on the other hand, predicts gas densities that are 2–3 orders of magnitude lower than those of the SG03 and SS73 models and a disc aspect ratio about one order of magnitude lower for SMBHs of $10^{7-8}$ M$_\odot$.
The spectra emitted by these disc models converge in the optical and UV bands to the template emission for the host galaxy (grey curve), which we employ in this paper as the AGN quiescent state.

\section{The initial pressure of the plasma ejection}
\label{app:radp}
In this appendix, we evaluate whether the disc ejections of plasma are initially gas pressure or radiation pressure dominated. We consider the total pressure as the sum of gas plus radiation pressure:
\begin{equation}
\label{Ptotal}
P = P_\mathrm{g} + P_\mathrm{r},
\end{equation}
where
\begin{align}
\label{Pg}
P_g & = 2 n_i k_B T, \\
\label{Pr}
P_r & = \frac{1}{3} a_r T^4.
\end{align}
In equation \ref{Pg}, $n_i$ denotes the number density of ions, and the factor of two corresponds to a fully ionised gas.
To determine which pressure component is dominant, we consider the parameter
\begin{equation}
\beta_r \equiv \frac{P_g}{P},
\end{equation}
where the cases $\beta_r \ll 1$ and $\beta_r \sim 1$ correspond to radiation pressure and gas pressure dominated regimes, respectively.
In terms of $\beta_r$, the pressure components are:
\begin{align}
\label{Pg_beta}
P_g & = \beta_r P, \\
\label{Pr_beta}
P_r & = (1 - \beta_r) P.
\end{align}
Combining equations \eqref{Ptotal}--\eqref{Pr_beta}, the parameter $\beta_r$ can be obtained by solving the quartic equation
\begin{equation}
\label{quartic}
A \beta_r^4 + \beta_r - 1 = 0,
\end{equation}
where $A$ is the dimensionless parameter
\begin{equation}
A = \frac{a_r P^3}{3 (2 n_i k_B)^4}.
\end{equation}
We estimate a lower limit for the typical values of the parameter $A$ using $P \approx E_\mathrm{0,th}/(3V_0)$, which we underestimate using 
$V_0 \approx \pi h_\mathrm{d}^3 / (2\theta_\mathrm{k}^2)$, where $V_0$ is an over-estimation for the initial volume of the disc eruption (see equation \ref{V0_approx}), and $E_\mathrm{0,th} = E_\mathrm{0}/2$, where $E_0$ is given by equation \ref{E_0ll}. Then we obtain:
\begin{align}
A &\sim \frac{1}{162} \frac{a_\mathrm{r} m_\mathrm{p}^3}{k_\mathrm{B}^4}
\frac{(\eta_\mathrm{w}\eta_\mathrm{in})^3}{n_\mathrm{i}}
\left(
\frac{\theta_\mathrm{k} G M_\bullet c}{h_\mathrm{d}}
\right)^6
\frac{1}{v_\mathrm{k}^{12}}\\
\nonumber
&\sim 8.229\times10^6 \left(
\frac{\eta_\mathrm{w}}{0.05}
\right)^3
\left(
\frac{\eta_\mathrm{in}}{0.1}
\right)^3
\left(
\frac{10^\mathrm{14}\, \mathrm{cm}^{-3}}{n_\mathrm{i}}
\right)\\
\nonumber
&\times \left(
\frac{\theta_\mathrm{k}}{10^\circ}
\right)^6
\left(
\frac{M_\bullet}{20 \mathrm{M}_\odot}
\right)^6
\left(
\frac{0.005}{h_\mathrm{d}/a}
\right)^6
\left(
\frac{10^4}{a/R_\mathrm{g}}
\right)^6 \\
\nonumber
&\times
\left(
\frac{10^7 \,\mathrm{M}_\odot}{M_\mathrm{S}}
\right)^6
\left(
\frac{200\,\mathrm{kms}^{-1}}{v_\mathrm{k}}
\right)^{12},
\end{align}
Considering values for $A$ within $(10^4, 10^{8})$, solutions to equation \ref{quartic} give values for $\beta_\mathrm{r}$ within $(10^{-5}, 0.1)$. Therefore, we can safely assume that the plasma eruptions discussed in the paper are initially radiation pressure dominated.
 
\bibliographystyle{apsrev4-2}
\bibliography{refs}

\providecommand{\noopsort}[1]{}\providecommand{\singleletter}[1]{#1}%
\begin{thebibliography}{100}%
\makeatletter
\providecommand \@ifxundefined [1]{%
 \@ifx{#1\undefined}
}%
\providecommand \@ifnum [1]{%
 \ifnum #1\expandafter \@firstoftwo
 \else \expandafter \@secondoftwo
 \fi
}%
\providecommand \@ifx [1]{%
 \ifx #1\expandafter \@firstoftwo
 \else \expandafter \@secondoftwo
 \fi
}%
\providecommand \natexlab [1]{#1}%
\providecommand \enquote  [1]{``#1''}%
\providecommand \bibnamefont  [1]{#1}%
\providecommand \bibfnamefont [1]{#1}%
\providecommand \citenamefont [1]{#1}%
\providecommand \href@noop [0]{\@secondoftwo}%
\providecommand \href [0]{\begingroup \@sanitize@url \@href}%
\providecommand \@href[1]{\@@startlink{#1}\@@href}%
\providecommand \@@href[1]{\endgroup#1\@@endlink}%
\providecommand \@sanitize@url [0]{\catcode `\\12\catcode `\$12\catcode `\&12\catcode `\#12\catcode `\^12\catcode `\_12\catcode `\%12\relax}%
\providecommand \@@startlink[1]{}%
\providecommand \@@endlink[0]{}%
\providecommand \url  [0]{\begingroup\@sanitize@url \@url }%
\providecommand \@url [1]{\endgroup\@href {#1}{\urlprefix }}%
\providecommand \urlprefix  [0]{URL }%
\providecommand \Eprint [0]{\href }%
\providecommand \doibase [0]{https://doi.org/}%
\providecommand \selectlanguage [0]{\@gobble}%
\providecommand \bibinfo  [0]{\@secondoftwo}%
\providecommand \bibfield  [0]{\@secondoftwo}%
\providecommand \translation [1]{[#1]}%
\providecommand \BibitemOpen [0]{}%
\providecommand \bibitemStop [0]{}%
\providecommand \bibitemNoStop [0]{.\EOS\space}%
\providecommand \EOS [0]{\spacefactor3000\relax}%
\providecommand \BibitemShut  [1]{\csname bibitem#1\endcsname}%
\let\auto@bib@innerbib\@empty
\bibitem [{\citenamefont {Abbott}\ \emph {et~al.}(2023)\citenamefont {Abbott}, \citenamefont {Abbott}, \citenamefont {Acernese}, \citenamefont {Ackley}, \citenamefont {Adams}, \citenamefont {Adhikari}, \citenamefont {Adhikari}, \citenamefont {Adya}, \citenamefont {Affeldt}, \citenamefont {Agarwal} \emph {et~al.}}]{bbh_o3}%
  \BibitemOpen
  \bibfield  {author} {\bibinfo {author} {\bibfnamefont {R.}~\bibnamefont {Abbott}}, \bibinfo {author} {\bibfnamefont {T.}~\bibnamefont {Abbott}}, \bibinfo {author} {\bibfnamefont {F.}~\bibnamefont {Acernese}}, \bibinfo {author} {\bibfnamefont {K.}~\bibnamefont {Ackley}}, \bibinfo {author} {\bibfnamefont {C.}~\bibnamefont {Adams}}, \bibinfo {author} {\bibfnamefont {N.}~\bibnamefont {Adhikari}}, \bibinfo {author} {\bibfnamefont {R.}~\bibnamefont {Adhikari}}, \bibinfo {author} {\bibfnamefont {V.}~\bibnamefont {Adya}}, \bibinfo {author} {\bibfnamefont {C.}~\bibnamefont {Affeldt}}, \bibinfo {author} {\bibfnamefont {D.}~\bibnamefont {Agarwal}}, \emph {et~al.},\ }\href@noop {} {\bibfield  {journal} {\bibinfo  {journal} {Physical Review X}\ }\textbf {\bibinfo {volume} {13}},\ \bibinfo {pages} {011048} (\bibinfo {year} {2023})}\BibitemShut {NoStop}%
\bibitem [{\citenamefont {Sadiq}\ \emph {et~al.}(2024)\citenamefont {Sadiq}, \citenamefont {Dent},\ and\ \citenamefont {Gieles}}]{Sadiq:2023zee}%
  \BibitemOpen
  \bibfield  {author} {\bibinfo {author} {\bibfnamefont {J.}~\bibnamefont {Sadiq}}, \bibinfo {author} {\bibfnamefont {T.}~\bibnamefont {Dent}},\ and\ \bibinfo {author} {\bibfnamefont {M.}~\bibnamefont {Gieles}},\ }\href {https://doi.org/10.3847/1538-4357/ad0ce6} {\bibfield  {journal} {\bibinfo  {journal} {Astrophys. J.}\ }\textbf {\bibinfo {volume} {960}},\ \bibinfo {pages} {65} (\bibinfo {year} {2024})},\ \Eprint {https://arxiv.org/abs/2307.12092} {arXiv:2307.12092 [astro-ph.HE]} \BibitemShut {NoStop}%
\bibitem [{\citenamefont {{Petrov}}\ \emph {et~al.}(2022)\citenamefont {{Petrov}}, \citenamefont {{Singer}}, \citenamefont {{Coughlin}}, \citenamefont {{Kumar}}, \citenamefont {{Almualla}}, \citenamefont {{Anand}}, \citenamefont {{Bulla}}, \citenamefont {{Dietrich}}, \citenamefont {{Foucart}},\ and\ \citenamefont {{Guessoum}}}]{petrov_2022}%
  \BibitemOpen
  \bibfield  {author} {\bibinfo {author} {\bibfnamefont {P.}~\bibnamefont {{Petrov}}}, \bibinfo {author} {\bibfnamefont {L.~P.}\ \bibnamefont {{Singer}}}, \bibinfo {author} {\bibfnamefont {M.~W.}\ \bibnamefont {{Coughlin}}}, \bibinfo {author} {\bibfnamefont {V.}~\bibnamefont {{Kumar}}}, \bibinfo {author} {\bibfnamefont {M.}~\bibnamefont {{Almualla}}}, \bibinfo {author} {\bibfnamefont {S.}~\bibnamefont {{Anand}}}, \bibinfo {author} {\bibfnamefont {M.}~\bibnamefont {{Bulla}}}, \bibinfo {author} {\bibfnamefont {T.}~\bibnamefont {{Dietrich}}}, \bibinfo {author} {\bibfnamefont {F.}~\bibnamefont {{Foucart}}},\ and\ \bibinfo {author} {\bibfnamefont {N.}~\bibnamefont {{Guessoum}}},\ }\href {https://doi.org/10.3847/1538-4357/ac366d} {\bibfield  {journal} {\bibinfo  {journal} {\apj}\ }\textbf {\bibinfo {volume} {924}},\ \bibinfo {eid} {54} (\bibinfo {year} {2022})},\ \Eprint {https://arxiv.org/abs/2108.07277} {arXiv:2108.07277 [astro-ph.HE]} \BibitemShut {NoStop}%
\bibitem [{\citenamefont {Abbott}\ \emph {et~al.}(2020)\citenamefont {Abbott}, \citenamefont {Abbott}, \citenamefont {Abbott}, \citenamefont {Abraham}, \citenamefont {Acernese}, \citenamefont {Ackley}, \citenamefont {Adams}, \citenamefont {Adya}, \citenamefont {Affeldt}, \citenamefont {Agathos} \emph {et~al.}}]{abbott2020prospects}%
  \BibitemOpen
  \bibfield  {author} {\bibinfo {author} {\bibfnamefont {B.~P.}\ \bibnamefont {Abbott}}, \bibinfo {author} {\bibfnamefont {R.}~\bibnamefont {Abbott}}, \bibinfo {author} {\bibfnamefont {T.}~\bibnamefont {Abbott}}, \bibinfo {author} {\bibfnamefont {S.}~\bibnamefont {Abraham}}, \bibinfo {author} {\bibfnamefont {F.}~\bibnamefont {Acernese}}, \bibinfo {author} {\bibfnamefont {K.}~\bibnamefont {Ackley}}, \bibinfo {author} {\bibfnamefont {C.}~\bibnamefont {Adams}}, \bibinfo {author} {\bibfnamefont {V.}~\bibnamefont {Adya}}, \bibinfo {author} {\bibfnamefont {C.}~\bibnamefont {Affeldt}}, \bibinfo {author} {\bibfnamefont {M.}~\bibnamefont {Agathos}}, \emph {et~al.},\ }\href@noop {} {\bibfield  {journal} {\bibinfo  {journal} {Living reviews in relativity}\ }\textbf {\bibinfo {volume} {23}},\ \bibinfo {pages} {1} (\bibinfo {year} {2020})}\BibitemShut {NoStop}%
\bibitem [{\citenamefont {{Ford}}\ and\ \citenamefont {{McKernan}}(2022)}]{bbh_chanel01}%
  \BibitemOpen
  \bibfield  {author} {\bibinfo {author} {\bibfnamefont {K.~E.~S.}\ \bibnamefont {{Ford}}}\ and\ \bibinfo {author} {\bibfnamefont {B.}~\bibnamefont {{McKernan}}},\ }\href {https://doi.org/10.1093/mnras/stac2861} {\bibfield  {journal} {\bibinfo  {journal} {\mnras}\ }\textbf {\bibinfo {volume} {517}},\ \bibinfo {pages} {5827} (\bibinfo {year} {2022})},\ \Eprint {https://arxiv.org/abs/2109.03212} {arXiv:2109.03212 [astro-ph.HE]} \BibitemShut {NoStop}%
\bibitem [{\citenamefont {{Gayathri}}\ \emph {et~al.}(2023)\citenamefont {{Gayathri}}, \citenamefont {{Wysocki}}, \citenamefont {{Yang}}, \citenamefont {{Delfavero}}, \citenamefont {{O'Shaughnessy}}, \citenamefont {{Haiman}}, \citenamefont {{Tagawa}},\ and\ \citenamefont {{Bartos}}}]{bbh_chanel02}%
  \BibitemOpen
  \bibfield  {author} {\bibinfo {author} {\bibfnamefont {V.}~\bibnamefont {{Gayathri}}}, \bibinfo {author} {\bibfnamefont {D.}~\bibnamefont {{Wysocki}}}, \bibinfo {author} {\bibfnamefont {Y.}~\bibnamefont {{Yang}}}, \bibinfo {author} {\bibfnamefont {V.}~\bibnamefont {{Delfavero}}}, \bibinfo {author} {\bibfnamefont {R.}~\bibnamefont {{O'Shaughnessy}}}, \bibinfo {author} {\bibfnamefont {Z.}~\bibnamefont {{Haiman}}}, \bibinfo {author} {\bibfnamefont {H.}~\bibnamefont {{Tagawa}}},\ and\ \bibinfo {author} {\bibfnamefont {I.}~\bibnamefont {{Bartos}}},\ }\href {https://doi.org/10.3847/2041-8213/acbfb8} {\bibfield  {journal} {\bibinfo  {journal} {\apjl}\ }\textbf {\bibinfo {volume} {945}},\ \bibinfo {eid} {L29} (\bibinfo {year} {2023})},\ \Eprint {https://arxiv.org/abs/2301.04187} {arXiv:2301.04187 [gr-qc]} \BibitemShut {NoStop}%
\bibitem [{\citenamefont {{Gayathri}}\ \emph {et~al.}(2021)\citenamefont {{Gayathri}}, \citenamefont {{Yang}}, \citenamefont {{Tagawa}}, \citenamefont {{Haiman}},\ and\ \citenamefont {{Bartos}}}]{bbh_chanel03}%
  \BibitemOpen
  \bibfield  {author} {\bibinfo {author} {\bibfnamefont {V.}~\bibnamefont {{Gayathri}}}, \bibinfo {author} {\bibfnamefont {Y.}~\bibnamefont {{Yang}}}, \bibinfo {author} {\bibfnamefont {H.}~\bibnamefont {{Tagawa}}}, \bibinfo {author} {\bibfnamefont {Z.}~\bibnamefont {{Haiman}}},\ and\ \bibinfo {author} {\bibfnamefont {I.}~\bibnamefont {{Bartos}}},\ }\href {https://doi.org/10.3847/2041-8213/ac2cc1} {\bibfield  {journal} {\bibinfo  {journal} {\apjl}\ }\textbf {\bibinfo {volume} {920}},\ \bibinfo {eid} {L42} (\bibinfo {year} {2021})},\ \Eprint {https://arxiv.org/abs/2104.10253} {arXiv:2104.10253 [gr-qc]} \BibitemShut {NoStop}%
\bibitem [{\citenamefont {{Connaughton}}\ \emph {et~al.}(2016)\citenamefont {{Connaughton}}, \citenamefont {{Burns}}, \citenamefont {{Goldstein}}, \citenamefont {{Blackburn}}, \citenamefont {{Briggs}}, \citenamefont {{Zhang}}, \citenamefont {{Camp}}, \citenamefont {{Christensen}}, \citenamefont {{Hui}}, \citenamefont {{Jenke}}, \citenamefont {{Littenberg}}, \citenamefont {{McEnery}}, \citenamefont {{Racusin}}, \citenamefont {{Shawhan}}, \citenamefont {{Singer}}, \citenamefont {{Veitch}}, \citenamefont {{Wilson-Hodge}}, \citenamefont {{Bhat}}, \citenamefont {{Bissaldi}}, \citenamefont {{Cleveland}}, \citenamefont {{Fitzpatrick}}, \citenamefont {{Giles}}, \citenamefont {{Gibby}}, \citenamefont {{von Kienlin}}, \citenamefont {{Kippen}}, \citenamefont {{McBreen}}, \citenamefont {{Mailyan}}, \citenamefont {{Meegan}}, \citenamefont {{Paciesas}}, \citenamefont {{Preece}}, \citenamefont {{Roberts}}, \citenamefont {{Sparke}}, \citenamefont {{Stanbro}}, \citenamefont {{Toelge}},\ and\ \citenamefont
  {{Veres}}}]{connaughton_2016}%
  \BibitemOpen
  \bibfield  {author} {\bibinfo {author} {\bibfnamefont {V.}~\bibnamefont {{Connaughton}}}, \bibinfo {author} {\bibfnamefont {E.}~\bibnamefont {{Burns}}}, \bibinfo {author} {\bibfnamefont {A.}~\bibnamefont {{Goldstein}}}, \bibinfo {author} {\bibfnamefont {L.}~\bibnamefont {{Blackburn}}}, \bibinfo {author} {\bibfnamefont {M.~S.}\ \bibnamefont {{Briggs}}}, \bibinfo {author} {\bibfnamefont {B.~B.}\ \bibnamefont {{Zhang}}}, \bibinfo {author} {\bibfnamefont {J.}~\bibnamefont {{Camp}}}, \bibinfo {author} {\bibfnamefont {N.}~\bibnamefont {{Christensen}}}, \bibinfo {author} {\bibfnamefont {C.~M.}\ \bibnamefont {{Hui}}}, \bibinfo {author} {\bibfnamefont {P.}~\bibnamefont {{Jenke}}}, \bibinfo {author} {\bibfnamefont {T.}~\bibnamefont {{Littenberg}}}, \bibinfo {author} {\bibfnamefont {J.~E.}\ \bibnamefont {{McEnery}}}, \bibinfo {author} {\bibfnamefont {J.}~\bibnamefont {{Racusin}}}, \bibinfo {author} {\bibfnamefont {P.}~\bibnamefont {{Shawhan}}}, \bibinfo {author} {\bibfnamefont {L.}~\bibnamefont {{Singer}}}, \bibinfo
  {author} {\bibfnamefont {J.}~\bibnamefont {{Veitch}}}, \bibinfo {author} {\bibfnamefont {C.~A.}\ \bibnamefont {{Wilson-Hodge}}}, \bibinfo {author} {\bibfnamefont {P.~N.}\ \bibnamefont {{Bhat}}}, \bibinfo {author} {\bibfnamefont {E.}~\bibnamefont {{Bissaldi}}}, \bibinfo {author} {\bibfnamefont {W.}~\bibnamefont {{Cleveland}}}, \bibinfo {author} {\bibfnamefont {G.}~\bibnamefont {{Fitzpatrick}}}, \bibinfo {author} {\bibfnamefont {M.~M.}\ \bibnamefont {{Giles}}}, \bibinfo {author} {\bibfnamefont {M.~H.}\ \bibnamefont {{Gibby}}}, \bibinfo {author} {\bibfnamefont {A.}~\bibnamefont {{von Kienlin}}}, \bibinfo {author} {\bibfnamefont {R.~M.}\ \bibnamefont {{Kippen}}}, \bibinfo {author} {\bibfnamefont {S.}~\bibnamefont {{McBreen}}}, \bibinfo {author} {\bibfnamefont {B.}~\bibnamefont {{Mailyan}}}, \bibinfo {author} {\bibfnamefont {C.~A.}\ \bibnamefont {{Meegan}}}, \bibinfo {author} {\bibfnamefont {W.~S.}\ \bibnamefont {{Paciesas}}}, \bibinfo {author} {\bibfnamefont {R.~D.}\ \bibnamefont {{Preece}}}, \bibinfo {author}
  {\bibfnamefont {O.~J.}\ \bibnamefont {{Roberts}}}, \bibinfo {author} {\bibfnamefont {L.}~\bibnamefont {{Sparke}}}, \bibinfo {author} {\bibfnamefont {M.}~\bibnamefont {{Stanbro}}}, \bibinfo {author} {\bibfnamefont {K.}~\bibnamefont {{Toelge}}},\ and\ \bibinfo {author} {\bibfnamefont {P.}~\bibnamefont {{Veres}}},\ }\href {https://doi.org/10.3847/2041-8205/826/1/L6} {\bibfield  {journal} {\bibinfo  {journal} {\apjl}\ }\textbf {\bibinfo {volume} {826}},\ \bibinfo {eid} {L6} (\bibinfo {year} {2016})},\ \Eprint {https://arxiv.org/abs/1602.03920} {arXiv:1602.03920 [astro-ph.HE]} \BibitemShut {NoStop}%
\bibitem [{\citenamefont {{Bagoly}}\ \emph {et~al.}(2016)\citenamefont {{Bagoly}}, \citenamefont {{Sz{\'e}csi}}, \citenamefont {{Bal{\'a}zs}}, \citenamefont {{Csabai}}, \citenamefont {{Horv{\'a}th}}, \citenamefont {{Dobos}}, \citenamefont {{Lichtenberger}},\ and\ \citenamefont {{T{\'o}th}}}]{Bagoly_2016}%
  \BibitemOpen
  \bibfield  {author} {\bibinfo {author} {\bibfnamefont {Z.}~\bibnamefont {{Bagoly}}}, \bibinfo {author} {\bibfnamefont {D.}~\bibnamefont {{Sz{\'e}csi}}}, \bibinfo {author} {\bibfnamefont {L.~G.}\ \bibnamefont {{Bal{\'a}zs}}}, \bibinfo {author} {\bibfnamefont {I.}~\bibnamefont {{Csabai}}}, \bibinfo {author} {\bibfnamefont {I.}~\bibnamefont {{Horv{\'a}th}}}, \bibinfo {author} {\bibfnamefont {L.}~\bibnamefont {{Dobos}}}, \bibinfo {author} {\bibfnamefont {J.}~\bibnamefont {{Lichtenberger}}},\ and\ \bibinfo {author} {\bibfnamefont {L.~V.}\ \bibnamefont {{T{\'o}th}}},\ }\href {https://doi.org/10.1051/0004-6361/201628569} {\bibfield  {journal} {\bibinfo  {journal} {\aap}\ }\textbf {\bibinfo {volume} {593}},\ \bibinfo {eid} {L10} (\bibinfo {year} {2016})},\ \Eprint {https://arxiv.org/abs/1603.06611} {arXiv:1603.06611 [astro-ph.HE]} \BibitemShut {NoStop}%
\bibitem [{\citenamefont {{Becerra}}\ \emph {et~al.}(2021)\citenamefont {{Becerra}}, \citenamefont {{Dichiara}}, \citenamefont {{Watson}}, \citenamefont {{Troja}}, \citenamefont {{Butler}}, \citenamefont {{Pereyra}}, \citenamefont {{Moreno M{\'e}ndez}}, \citenamefont {{De Colle}}, \citenamefont {{Lee}}, \citenamefont {{Kutyrev}},\ and\ \citenamefont {{L{\'o}pez}}}]{becerra_2021}%
  \BibitemOpen
  \bibfield  {author} {\bibinfo {author} {\bibfnamefont {R.~L.}\ \bibnamefont {{Becerra}}}, \bibinfo {author} {\bibfnamefont {S.}~\bibnamefont {{Dichiara}}}, \bibinfo {author} {\bibfnamefont {A.~M.}\ \bibnamefont {{Watson}}}, \bibinfo {author} {\bibfnamefont {E.}~\bibnamefont {{Troja}}}, \bibinfo {author} {\bibfnamefont {N.~R.}\ \bibnamefont {{Butler}}}, \bibinfo {author} {\bibfnamefont {M.}~\bibnamefont {{Pereyra}}}, \bibinfo {author} {\bibfnamefont {E.}~\bibnamefont {{Moreno M{\'e}ndez}}}, \bibinfo {author} {\bibfnamefont {F.}~\bibnamefont {{De Colle}}}, \bibinfo {author} {\bibfnamefont {W.~H.}\ \bibnamefont {{Lee}}}, \bibinfo {author} {\bibfnamefont {A.~S.}\ \bibnamefont {{Kutyrev}}},\ and\ \bibinfo {author} {\bibfnamefont {K.~O.~C.}\ \bibnamefont {{L{\'o}pez}}},\ }\href {https://doi.org/10.1093/mnras/stab2086} {\bibfield  {journal} {\bibinfo  {journal} {MNRAS}\ }\textbf {\bibinfo {volume} {507}},\ \bibinfo {pages} {1401} (\bibinfo {year} {2021})},\ \Eprint {https://arxiv.org/abs/2106.15075}
  {arXiv:2106.15075 [astro-ph.HE]} \BibitemShut {NoStop}%
\bibitem [{\citenamefont {{Graham}}\ \emph {et~al.}(2020)\citenamefont {{Graham}}, \citenamefont {{Ford}}, \citenamefont {{McKernan}}, \citenamefont {{Ross}}, \citenamefont {{Stern}}, \citenamefont {{Burdge}}, \citenamefont {{Coughlin}}, \citenamefont {{Djorgovski}}, \citenamefont {{Drake}}, \citenamefont {{Duev}}, \citenamefont {{Kasliwal}}, \citenamefont {{Mahabal}}, \citenamefont {{van Velzen}}, \citenamefont {{Belecki}}, \citenamefont {{Bellm}}, \citenamefont {{Burruss}}, \citenamefont {{Cenko}}, \citenamefont {{Cunningham}}, \citenamefont {{Helou}}, \citenamefont {{Kulkarni}}, \citenamefont {{Masci}}, \citenamefont {{Prince}}, \citenamefont {{Reiley}}, \citenamefont {{Rodriguez}}, \citenamefont {{Rusholme}}, \citenamefont {{Smith}},\ and\ \citenamefont {{Soumagnac}}}]{Graham_2020}%
  \BibitemOpen
  \bibfield  {author} {\bibinfo {author} {\bibfnamefont {M.~J.}\ \bibnamefont {{Graham}}}, \bibinfo {author} {\bibfnamefont {K.~E.~S.}\ \bibnamefont {{Ford}}}, \bibinfo {author} {\bibfnamefont {B.}~\bibnamefont {{McKernan}}}, \bibinfo {author} {\bibfnamefont {N.~P.}\ \bibnamefont {{Ross}}}, \bibinfo {author} {\bibfnamefont {D.}~\bibnamefont {{Stern}}}, \bibinfo {author} {\bibfnamefont {K.}~\bibnamefont {{Burdge}}}, \bibinfo {author} {\bibfnamefont {M.}~\bibnamefont {{Coughlin}}}, \bibinfo {author} {\bibfnamefont {S.~G.}\ \bibnamefont {{Djorgovski}}}, \bibinfo {author} {\bibfnamefont {A.~J.}\ \bibnamefont {{Drake}}}, \bibinfo {author} {\bibfnamefont {D.}~\bibnamefont {{Duev}}}, \bibinfo {author} {\bibfnamefont {M.}~\bibnamefont {{Kasliwal}}}, \bibinfo {author} {\bibfnamefont {A.~A.}\ \bibnamefont {{Mahabal}}}, \bibinfo {author} {\bibfnamefont {S.}~\bibnamefont {{van Velzen}}}, \bibinfo {author} {\bibfnamefont {J.}~\bibnamefont {{Belecki}}}, \bibinfo {author} {\bibfnamefont {E.~C.}\ \bibnamefont {{Bellm}}},
  \bibinfo {author} {\bibfnamefont {R.}~\bibnamefont {{Burruss}}}, \bibinfo {author} {\bibfnamefont {S.~B.}\ \bibnamefont {{Cenko}}}, \bibinfo {author} {\bibfnamefont {V.}~\bibnamefont {{Cunningham}}}, \bibinfo {author} {\bibfnamefont {G.}~\bibnamefont {{Helou}}}, \bibinfo {author} {\bibfnamefont {S.~R.}\ \bibnamefont {{Kulkarni}}}, \bibinfo {author} {\bibfnamefont {F.~J.}\ \bibnamefont {{Masci}}}, \bibinfo {author} {\bibfnamefont {T.}~\bibnamefont {{Prince}}}, \bibinfo {author} {\bibfnamefont {D.}~\bibnamefont {{Reiley}}}, \bibinfo {author} {\bibfnamefont {H.}~\bibnamefont {{Rodriguez}}}, \bibinfo {author} {\bibfnamefont {B.}~\bibnamefont {{Rusholme}}}, \bibinfo {author} {\bibfnamefont {R.~M.}\ \bibnamefont {{Smith}}},\ and\ \bibinfo {author} {\bibfnamefont {M.~T.}\ \bibnamefont {{Soumagnac}}},\ }\href {https://doi.org/10.1103/PhysRevLett.124.251102} {\bibfield  {journal} {\bibinfo  {journal} {\prl}\ }\textbf {\bibinfo {volume} {124}},\ \bibinfo {eid} {251102} (\bibinfo {year} {2020})},\ \Eprint
  {https://arxiv.org/abs/2006.14122} {arXiv:2006.14122 [astro-ph.HE]} \BibitemShut {NoStop}%
\bibitem [{\citenamefont {{Ohgami}}\ \emph {et~al.}(2023)\citenamefont {{Ohgami}}, \citenamefont {{Becerra Gonz{\'a}lez}}, \citenamefont {{Tominaga}}, \citenamefont {{Morokuma}}, \citenamefont {{Utsumi}}, \citenamefont {{Niino}}, \citenamefont {{Tanaka}}, \citenamefont {{Banerjee}}, \citenamefont {{Poidevin}}, \citenamefont {{Acosta-Pulido}}, \citenamefont {{P{\'e}rez-Fournon}}, \citenamefont {{Mu{\~n}oz-Darias}}, \citenamefont {{Akitaya}}, \citenamefont {{Yanagisawa}}, \citenamefont {{Sasada}}, \citenamefont {{Yoshida}}, \citenamefont {{Simunovic}}, \citenamefont {{Ohsawa}}, \citenamefont {{Tanaka}}, \citenamefont {{Terai}}, \citenamefont {{Takagi}},\ and\ \citenamefont {{J-GEM Collaboration}}}]{ohgami_2023}%
  \BibitemOpen
  \bibfield  {author} {\bibinfo {author} {\bibfnamefont {T.}~\bibnamefont {{Ohgami}}}, \bibinfo {author} {\bibfnamefont {J.}~\bibnamefont {{Becerra Gonz{\'a}lez}}}, \bibinfo {author} {\bibfnamefont {N.}~\bibnamefont {{Tominaga}}}, \bibinfo {author} {\bibfnamefont {T.}~\bibnamefont {{Morokuma}}}, \bibinfo {author} {\bibfnamefont {Y.}~\bibnamefont {{Utsumi}}}, \bibinfo {author} {\bibfnamefont {Y.}~\bibnamefont {{Niino}}}, \bibinfo {author} {\bibfnamefont {M.}~\bibnamefont {{Tanaka}}}, \bibinfo {author} {\bibfnamefont {S.}~\bibnamefont {{Banerjee}}}, \bibinfo {author} {\bibfnamefont {F.}~\bibnamefont {{Poidevin}}}, \bibinfo {author} {\bibfnamefont {J.~A.}\ \bibnamefont {{Acosta-Pulido}}}, \bibinfo {author} {\bibfnamefont {I.}~\bibnamefont {{P{\'e}rez-Fournon}}}, \bibinfo {author} {\bibfnamefont {T.}~\bibnamefont {{Mu{\~n}oz-Darias}}}, \bibinfo {author} {\bibfnamefont {H.}~\bibnamefont {{Akitaya}}}, \bibinfo {author} {\bibfnamefont {K.}~\bibnamefont {{Yanagisawa}}}, \bibinfo {author} {\bibfnamefont
  {M.}~\bibnamefont {{Sasada}}}, \bibinfo {author} {\bibfnamefont {M.}~\bibnamefont {{Yoshida}}}, \bibinfo {author} {\bibfnamefont {M.}~\bibnamefont {{Simunovic}}}, \bibinfo {author} {\bibfnamefont {R.}~\bibnamefont {{Ohsawa}}}, \bibinfo {author} {\bibfnamefont {I.}~\bibnamefont {{Tanaka}}}, \bibinfo {author} {\bibfnamefont {T.}~\bibnamefont {{Terai}}}, \bibinfo {author} {\bibfnamefont {Y.}~\bibnamefont {{Takagi}}},\ and\ \bibinfo {author} {\bibnamefont {{J-GEM Collaboration}}},\ }\href {https://doi.org/10.3847/1538-4357/acbd42} {\bibfield  {journal} {\bibinfo  {journal} {\apj}\ }\textbf {\bibinfo {volume} {947}},\ \bibinfo {eid} {9} (\bibinfo {year} {2023})},\ \Eprint {https://arxiv.org/abs/2302.09269} {arXiv:2302.09269 [astro-ph.HE]} \BibitemShut {NoStop}%
\bibitem [{\citenamefont {{Santos}}\ \emph {et~al.}(2024)\citenamefont {{Santos}}, \citenamefont {{Kilpatrick}}, \citenamefont {{Bom}}, \citenamefont {{Darc}}, \citenamefont {{Herpich}}, \citenamefont {{Lacerda}}, \citenamefont {{Sartori}}, \citenamefont {{Alvarez-Candal}}, \citenamefont {{Mendes de Oliveira}}, \citenamefont {{Kanaan}}, \citenamefont {{Ribeiro}},\ and\ \citenamefont {{Schoenell}}}]{Santos_2024}%
  \BibitemOpen
  \bibfield  {author} {\bibinfo {author} {\bibfnamefont {A.}~\bibnamefont {{Santos}}}, \bibinfo {author} {\bibfnamefont {C.~D.}\ \bibnamefont {{Kilpatrick}}}, \bibinfo {author} {\bibfnamefont {C.~R.}\ \bibnamefont {{Bom}}}, \bibinfo {author} {\bibfnamefont {P.}~\bibnamefont {{Darc}}}, \bibinfo {author} {\bibfnamefont {F.~R.}\ \bibnamefont {{Herpich}}}, \bibinfo {author} {\bibfnamefont {E.~A.~D.}\ \bibnamefont {{Lacerda}}}, \bibinfo {author} {\bibfnamefont {M.~J.}\ \bibnamefont {{Sartori}}}, \bibinfo {author} {\bibfnamefont {A.}~\bibnamefont {{Alvarez-Candal}}}, \bibinfo {author} {\bibfnamefont {C.}~\bibnamefont {{Mendes de Oliveira}}}, \bibinfo {author} {\bibfnamefont {A.}~\bibnamefont {{Kanaan}}}, \bibinfo {author} {\bibfnamefont {T.}~\bibnamefont {{Ribeiro}}},\ and\ \bibinfo {author} {\bibfnamefont {W.}~\bibnamefont {{Schoenell}}},\ }\href {https://doi.org/10.1093/mnras/stae466} {\bibfield  {journal} {\bibinfo  {journal} {\mnras}\ }\textbf {\bibinfo {volume} {529}},\ \bibinfo {pages} {59} (\bibinfo {year}
  {2024})},\ \Eprint {https://arxiv.org/abs/2312.15057} {arXiv:2312.15057 [astro-ph.IM]} \BibitemShut {NoStop}%
\bibitem [{\citenamefont {{Graham}}\ \emph {et~al.}(2023)\citenamefont {{Graham}}, \citenamefont {{McKernan}}, \citenamefont {{Ford}}, \citenamefont {{Stern}}, \citenamefont {{Djorgovski}}, \citenamefont {{Coughlin}}, \citenamefont {{Burdge}}, \citenamefont {{Bellm}}, \citenamefont {{Helou}}, \citenamefont {{Mahabal}}, \citenamefont {{Masci}}, \citenamefont {{Purdum}}, \citenamefont {{Rosnet}},\ and\ \citenamefont {{Rusholme}}}]{Graham_2023}%
  \BibitemOpen
  \bibfield  {author} {\bibinfo {author} {\bibfnamefont {M.~J.}\ \bibnamefont {{Graham}}}, \bibinfo {author} {\bibfnamefont {B.}~\bibnamefont {{McKernan}}}, \bibinfo {author} {\bibfnamefont {K.~E.~S.}\ \bibnamefont {{Ford}}}, \bibinfo {author} {\bibfnamefont {D.}~\bibnamefont {{Stern}}}, \bibinfo {author} {\bibfnamefont {S.~G.}\ \bibnamefont {{Djorgovski}}}, \bibinfo {author} {\bibfnamefont {M.}~\bibnamefont {{Coughlin}}}, \bibinfo {author} {\bibfnamefont {K.~B.}\ \bibnamefont {{Burdge}}}, \bibinfo {author} {\bibfnamefont {E.~C.}\ \bibnamefont {{Bellm}}}, \bibinfo {author} {\bibfnamefont {G.}~\bibnamefont {{Helou}}}, \bibinfo {author} {\bibfnamefont {A.~A.}\ \bibnamefont {{Mahabal}}}, \bibinfo {author} {\bibfnamefont {F.~J.}\ \bibnamefont {{Masci}}}, \bibinfo {author} {\bibfnamefont {J.}~\bibnamefont {{Purdum}}}, \bibinfo {author} {\bibfnamefont {P.}~\bibnamefont {{Rosnet}}},\ and\ \bibinfo {author} {\bibfnamefont {B.}~\bibnamefont {{Rusholme}}},\ }\href {https://doi.org/10.3847/1538-4357/aca480} {\bibfield
  {journal} {\bibinfo  {journal} {\apj}\ }\textbf {\bibinfo {volume} {942}},\ \bibinfo {eid} {99} (\bibinfo {year} {2023})},\ \Eprint {https://arxiv.org/abs/2209.13004} {arXiv:2209.13004 [astro-ph.HE]} \BibitemShut {NoStop}%
\bibitem [{\citenamefont {{Palmese}}\ \emph {et~al.}(2021)\citenamefont {{Palmese}}, \citenamefont {{Fishbach}}, \citenamefont {{Burke}}, \citenamefont {{Annis}},\ and\ \citenamefont {{Liu}}}]{Palmese_2021}%
  \BibitemOpen
  \bibfield  {author} {\bibinfo {author} {\bibfnamefont {A.}~\bibnamefont {{Palmese}}}, \bibinfo {author} {\bibfnamefont {M.}~\bibnamefont {{Fishbach}}}, \bibinfo {author} {\bibfnamefont {C.~J.}\ \bibnamefont {{Burke}}}, \bibinfo {author} {\bibfnamefont {J.}~\bibnamefont {{Annis}}},\ and\ \bibinfo {author} {\bibfnamefont {X.}~\bibnamefont {{Liu}}},\ }\href {https://doi.org/10.3847/2041-8213/ac0883} {\bibfield  {journal} {\bibinfo  {journal} {\apjl}\ }\textbf {\bibinfo {volume} {914}},\ \bibinfo {eid} {L34} (\bibinfo {year} {2021})},\ \Eprint {https://arxiv.org/abs/2103.16069} {arXiv:2103.16069 [astro-ph.HE]} \BibitemShut {NoStop}%
\bibitem [{\citenamefont {Santini}\ \emph {et~al.}(2023)\citenamefont {Santini}, \citenamefont {Gerosa}, \citenamefont {Cotesta},\ and\ \citenamefont {Berti}}]{Santini_2023}%
  \BibitemOpen
  \bibfield  {author} {\bibinfo {author} {\bibfnamefont {A.}~\bibnamefont {Santini}}, \bibinfo {author} {\bibfnamefont {D.}~\bibnamefont {Gerosa}}, \bibinfo {author} {\bibfnamefont {R.}~\bibnamefont {Cotesta}},\ and\ \bibinfo {author} {\bibfnamefont {E.}~\bibnamefont {Berti}},\ }\href {https://doi.org/10.1103/PhysRevD.108.083033} {\bibfield  {journal} {\bibinfo  {journal} {Phys. Rev. D}\ }\textbf {\bibinfo {volume} {108}},\ \bibinfo {pages} {083033} (\bibinfo {year} {2023})}\BibitemShut {NoStop}%
\bibitem [{\citenamefont {{Morton}}\ \emph {et~al.}(2023)\citenamefont {{Morton}}, \citenamefont {{Rinaldi}}, \citenamefont {{Torres-Orjuela}}, \citenamefont {{Derdzinski}}, \citenamefont {{Vaccaro}},\ and\ \citenamefont {{Del Pozzo}}}]{Morton_2023}%
  \BibitemOpen
  \bibfield  {author} {\bibinfo {author} {\bibfnamefont {S.~L.}\ \bibnamefont {{Morton}}}, \bibinfo {author} {\bibfnamefont {S.}~\bibnamefont {{Rinaldi}}}, \bibinfo {author} {\bibfnamefont {A.}~\bibnamefont {{Torres-Orjuela}}}, \bibinfo {author} {\bibfnamefont {A.}~\bibnamefont {{Derdzinski}}}, \bibinfo {author} {\bibfnamefont {M.~P.}\ \bibnamefont {{Vaccaro}}},\ and\ \bibinfo {author} {\bibfnamefont {W.}~\bibnamefont {{Del Pozzo}}},\ }\href {https://doi.org/10.1103/PhysRevD.108.123039} {\bibfield  {journal} {\bibinfo  {journal} {\prd}\ }\textbf {\bibinfo {volume} {108}},\ \bibinfo {eid} {123039} (\bibinfo {year} {2023})},\ \Eprint {https://arxiv.org/abs/2310.16025} {arXiv:2310.16025 [gr-qc]} \BibitemShut {NoStop}%
\bibitem [{\citenamefont {{Veronesi}}\ \emph {et~al.}(2025)\citenamefont {{Veronesi}}, \citenamefont {{van Velzen}},\ and\ \citenamefont {{Rossi}}}]{Veronesi_2025}%
  \BibitemOpen
  \bibfield  {author} {\bibinfo {author} {\bibfnamefont {N.}~\bibnamefont {{Veronesi}}}, \bibinfo {author} {\bibfnamefont {S.}~\bibnamefont {{van Velzen}}},\ and\ \bibinfo {author} {\bibfnamefont {E.~M.}\ \bibnamefont {{Rossi}}},\ }\href {https://doi.org/10.1093/mnras/stae2787} {\bibfield  {journal} {\bibinfo  {journal} {\mnras}\ }\textbf {\bibinfo {volume} {536}},\ \bibinfo {pages} {3112} (\bibinfo {year} {2025})},\ \Eprint {https://arxiv.org/abs/2405.05318} {arXiv:2405.05318 [astro-ph.HE]} \BibitemShut {NoStop}%
\bibitem [{\citenamefont {{Perna}}\ \emph {et~al.}(2016)\citenamefont {{Perna}}, \citenamefont {{Lazzati}},\ and\ \citenamefont {{Giacomazzo}}}]{Perna_2016}%
  \BibitemOpen
  \bibfield  {author} {\bibinfo {author} {\bibfnamefont {R.}~\bibnamefont {{Perna}}}, \bibinfo {author} {\bibfnamefont {D.}~\bibnamefont {{Lazzati}}},\ and\ \bibinfo {author} {\bibfnamefont {B.}~\bibnamefont {{Giacomazzo}}},\ }\href {https://doi.org/10.3847/2041-8205/821/1/L18} {\bibfield  {journal} {\bibinfo  {journal} {\apjl}\ }\textbf {\bibinfo {volume} {821}},\ \bibinfo {eid} {L18} (\bibinfo {year} {2016})},\ \Eprint {https://arxiv.org/abs/1602.05140} {arXiv:1602.05140 [astro-ph.HE]} \BibitemShut {NoStop}%
\bibitem [{\citenamefont {{Janiuk}}\ \emph {et~al.}(2017)\citenamefont {{Janiuk}}, \citenamefont {{Bejger}}, \citenamefont {{Charzy{\'n}ski}},\ and\ \citenamefont {{Sukova}}}]{Janiuk_2017}%
  \BibitemOpen
  \bibfield  {author} {\bibinfo {author} {\bibfnamefont {A.}~\bibnamefont {{Janiuk}}}, \bibinfo {author} {\bibfnamefont {M.}~\bibnamefont {{Bejger}}}, \bibinfo {author} {\bibfnamefont {S.}~\bibnamefont {{Charzy{\'n}ski}}},\ and\ \bibinfo {author} {\bibfnamefont {P.}~\bibnamefont {{Sukova}}},\ }\href {https://doi.org/10.1016/j.newast.2016.08.002} {\bibfield  {journal} {\bibinfo  {journal} {\na}\ }\textbf {\bibinfo {volume} {51}},\ \bibinfo {pages} {7} (\bibinfo {year} {2017})},\ \Eprint {https://arxiv.org/abs/1604.07132} {arXiv:1604.07132 [astro-ph.HE]} \BibitemShut {NoStop}%
\bibitem [{\citenamefont {{McKernan}}\ \emph {et~al.}(2019)\citenamefont {{McKernan}}, \citenamefont {{Ford}}, \citenamefont {{Bartos}}, \citenamefont {{Graham}}, \citenamefont {{Lyra}}, \citenamefont {{Marka}}, \citenamefont {{Marka}}, \citenamefont {{Ross}}, \citenamefont {{Stern}},\ and\ \citenamefont {{Yang}}}]{McKernan_2019}%
  \BibitemOpen
  \bibfield  {author} {\bibinfo {author} {\bibfnamefont {B.}~\bibnamefont {{McKernan}}}, \bibinfo {author} {\bibfnamefont {K.~E.~S.}\ \bibnamefont {{Ford}}}, \bibinfo {author} {\bibfnamefont {I.}~\bibnamefont {{Bartos}}}, \bibinfo {author} {\bibfnamefont {M.~J.}\ \bibnamefont {{Graham}}}, \bibinfo {author} {\bibfnamefont {W.}~\bibnamefont {{Lyra}}}, \bibinfo {author} {\bibfnamefont {S.}~\bibnamefont {{Marka}}}, \bibinfo {author} {\bibfnamefont {Z.}~\bibnamefont {{Marka}}}, \bibinfo {author} {\bibfnamefont {N.~P.}\ \bibnamefont {{Ross}}}, \bibinfo {author} {\bibfnamefont {D.}~\bibnamefont {{Stern}}},\ and\ \bibinfo {author} {\bibfnamefont {Y.}~\bibnamefont {{Yang}}},\ }\href {https://doi.org/10.3847/2041-8213/ab4886} {\bibfield  {journal} {\bibinfo  {journal} {\apjl}\ }\textbf {\bibinfo {volume} {884}},\ \bibinfo {eid} {L50} (\bibinfo {year} {2019})},\ \Eprint {https://arxiv.org/abs/1907.03746} {arXiv:1907.03746 [astro-ph.HE]} \BibitemShut {NoStop}%
\bibitem [{\citenamefont {{Wang}}\ \emph {et~al.}(2021{\natexlab{a}})\citenamefont {{Wang}}, \citenamefont {{Liu}}, \citenamefont {{Ho}}, \citenamefont {{Li}},\ and\ \citenamefont {{Du}}}]{Wang_2021b}%
  \BibitemOpen
  \bibfield  {author} {\bibinfo {author} {\bibfnamefont {J.-M.}\ \bibnamefont {{Wang}}}, \bibinfo {author} {\bibfnamefont {J.-R.}\ \bibnamefont {{Liu}}}, \bibinfo {author} {\bibfnamefont {L.~C.}\ \bibnamefont {{Ho}}}, \bibinfo {author} {\bibfnamefont {Y.-R.}\ \bibnamefont {{Li}}},\ and\ \bibinfo {author} {\bibfnamefont {P.}~\bibnamefont {{Du}}},\ }\href {https://doi.org/10.3847/2041-8213/ac0b46} {\bibfield  {journal} {\bibinfo  {journal} {\apjl}\ }\textbf {\bibinfo {volume} {916}},\ \bibinfo {eid} {L17} (\bibinfo {year} {2021}{\natexlab{a}})},\ \Eprint {https://arxiv.org/abs/2106.07334} {arXiv:2106.07334 [astro-ph.HE]} \BibitemShut {NoStop}%
\bibitem [{\citenamefont {{Kimura}}\ \emph {et~al.}(2021)\citenamefont {{Kimura}}, \citenamefont {{Murase}},\ and\ \citenamefont {{Bartos}}}]{Kimura_2021}%
  \BibitemOpen
  \bibfield  {author} {\bibinfo {author} {\bibfnamefont {S.~S.}\ \bibnamefont {{Kimura}}}, \bibinfo {author} {\bibfnamefont {K.}~\bibnamefont {{Murase}}},\ and\ \bibinfo {author} {\bibfnamefont {I.}~\bibnamefont {{Bartos}}},\ }\href {https://doi.org/10.3847/1538-4357/ac0535} {\bibfield  {journal} {\bibinfo  {journal} {\apj}\ }\textbf {\bibinfo {volume} {916}},\ \bibinfo {eid} {111} (\bibinfo {year} {2021})},\ \Eprint {https://arxiv.org/abs/2103.02461} {arXiv:2103.02461 [astro-ph.HE]} \BibitemShut {NoStop}%
\bibitem [{\citenamefont {{Tagawa}}\ \emph {et~al.}(2023{\natexlab{a}})\citenamefont {{Tagawa}}, \citenamefont {{Kimura}}, \citenamefont {{Haiman}}, \citenamefont {{Perna}},\ and\ \citenamefont {{Bartos}}}]{Tagawa_2023}%
  \BibitemOpen
  \bibfield  {author} {\bibinfo {author} {\bibfnamefont {H.}~\bibnamefont {{Tagawa}}}, \bibinfo {author} {\bibfnamefont {S.~S.}\ \bibnamefont {{Kimura}}}, \bibinfo {author} {\bibfnamefont {Z.}~\bibnamefont {{Haiman}}}, \bibinfo {author} {\bibfnamefont {R.}~\bibnamefont {{Perna}}},\ and\ \bibinfo {author} {\bibfnamefont {I.}~\bibnamefont {{Bartos}}},\ }\href {https://doi.org/10.3847/1538-4357/acc4bb} {\bibfield  {journal} {\bibinfo  {journal} {\apj}\ }\textbf {\bibinfo {volume} {950}},\ \bibinfo {eid} {13} (\bibinfo {year} {2023}{\natexlab{a}})},\ \Eprint {https://arxiv.org/abs/2301.07111} {arXiv:2301.07111 [astro-ph.HE]} \BibitemShut {NoStop}%
\bibitem [{\citenamefont {{Rodr{\'\i}guez-Ram{\'\i}rez}}\ \emph {et~al.}(2024)\citenamefont {{Rodr{\'\i}guez-Ram{\'\i}rez}}, \citenamefont {{Bom}}, \citenamefont {{Fraga}},\ and\ \citenamefont {{Nemmen}}}]{Rodriguez-Ramirez_2024}%
  \BibitemOpen
  \bibfield  {author} {\bibinfo {author} {\bibfnamefont {J.~C.}\ \bibnamefont {{Rodr{\'\i}guez-Ram{\'\i}rez}}}, \bibinfo {author} {\bibfnamefont {C.~R.}\ \bibnamefont {{Bom}}}, \bibinfo {author} {\bibfnamefont {B.}~\bibnamefont {{Fraga}}},\ and\ \bibinfo {author} {\bibfnamefont {R.}~\bibnamefont {{Nemmen}}},\ }\href {https://doi.org/10.1093/mnras/stad3575} {\bibfield  {journal} {\bibinfo  {journal} {\mnras}\ }\textbf {\bibinfo {volume} {527}},\ \bibinfo {pages} {6076} (\bibinfo {year} {2024})},\ \Eprint {https://arxiv.org/abs/2304.10567} {arXiv:2304.10567 [astro-ph.HE]} \BibitemShut {NoStop}%
\bibitem [{\citenamefont {{Chen}}\ and\ \citenamefont {{Dai}}(2024)}]{Chen_2024}%
  \BibitemOpen
  \bibfield  {author} {\bibinfo {author} {\bibfnamefont {K.}~\bibnamefont {{Chen}}}\ and\ \bibinfo {author} {\bibfnamefont {Z.-G.}\ \bibnamefont {{Dai}}},\ }\href {https://doi.org/10.3847/1538-4357/ad0dfd} {\bibfield  {journal} {\bibinfo  {journal} {\apj}\ }\textbf {\bibinfo {volume} {961}},\ \bibinfo {eid} {206} (\bibinfo {year} {2024})},\ \Eprint {https://arxiv.org/abs/2311.10518} {arXiv:2311.10518 [astro-ph.HE]} \BibitemShut {NoStop}%
\bibitem [{\citenamefont {{Gayathri}}\ \emph {et~al.}(2020)\citenamefont {{Gayathri}}, \citenamefont {{Healy}}, \citenamefont {{Lange}}, \citenamefont {{O'Brien}}, \citenamefont {{Szczepanczyk}}, \citenamefont {{Bartos}}, \citenamefont {{Campanelli}}, \citenamefont {{Klimenko}}, \citenamefont {{Lousto}},\ and\ \citenamefont {{O'Shaughnessy}}}]{Gayathri_2020}%
  \BibitemOpen
  \bibfield  {author} {\bibinfo {author} {\bibfnamefont {V.}~\bibnamefont {{Gayathri}}}, \bibinfo {author} {\bibfnamefont {J.}~\bibnamefont {{Healy}}}, \bibinfo {author} {\bibfnamefont {J.}~\bibnamefont {{Lange}}}, \bibinfo {author} {\bibfnamefont {B.}~\bibnamefont {{O'Brien}}}, \bibinfo {author} {\bibfnamefont {M.}~\bibnamefont {{Szczepanczyk}}}, \bibinfo {author} {\bibfnamefont {I.}~\bibnamefont {{Bartos}}}, \bibinfo {author} {\bibfnamefont {M.}~\bibnamefont {{Campanelli}}}, \bibinfo {author} {\bibfnamefont {S.}~\bibnamefont {{Klimenko}}}, \bibinfo {author} {\bibfnamefont {C.}~\bibnamefont {{Lousto}}},\ and\ \bibinfo {author} {\bibfnamefont {R.}~\bibnamefont {{O'Shaughnessy}}},\ }\href@noop {} {\bibfield  {journal} {\bibinfo  {journal} {arXiv e-prints}\ ,\ \bibinfo {eid} {arXiv:2009.14247}} (\bibinfo {year} {2020})},\ \Eprint {https://arxiv.org/abs/2009.14247} {arXiv:2009.14247 [astro-ph.HE]} \BibitemShut {NoStop}%
\bibitem [{\citenamefont {{Haster}}(2020)}]{Haster_2020}%
  \BibitemOpen
  \bibfield  {author} {\bibinfo {author} {\bibfnamefont {C.-J.}\ \bibnamefont {{Haster}}},\ }\href {https://doi.org/10.3847/2515-5172/abcb99} {\bibfield  {journal} {\bibinfo  {journal} {Research Notes of the American Astronomical Society}\ }\textbf {\bibinfo {volume} {4}},\ \bibinfo {eid} {209} (\bibinfo {year} {2020})}\BibitemShut {NoStop}%
\bibitem [{\citenamefont {{Bom}}\ and\ \citenamefont {{Palmese}}(2023)}]{Bom_2023}%
  \BibitemOpen
  \bibfield  {author} {\bibinfo {author} {\bibfnamefont {C.~R.}\ \bibnamefont {{Bom}}}\ and\ \bibinfo {author} {\bibfnamefont {A.}~\bibnamefont {{Palmese}}},\ }\href {https://doi.org/10.48550/arXiv.2307.01330} {\bibfield  {journal} {\bibinfo  {journal} {arXiv e-prints}\ ,\ \bibinfo {eid} {arXiv:2307.01330}} (\bibinfo {year} {2023})},\ \Eprint {https://arxiv.org/abs/2307.01330} {arXiv:2307.01330 [astro-ph.CO]} \BibitemShut {NoStop}%
\bibitem [{\citenamefont {{Alfradique}}\ \emph {et~al.}(2024)\citenamefont {{Alfradique}}, \citenamefont {{Bom}}, \citenamefont {{Palmese}}, \citenamefont {{Teixeira}}, \citenamefont {{Santana-Silva}}, \citenamefont {{Drlica-Wagner}}, \citenamefont {{Riley}}, \citenamefont {{Mart{\'\i}nez-V{\'a}zquez}}, \citenamefont {{Sand}}, \citenamefont {{Stringfellow}}, \citenamefont {{Medina}}, \citenamefont {{Carballo-Bello}}, \citenamefont {{Choi}}, \citenamefont {{Esteves}}, \citenamefont {{Limberg}}, \citenamefont {{Mutlu-Pakdil}}, \citenamefont {{No{\"e}l}}, \citenamefont {{Pace}}, \citenamefont {{Sakowska}},\ and\ \citenamefont {{Wu}}}]{Alfradique_2024}%
  \BibitemOpen
  \bibfield  {author} {\bibinfo {author} {\bibfnamefont {V.}~\bibnamefont {{Alfradique}}}, \bibinfo {author} {\bibfnamefont {C.~R.}\ \bibnamefont {{Bom}}}, \bibinfo {author} {\bibfnamefont {A.}~\bibnamefont {{Palmese}}}, \bibinfo {author} {\bibfnamefont {G.}~\bibnamefont {{Teixeira}}}, \bibinfo {author} {\bibfnamefont {L.}~\bibnamefont {{Santana-Silva}}}, \bibinfo {author} {\bibfnamefont {A.}~\bibnamefont {{Drlica-Wagner}}}, \bibinfo {author} {\bibfnamefont {A.~H.}\ \bibnamefont {{Riley}}}, \bibinfo {author} {\bibfnamefont {C.~E.}\ \bibnamefont {{Mart{\'\i}nez-V{\'a}zquez}}}, \bibinfo {author} {\bibfnamefont {D.~J.}\ \bibnamefont {{Sand}}}, \bibinfo {author} {\bibfnamefont {G.~S.}\ \bibnamefont {{Stringfellow}}}, \bibinfo {author} {\bibfnamefont {G.~E.}\ \bibnamefont {{Medina}}}, \bibinfo {author} {\bibfnamefont {J.~A.}\ \bibnamefont {{Carballo-Bello}}}, \bibinfo {author} {\bibfnamefont {Y.}~\bibnamefont {{Choi}}}, \bibinfo {author} {\bibfnamefont {J.}~\bibnamefont {{Esteves}}}, \bibinfo {author}
  {\bibfnamefont {G.}~\bibnamefont {{Limberg}}}, \bibinfo {author} {\bibfnamefont {B.}~\bibnamefont {{Mutlu-Pakdil}}}, \bibinfo {author} {\bibfnamefont {N.~E.~D.}\ \bibnamefont {{No{\"e}l}}}, \bibinfo {author} {\bibfnamefont {A.~B.}\ \bibnamefont {{Pace}}}, \bibinfo {author} {\bibfnamefont {J.~D.}\ \bibnamefont {{Sakowska}}},\ and\ \bibinfo {author} {\bibfnamefont {J.~F.}\ \bibnamefont {{Wu}}},\ }\href {https://doi.org/10.1093/mnras/stae086} {\bibfield  {journal} {\bibinfo  {journal} {\mnras}\ }\textbf {\bibinfo {volume} {528}},\ \bibinfo {pages} {3249} (\bibinfo {year} {2024})},\ \Eprint {https://arxiv.org/abs/2310.13695} {arXiv:2310.13695 [astro-ph.CO]} \BibitemShut {NoStop}%
\bibitem [{\citenamefont {{Morris}}(1993)}]{Morris_1993}%
  \BibitemOpen
  \bibfield  {author} {\bibinfo {author} {\bibfnamefont {M.}~\bibnamefont {{Morris}}},\ }\href {https://doi.org/10.1086/172607} {\bibfield  {journal} {\bibinfo  {journal} {\apj}\ }\textbf {\bibinfo {volume} {408}},\ \bibinfo {pages} {496} (\bibinfo {year} {1993})}\BibitemShut {NoStop}%
\bibitem [{\citenamefont {{Miralda-Escud{\'e}}}\ and\ \citenamefont {{Gould}}(2000)}]{Miralda_2000}%
  \BibitemOpen
  \bibfield  {author} {\bibinfo {author} {\bibfnamefont {J.}~\bibnamefont {{Miralda-Escud{\'e}}}}\ and\ \bibinfo {author} {\bibfnamefont {A.}~\bibnamefont {{Gould}}},\ }\href {https://doi.org/10.1086/317837} {\bibfield  {journal} {\bibinfo  {journal} {\apj}\ }\textbf {\bibinfo {volume} {545}},\ \bibinfo {pages} {847} (\bibinfo {year} {2000})},\ \Eprint {https://arxiv.org/abs/astro-ph/0003269} {arXiv:astro-ph/0003269 [astro-ph]} \BibitemShut {NoStop}%
\bibitem [{\citenamefont {{Hailey}}\ \emph {et~al.}(2018)\citenamefont {{Hailey}}, \citenamefont {{Mori}}, \citenamefont {{Bauer}}, \citenamefont {{Berkowitz}}, \citenamefont {{Hong}},\ and\ \citenamefont {{Hord}}}]{Hailey_2018}%
  \BibitemOpen
  \bibfield  {author} {\bibinfo {author} {\bibfnamefont {C.~J.}\ \bibnamefont {{Hailey}}}, \bibinfo {author} {\bibfnamefont {K.}~\bibnamefont {{Mori}}}, \bibinfo {author} {\bibfnamefont {F.~E.}\ \bibnamefont {{Bauer}}}, \bibinfo {author} {\bibfnamefont {M.~E.}\ \bibnamefont {{Berkowitz}}}, \bibinfo {author} {\bibfnamefont {J.}~\bibnamefont {{Hong}}},\ and\ \bibinfo {author} {\bibfnamefont {B.~J.}\ \bibnamefont {{Hord}}},\ }\href {https://doi.org/10.1038/nature25029} {\bibfield  {journal} {\bibinfo  {journal} {\nat}\ }\textbf {\bibinfo {volume} {556}},\ \bibinfo {pages} {70} (\bibinfo {year} {2018})}\BibitemShut {NoStop}%
\bibitem [{\citenamefont {{McKernan}}\ \emph {et~al.}(2012)\citenamefont {{McKernan}}, \citenamefont {{Ford}}, \citenamefont {{Lyra}},\ and\ \citenamefont {{Perets}}}]{McKernan_2012}%
  \BibitemOpen
  \bibfield  {author} {\bibinfo {author} {\bibfnamefont {B.}~\bibnamefont {{McKernan}}}, \bibinfo {author} {\bibfnamefont {K.~E.~S.}\ \bibnamefont {{Ford}}}, \bibinfo {author} {\bibfnamefont {W.}~\bibnamefont {{Lyra}}},\ and\ \bibinfo {author} {\bibfnamefont {H.~B.}\ \bibnamefont {{Perets}}},\ }\href {https://doi.org/10.1111/j.1365-2966.2012.21486.x} {\bibfield  {journal} {\bibinfo  {journal} {\mnras}\ }\textbf {\bibinfo {volume} {425}},\ \bibinfo {pages} {460} (\bibinfo {year} {2012})},\ \Eprint {https://arxiv.org/abs/1206.2309} {arXiv:1206.2309 [astro-ph.GA]} \BibitemShut {NoStop}%
\bibitem [{\citenamefont {{Bartos}}\ \emph {et~al.}(2017)\citenamefont {{Bartos}}, \citenamefont {{Kocsis}}, \citenamefont {{Haiman}},\ and\ \citenamefont {{M{\'a}rka}}}]{Bartos_2017}%
  \BibitemOpen
  \bibfield  {author} {\bibinfo {author} {\bibfnamefont {I.}~\bibnamefont {{Bartos}}}, \bibinfo {author} {\bibfnamefont {B.}~\bibnamefont {{Kocsis}}}, \bibinfo {author} {\bibfnamefont {Z.}~\bibnamefont {{Haiman}}},\ and\ \bibinfo {author} {\bibfnamefont {S.}~\bibnamefont {{M{\'a}rka}}},\ }\href {https://doi.org/10.3847/1538-4357/835/2/165} {\bibfield  {journal} {\bibinfo  {journal} {\apj}\ }\textbf {\bibinfo {volume} {835}},\ \bibinfo {eid} {165} (\bibinfo {year} {2017})},\ \Eprint {https://arxiv.org/abs/1602.03831} {arXiv:1602.03831 [astro-ph.HE]} \BibitemShut {NoStop}%
\bibitem [{\citenamefont {{Tagawa}}\ \emph {et~al.}(2020)\citenamefont {{Tagawa}}, \citenamefont {{Haiman}},\ and\ \citenamefont {{Kocsis}}}]{Tagawa_2020}%
  \BibitemOpen
  \bibfield  {author} {\bibinfo {author} {\bibfnamefont {H.}~\bibnamefont {{Tagawa}}}, \bibinfo {author} {\bibfnamefont {Z.}~\bibnamefont {{Haiman}}},\ and\ \bibinfo {author} {\bibfnamefont {B.}~\bibnamefont {{Kocsis}}},\ }\href {https://doi.org/10.3847/1538-4357/ab9b8c} {\bibfield  {journal} {\bibinfo  {journal} {\apj}\ }\textbf {\bibinfo {volume} {898}},\ \bibinfo {eid} {25} (\bibinfo {year} {2020})},\ \Eprint {https://arxiv.org/abs/1912.08218} {arXiv:1912.08218 [astro-ph.GA]} \BibitemShut {NoStop}%
\bibitem [{\citenamefont {{Rowan}}\ \emph {et~al.}(2023)\citenamefont {{Rowan}}, \citenamefont {{Boekholt}}, \citenamefont {{Kocsis}},\ and\ \citenamefont {{Haiman}}}]{Rowan_2023}%
  \BibitemOpen
  \bibfield  {author} {\bibinfo {author} {\bibfnamefont {C.}~\bibnamefont {{Rowan}}}, \bibinfo {author} {\bibfnamefont {T.}~\bibnamefont {{Boekholt}}}, \bibinfo {author} {\bibfnamefont {B.}~\bibnamefont {{Kocsis}}},\ and\ \bibinfo {author} {\bibfnamefont {Z.}~\bibnamefont {{Haiman}}},\ }\href {https://doi.org/10.1093/mnras/stad1926} {\bibfield  {journal} {\bibinfo  {journal} {\mnras}\ }\textbf {\bibinfo {volume} {524}},\ \bibinfo {pages} {2770} (\bibinfo {year} {2023})},\ \Eprint {https://arxiv.org/abs/2212.06133} {arXiv:2212.06133 [astro-ph.GA]} \BibitemShut {NoStop}%
\bibitem [{\citenamefont {{Whitehead}}\ \emph {et~al.}(2024)\citenamefont {{Whitehead}}, \citenamefont {{Rowan}}, \citenamefont {{Boekholt}},\ and\ \citenamefont {{Kocsis}}}]{Whitehead_2024}%
  \BibitemOpen
  \bibfield  {author} {\bibinfo {author} {\bibfnamefont {H.}~\bibnamefont {{Whitehead}}}, \bibinfo {author} {\bibfnamefont {C.}~\bibnamefont {{Rowan}}}, \bibinfo {author} {\bibfnamefont {T.}~\bibnamefont {{Boekholt}}},\ and\ \bibinfo {author} {\bibfnamefont {B.}~\bibnamefont {{Kocsis}}},\ }\href {https://doi.org/10.1093/mnras/stae1430} {\bibfield  {journal} {\bibinfo  {journal} {\mnras}\ }\textbf {\bibinfo {volume} {531}},\ \bibinfo {pages} {4656} (\bibinfo {year} {2024})},\ \Eprint {https://arxiv.org/abs/2309.11561} {arXiv:2309.11561} \BibitemShut {NoStop}%
\bibitem [{\citenamefont {{Dittmann}}\ \emph {et~al.}(2024)\citenamefont {{Dittmann}}, \citenamefont {{Dempsey}},\ and\ \citenamefont {{Li}}}]{Dittmann_2024}%
  \BibitemOpen
  \bibfield  {author} {\bibinfo {author} {\bibfnamefont {A.~J.}\ \bibnamefont {{Dittmann}}}, \bibinfo {author} {\bibfnamefont {A.~M.}\ \bibnamefont {{Dempsey}}},\ and\ \bibinfo {author} {\bibfnamefont {H.}~\bibnamefont {{Li}}},\ }\href {https://doi.org/10.3847/1538-4357/ad23ce} {\bibfield  {journal} {\bibinfo  {journal} {\apj}\ }\textbf {\bibinfo {volume} {964}},\ \bibinfo {eid} {61} (\bibinfo {year} {2024})},\ \Eprint {https://arxiv.org/abs/2310.03832} {arXiv:2310.03832 [astro-ph.HE]} \BibitemShut {NoStop}%
\bibitem [{\citenamefont {{Calcino}}\ \emph {et~al.}(2024)\citenamefont {{Calcino}}, \citenamefont {{Dempsey}}, \citenamefont {{Dittmann}},\ and\ \citenamefont {{Li}}}]{Calcino_2024}%
  \BibitemOpen
  \bibfield  {author} {\bibinfo {author} {\bibfnamefont {J.}~\bibnamefont {{Calcino}}}, \bibinfo {author} {\bibfnamefont {A.~M.}\ \bibnamefont {{Dempsey}}}, \bibinfo {author} {\bibfnamefont {A.~J.}\ \bibnamefont {{Dittmann}}},\ and\ \bibinfo {author} {\bibfnamefont {H.}~\bibnamefont {{Li}}},\ }\href {https://doi.org/10.3847/1538-4357/ad4a53} {\bibfield  {journal} {\bibinfo  {journal} {\apj}\ }\textbf {\bibinfo {volume} {970}},\ \bibinfo {eid} {107} (\bibinfo {year} {2024})},\ \Eprint {https://arxiv.org/abs/2311.13727} {arXiv:2311.13727 [astro-ph.HE]} \BibitemShut {NoStop}%
\bibitem [{\citenamefont {{Callister}}\ \emph {et~al.}(2021)\citenamefont {{Callister}}, \citenamefont {{Haster}}, \citenamefont {{Ng}}, \citenamefont {{Vitale}},\ and\ \citenamefont {{Farr}}}]{Callister_2021}%
  \BibitemOpen
  \bibfield  {author} {\bibinfo {author} {\bibfnamefont {T.~A.}\ \bibnamefont {{Callister}}}, \bibinfo {author} {\bibfnamefont {C.-J.}\ \bibnamefont {{Haster}}}, \bibinfo {author} {\bibfnamefont {K.~K.~Y.}\ \bibnamefont {{Ng}}}, \bibinfo {author} {\bibfnamefont {S.}~\bibnamefont {{Vitale}}},\ and\ \bibinfo {author} {\bibfnamefont {W.~M.}\ \bibnamefont {{Farr}}},\ }\href {https://doi.org/10.3847/2041-8213/ac2ccc} {\bibfield  {journal} {\bibinfo  {journal} {\apjl}\ }\textbf {\bibinfo {volume} {922}},\ \bibinfo {eid} {L5} (\bibinfo {year} {2021})},\ \Eprint {https://arxiv.org/abs/2106.00521} {arXiv:2106.00521 [astro-ph.HE]} \BibitemShut {NoStop}%
\bibitem [{\citenamefont {{Kimball}}\ \emph {et~al.}(2020)\citenamefont {{Kimball}}, \citenamefont {{Berry}},\ and\ \citenamefont {{Kalogera}}}]{Kimball_2020}%
  \BibitemOpen
  \bibfield  {author} {\bibinfo {author} {\bibfnamefont {C.}~\bibnamefont {{Kimball}}}, \bibinfo {author} {\bibfnamefont {C.}~\bibnamefont {{Berry}}},\ and\ \bibinfo {author} {\bibfnamefont {V.}~\bibnamefont {{Kalogera}}},\ }\href {https://doi.org/10.3847/2515-5172/ab66be} {\bibfield  {journal} {\bibinfo  {journal} {Research Notes of the American Astronomical Society}\ }\textbf {\bibinfo {volume} {4}},\ \bibinfo {eid} {2} (\bibinfo {year} {2020})},\ \Eprint {https://arxiv.org/abs/1903.07813} {arXiv:1903.07813 [astro-ph.HE]} \BibitemShut {NoStop}%
\bibitem [{\citenamefont {Bellm}\ \emph {et~al.}(2018)\citenamefont {Bellm}, \citenamefont {Kulkarni}, \citenamefont {Graham}, \citenamefont {Dekany}, \citenamefont {Smith}, \citenamefont {Riddle}, \citenamefont {Masci}, \citenamefont {Helou}, \citenamefont {Prince}, \citenamefont {Adams} \emph {et~al.}}]{ZTF_2018}%
  \BibitemOpen
  \bibfield  {author} {\bibinfo {author} {\bibfnamefont {E.~C.}\ \bibnamefont {Bellm}}, \bibinfo {author} {\bibfnamefont {S.~R.}\ \bibnamefont {Kulkarni}}, \bibinfo {author} {\bibfnamefont {M.~J.}\ \bibnamefont {Graham}}, \bibinfo {author} {\bibfnamefont {R.}~\bibnamefont {Dekany}}, \bibinfo {author} {\bibfnamefont {R.~M.}\ \bibnamefont {Smith}}, \bibinfo {author} {\bibfnamefont {R.}~\bibnamefont {Riddle}}, \bibinfo {author} {\bibfnamefont {F.~J.}\ \bibnamefont {Masci}}, \bibinfo {author} {\bibfnamefont {G.}~\bibnamefont {Helou}}, \bibinfo {author} {\bibfnamefont {T.~A.}\ \bibnamefont {Prince}}, \bibinfo {author} {\bibfnamefont {S.~M.}\ \bibnamefont {Adams}}, \emph {et~al.},\ }\href@noop {} {\bibfield  {journal} {\bibinfo  {journal} {Publications of the Astronomical Society of the Pacific}\ }\textbf {\bibinfo {volume} {131}},\ \bibinfo {pages} {018002} (\bibinfo {year} {2018})}\BibitemShut {NoStop}%
\bibitem [{\citenamefont {{De Paolis}}\ \emph {et~al.}(2020)\citenamefont {{De Paolis}}, \citenamefont {{Nucita}}, \citenamefont {{Strafella}}, \citenamefont {{Licchelli}},\ and\ \citenamefont {{Ingrosso}}}]{DePaolis_2020}%
  \BibitemOpen
  \bibfield  {author} {\bibinfo {author} {\bibfnamefont {F.}~\bibnamefont {{De Paolis}}}, \bibinfo {author} {\bibfnamefont {A.~A.}\ \bibnamefont {{Nucita}}}, \bibinfo {author} {\bibfnamefont {F.}~\bibnamefont {{Strafella}}}, \bibinfo {author} {\bibfnamefont {D.}~\bibnamefont {{Licchelli}}},\ and\ \bibinfo {author} {\bibfnamefont {G.}~\bibnamefont {{Ingrosso}}},\ }\href {https://doi.org/10.1093/mnrasl/slaa140} {\bibfield  {journal} {\bibinfo  {journal} {\mnras}\ }\textbf {\bibinfo {volume} {499}},\ \bibinfo {pages} {L87} (\bibinfo {year} {2020})},\ \Eprint {https://arxiv.org/abs/2008.02692} {arXiv:2008.02692 [astro-ph.GA]} \BibitemShut {NoStop}%
\bibitem [{\citenamefont {{Ashton}}\ \emph {et~al.}(2021)\citenamefont {{Ashton}}, \citenamefont {{Ackley}}, \citenamefont {{Hernandez}},\ and\ \citenamefont {{Piotrzkowski}}}]{Ashton_2021}%
  \BibitemOpen
  \bibfield  {author} {\bibinfo {author} {\bibfnamefont {G.}~\bibnamefont {{Ashton}}}, \bibinfo {author} {\bibfnamefont {K.}~\bibnamefont {{Ackley}}}, \bibinfo {author} {\bibfnamefont {I.~M.}\ \bibnamefont {{Hernandez}}},\ and\ \bibinfo {author} {\bibfnamefont {B.}~\bibnamefont {{Piotrzkowski}}},\ }\href {https://doi.org/10.1088/1361-6382/ac33bb} {\bibfield  {journal} {\bibinfo  {journal} {Classical and Quantum Gravity}\ }\textbf {\bibinfo {volume} {38}},\ \bibinfo {eid} {235004} (\bibinfo {year} {2021})},\ \Eprint {https://arxiv.org/abs/2009.12346} {arXiv:2009.12346 [astro-ph.HE]} \BibitemShut {NoStop}%
\bibitem [{\citenamefont {{Vanden Berk}}\ \emph {et~al.}(2004)\citenamefont {{Vanden Berk}}, \citenamefont {{Wilhite}}, \citenamefont {{Kron}}, \citenamefont {{Anderson}}, \citenamefont {{Brunner}}, \citenamefont {{Hall}}, \citenamefont {{Ivezi{\'c}}}, \citenamefont {{Richards}}, \citenamefont {{Schneider}}, \citenamefont {{York}}, \citenamefont {{Brinkmann}}, \citenamefont {{Lamb}}, \citenamefont {{Nichol}},\ and\ \citenamefont {{Schlegel}}}]{VandenBerk_2004}%
  \BibitemOpen
  \bibfield  {author} {\bibinfo {author} {\bibfnamefont {D.~E.}\ \bibnamefont {{Vanden Berk}}}, \bibinfo {author} {\bibfnamefont {B.~C.}\ \bibnamefont {{Wilhite}}}, \bibinfo {author} {\bibfnamefont {R.~G.}\ \bibnamefont {{Kron}}}, \bibinfo {author} {\bibfnamefont {S.~F.}\ \bibnamefont {{Anderson}}}, \bibinfo {author} {\bibfnamefont {R.~J.}\ \bibnamefont {{Brunner}}}, \bibinfo {author} {\bibfnamefont {P.~B.}\ \bibnamefont {{Hall}}}, \bibinfo {author} {\bibfnamefont {{\v{Z}}.}~\bibnamefont {{Ivezi{\'c}}}}, \bibinfo {author} {\bibfnamefont {G.~T.}\ \bibnamefont {{Richards}}}, \bibinfo {author} {\bibfnamefont {D.~P.}\ \bibnamefont {{Schneider}}}, \bibinfo {author} {\bibfnamefont {D.~G.}\ \bibnamefont {{York}}}, \bibinfo {author} {\bibfnamefont {J.~V.}\ \bibnamefont {{Brinkmann}}}, \bibinfo {author} {\bibfnamefont {D.~Q.}\ \bibnamefont {{Lamb}}}, \bibinfo {author} {\bibfnamefont {R.~C.}\ \bibnamefont {{Nichol}}},\ and\ \bibinfo {author} {\bibfnamefont {D.~J.}\ \bibnamefont {{Schlegel}}},\ }\href
  {https://doi.org/10.1086/380563} {\bibfield  {journal} {\bibinfo  {journal} {\apj}\ }\textbf {\bibinfo {volume} {601}},\ \bibinfo {pages} {692} (\bibinfo {year} {2004})},\ \Eprint {https://arxiv.org/abs/astro-ph/0310336} {arXiv:astro-ph/0310336 [astro-ph]} \BibitemShut {NoStop}%
\bibitem [{\citenamefont {{Yu}}\ \emph {et~al.}(2022)\citenamefont {{Yu}}, \citenamefont {{Richards}}, \citenamefont {{Vogeley}}, \citenamefont {{Moreno}},\ and\ \citenamefont {{Graham}}}]{Yu_2022}%
  \BibitemOpen
  \bibfield  {author} {\bibinfo {author} {\bibfnamefont {W.}~\bibnamefont {{Yu}}}, \bibinfo {author} {\bibfnamefont {G.~T.}\ \bibnamefont {{Richards}}}, \bibinfo {author} {\bibfnamefont {M.~S.}\ \bibnamefont {{Vogeley}}}, \bibinfo {author} {\bibfnamefont {J.}~\bibnamefont {{Moreno}}},\ and\ \bibinfo {author} {\bibfnamefont {M.~J.}\ \bibnamefont {{Graham}}},\ }\href {https://doi.org/10.3847/1538-4357/ac8351} {\bibfield  {journal} {\bibinfo  {journal} {\apj}\ }\textbf {\bibinfo {volume} {936}},\ \bibinfo {eid} {132} (\bibinfo {year} {2022})},\ \Eprint {https://arxiv.org/abs/2201.08943} {arXiv:2201.08943 [astro-ph.GA]} \BibitemShut {NoStop}%
\bibitem [{\citenamefont {{Ross}}\ \emph {et~al.}(2018)\citenamefont {{Ross}}, \citenamefont {{Ford}}, \citenamefont {{Graham}}, \citenamefont {{McKernan}}, \citenamefont {{Stern}}, \citenamefont {{Meisner}}, \citenamefont {{Assef}}, \citenamefont {{Dey}}, \citenamefont {{Drake}}, \citenamefont {{Jun}},\ and\ \citenamefont {{Lang}}}]{Ross_2018}%
  \BibitemOpen
  \bibfield  {author} {\bibinfo {author} {\bibfnamefont {N.~P.}\ \bibnamefont {{Ross}}}, \bibinfo {author} {\bibfnamefont {K.~E.~S.}\ \bibnamefont {{Ford}}}, \bibinfo {author} {\bibfnamefont {M.}~\bibnamefont {{Graham}}}, \bibinfo {author} {\bibfnamefont {B.}~\bibnamefont {{McKernan}}}, \bibinfo {author} {\bibfnamefont {D.}~\bibnamefont {{Stern}}}, \bibinfo {author} {\bibfnamefont {A.~M.}\ \bibnamefont {{Meisner}}}, \bibinfo {author} {\bibfnamefont {R.~J.}\ \bibnamefont {{Assef}}}, \bibinfo {author} {\bibfnamefont {A.}~\bibnamefont {{Dey}}}, \bibinfo {author} {\bibfnamefont {A.~J.}\ \bibnamefont {{Drake}}}, \bibinfo {author} {\bibfnamefont {H.~D.}\ \bibnamefont {{Jun}}},\ and\ \bibinfo {author} {\bibfnamefont {D.}~\bibnamefont {{Lang}}},\ }\href {https://doi.org/10.1093/mnras/sty2002} {\bibfield  {journal} {\bibinfo  {journal} {\mnras}\ }\textbf {\bibinfo {volume} {480}},\ \bibinfo {pages} {4468} (\bibinfo {year} {2018})},\ \Eprint {https://arxiv.org/abs/1805.06921} {arXiv:1805.06921 [astro-ph.GA]}
  \BibitemShut {NoStop}%
\bibitem [{\citenamefont {{de Gouveia Dal Pino}}\ \emph {et~al.}(2010)\citenamefont {{de Gouveia Dal Pino}}, \citenamefont {{Piovezan}},\ and\ \citenamefont {{Kadowaki}}}]{deGouveia_2010}%
  \BibitemOpen
  \bibfield  {author} {\bibinfo {author} {\bibfnamefont {E.~M.}\ \bibnamefont {{de Gouveia Dal Pino}}}, \bibinfo {author} {\bibfnamefont {P.~P.}\ \bibnamefont {{Piovezan}}},\ and\ \bibinfo {author} {\bibfnamefont {L.~H.~S.}\ \bibnamefont {{Kadowaki}}},\ }\href {https://doi.org/10.1051/0004-6361/200913462} {\bibfield  {journal} {\bibinfo  {journal} {\aap}\ }\textbf {\bibinfo {volume} {518}},\ \bibinfo {eid} {A5} (\bibinfo {year} {2010})},\ \Eprint {https://arxiv.org/abs/1005.3067} {arXiv:1005.3067 [astro-ph.HE]} \BibitemShut {NoStop}%
\bibitem [{\citenamefont {{Scepi}}\ \emph {et~al.}(2021)\citenamefont {{Scepi}}, \citenamefont {{Begelman}},\ and\ \citenamefont {{Dexter}}}]{Scepi_2021}%
  \BibitemOpen
  \bibfield  {author} {\bibinfo {author} {\bibfnamefont {N.}~\bibnamefont {{Scepi}}}, \bibinfo {author} {\bibfnamefont {M.~C.}\ \bibnamefont {{Begelman}}},\ and\ \bibinfo {author} {\bibfnamefont {J.}~\bibnamefont {{Dexter}}},\ }\href {https://doi.org/10.1093/mnrasl/slab002} {\bibfield  {journal} {\bibinfo  {journal} {\mnras}\ }\textbf {\bibinfo {volume} {502}},\ \bibinfo {pages} {L50} (\bibinfo {year} {2021})},\ \Eprint {https://arxiv.org/abs/2011.01954} {arXiv:2011.01954 [astro-ph.HE]} \BibitemShut {NoStop}%
\bibitem [{\citenamefont {{Grishin}}\ \emph {et~al.}(2021)\citenamefont {{Grishin}}, \citenamefont {{Bobrick}}, \citenamefont {{Hirai}}, \citenamefont {{Mandel}},\ and\ \citenamefont {{Perets}}}]{Grishin_2021}%
  \BibitemOpen
  \bibfield  {author} {\bibinfo {author} {\bibfnamefont {E.}~\bibnamefont {{Grishin}}}, \bibinfo {author} {\bibfnamefont {A.}~\bibnamefont {{Bobrick}}}, \bibinfo {author} {\bibfnamefont {R.}~\bibnamefont {{Hirai}}}, \bibinfo {author} {\bibfnamefont {I.}~\bibnamefont {{Mandel}}},\ and\ \bibinfo {author} {\bibfnamefont {H.~B.}\ \bibnamefont {{Perets}}},\ }\href {https://doi.org/10.1093/mnras/stab1957} {\bibfield  {journal} {\bibinfo  {journal} {\mnras}\ }\textbf {\bibinfo {volume} {507}},\ \bibinfo {pages} {156} (\bibinfo {year} {2021})},\ \Eprint {https://arxiv.org/abs/2105.09953} {arXiv:2105.09953 [astro-ph.HE]} \BibitemShut {NoStop}%
\bibitem [{\citenamefont {{Chan}}\ \emph {et~al.}(2019)\citenamefont {{Chan}}, \citenamefont {{Piran}}, \citenamefont {{Krolik}},\ and\ \citenamefont {{Saban}}}]{Chan_2019}%
  \BibitemOpen
  \bibfield  {author} {\bibinfo {author} {\bibfnamefont {C.-H.}\ \bibnamefont {{Chan}}}, \bibinfo {author} {\bibfnamefont {T.}~\bibnamefont {{Piran}}}, \bibinfo {author} {\bibfnamefont {J.~H.}\ \bibnamefont {{Krolik}}},\ and\ \bibinfo {author} {\bibfnamefont {D.}~\bibnamefont {{Saban}}},\ }\href {https://doi.org/10.3847/1538-4357/ab2b40} {\bibfield  {journal} {\bibinfo  {journal} {\apj}\ }\textbf {\bibinfo {volume} {881}},\ \bibinfo {eid} {113} (\bibinfo {year} {2019})},\ \Eprint {https://arxiv.org/abs/1904.12261} {arXiv:1904.12261 [astro-ph.HE]} \BibitemShut {NoStop}%
\bibitem [{\citenamefont {{Wang}}\ \emph {et~al.}(2021{\natexlab{b}})\citenamefont {{Wang}}, \citenamefont {{Liu}}, \citenamefont {{Ho}},\ and\ \citenamefont {{Du}}}]{Wang_2021a}%
  \BibitemOpen
  \bibfield  {author} {\bibinfo {author} {\bibfnamefont {J.-M.}\ \bibnamefont {{Wang}}}, \bibinfo {author} {\bibfnamefont {J.-R.}\ \bibnamefont {{Liu}}}, \bibinfo {author} {\bibfnamefont {L.~C.}\ \bibnamefont {{Ho}}},\ and\ \bibinfo {author} {\bibfnamefont {P.}~\bibnamefont {{Du}}},\ }\href {https://doi.org/10.3847/2041-8213/abee81} {\bibfield  {journal} {\bibinfo  {journal} {\apjl}\ }\textbf {\bibinfo {volume} {911}},\ \bibinfo {eid} {L14} (\bibinfo {year} {2021}{\natexlab{b}})},\ \Eprint {https://arxiv.org/abs/2103.07708} {arXiv:2103.07708 [astro-ph.HE]} \BibitemShut {NoStop}%
\bibitem [{\citenamefont {{Tagawa}}\ \emph {et~al.}(2023{\natexlab{b}})\citenamefont {{Tagawa}}, \citenamefont {{Kimura}}, \citenamefont {{Haiman}}, \citenamefont {{Perna}},\ and\ \citenamefont {{Bartos}}}]{Tagawa_2023b}%
  \BibitemOpen
  \bibfield  {author} {\bibinfo {author} {\bibfnamefont {H.}~\bibnamefont {{Tagawa}}}, \bibinfo {author} {\bibfnamefont {S.~S.}\ \bibnamefont {{Kimura}}}, \bibinfo {author} {\bibfnamefont {Z.}~\bibnamefont {{Haiman}}}, \bibinfo {author} {\bibfnamefont {R.}~\bibnamefont {{Perna}}},\ and\ \bibinfo {author} {\bibfnamefont {I.}~\bibnamefont {{Bartos}}},\ }\href {https://doi.org/10.48550/arXiv.2310.18392} {\bibfield  {journal} {\bibinfo  {journal} {arXiv e-prints}\ ,\ \bibinfo {eid} {arXiv:2310.18392}} (\bibinfo {year} {2023}{\natexlab{b}})},\ \Eprint {https://arxiv.org/abs/2310.18392} {arXiv:2310.18392 [astro-ph.HE]} \BibitemShut {NoStop}%
\bibitem [{\citenamefont {{Chen}}\ \emph {et~al.}(2023)\citenamefont {{Chen}}, \citenamefont {{Ren}},\ and\ \citenamefont {{Dai}}}]{Chen_2023}%
  \BibitemOpen
  \bibfield  {author} {\bibinfo {author} {\bibfnamefont {K.}~\bibnamefont {{Chen}}}, \bibinfo {author} {\bibfnamefont {J.}~\bibnamefont {{Ren}}},\ and\ \bibinfo {author} {\bibfnamefont {Z.-G.}\ \bibnamefont {{Dai}}},\ }\href {https://doi.org/10.3847/1538-4357/acc45f} {\bibfield  {journal} {\bibinfo  {journal} {\apj}\ }\textbf {\bibinfo {volume} {948}},\ \bibinfo {eid} {136} (\bibinfo {year} {2023})},\ \Eprint {https://arxiv.org/abs/2303.07639} {arXiv:2303.07639 [astro-ph.HE]} \BibitemShut {NoStop}%
\bibitem [{Note1()}]{Note1}%
  \BibitemOpen
  \bibinfo {note} {The mass of the remnant is typically slightly smaller than the mass of the BBH due to the release of gravitational waves at coalescence. Nevertheless, for practical purposes proposed in this work, we do not make a distinction between the masses of the BBH and the remnant.}\BibitemShut {Stop}%
\bibitem [{\citenamefont {{Campanelli}}\ \emph {et~al.}(2007)\citenamefont {{Campanelli}}, \citenamefont {{Lousto}}, \citenamefont {{Zlochower}},\ and\ \citenamefont {{Merritt}}}]{Campanelli_2007}%
  \BibitemOpen
  \bibfield  {author} {\bibinfo {author} {\bibfnamefont {M.}~\bibnamefont {{Campanelli}}}, \bibinfo {author} {\bibfnamefont {C.~O.}\ \bibnamefont {{Lousto}}}, \bibinfo {author} {\bibfnamefont {Y.}~\bibnamefont {{Zlochower}}},\ and\ \bibinfo {author} {\bibfnamefont {D.}~\bibnamefont {{Merritt}}},\ }\href {https://doi.org/10.1103/PhysRevLett.98.231102} {\bibfield  {journal} {\bibinfo  {journal} {\prl}\ }\textbf {\bibinfo {volume} {98}},\ \bibinfo {eid} {231102} (\bibinfo {year} {2007})},\ \Eprint {https://arxiv.org/abs/gr-qc/0702133} {arXiv:gr-qc/0702133 [gr-qc]} \BibitemShut {NoStop}%
\bibitem [{\citenamefont {{Lousto}}\ \emph {et~al.}(2010)\citenamefont {{Lousto}}, \citenamefont {{Campanelli}}, \citenamefont {{Zlochower}},\ and\ \citenamefont {{Nakano}}}]{Lousto_2010}%
  \BibitemOpen
  \bibfield  {author} {\bibinfo {author} {\bibfnamefont {C.~O.}\ \bibnamefont {{Lousto}}}, \bibinfo {author} {\bibfnamefont {M.}~\bibnamefont {{Campanelli}}}, \bibinfo {author} {\bibfnamefont {Y.}~\bibnamefont {{Zlochower}}},\ and\ \bibinfo {author} {\bibfnamefont {H.}~\bibnamefont {{Nakano}}},\ }\href {https://doi.org/10.1088/0264-9381/27/11/114006} {\bibfield  {journal} {\bibinfo  {journal} {Classical and Quantum Gravity}\ }\textbf {\bibinfo {volume} {27}},\ \bibinfo {eid} {114006} (\bibinfo {year} {2010})},\ \Eprint {https://arxiv.org/abs/0904.3541} {arXiv:0904.3541 [gr-qc]} \BibitemShut {NoStop}%
\bibitem [{\citenamefont {{Lousto}}\ \emph {et~al.}(2012)\citenamefont {{Lousto}}, \citenamefont {{Zlochower}}, \citenamefont {{Dotti}},\ and\ \citenamefont {{Volonteri}}}]{Lousto_2012}%
  \BibitemOpen
  \bibfield  {author} {\bibinfo {author} {\bibfnamefont {C.~O.}\ \bibnamefont {{Lousto}}}, \bibinfo {author} {\bibfnamefont {Y.}~\bibnamefont {{Zlochower}}}, \bibinfo {author} {\bibfnamefont {M.}~\bibnamefont {{Dotti}}},\ and\ \bibinfo {author} {\bibfnamefont {M.}~\bibnamefont {{Volonteri}}},\ }\href {https://doi.org/10.1103/PhysRevD.85.084015} {\bibfield  {journal} {\bibinfo  {journal} {\prd}\ }\textbf {\bibinfo {volume} {85}},\ \bibinfo {eid} {084015} (\bibinfo {year} {2012})},\ \Eprint {https://arxiv.org/abs/1201.1923} {arXiv:1201.1923 [gr-qc]} \BibitemShut {NoStop}%
\bibitem [{\citenamefont {{Varma}}\ \emph {et~al.}(2022)\citenamefont {{Varma}}, \citenamefont {{Biscoveanu}}, \citenamefont {{Islam}}, \citenamefont {{Shaik}}, \citenamefont {{Haster}}, \citenamefont {{Isi}}, \citenamefont {{Farr}}, \citenamefont {{Field}},\ and\ \citenamefont {{Vitale}}}]{Varma_2022}%
  \BibitemOpen
  \bibfield  {author} {\bibinfo {author} {\bibfnamefont {V.}~\bibnamefont {{Varma}}}, \bibinfo {author} {\bibfnamefont {S.}~\bibnamefont {{Biscoveanu}}}, \bibinfo {author} {\bibfnamefont {T.}~\bibnamefont {{Islam}}}, \bibinfo {author} {\bibfnamefont {F.~H.}\ \bibnamefont {{Shaik}}}, \bibinfo {author} {\bibfnamefont {C.-J.}\ \bibnamefont {{Haster}}}, \bibinfo {author} {\bibfnamefont {M.}~\bibnamefont {{Isi}}}, \bibinfo {author} {\bibfnamefont {W.~M.}\ \bibnamefont {{Farr}}}, \bibinfo {author} {\bibfnamefont {S.~E.}\ \bibnamefont {{Field}}},\ and\ \bibinfo {author} {\bibfnamefont {S.}~\bibnamefont {{Vitale}}},\ }\href {https://doi.org/10.1103/PhysRevLett.128.191102} {\bibfield  {journal} {\bibinfo  {journal} {\prl}\ }\textbf {\bibinfo {volume} {128}},\ \bibinfo {eid} {191102} (\bibinfo {year} {2022})},\ \Eprint {https://arxiv.org/abs/2201.01302} {arXiv:2201.01302 [astro-ph.HE]} \BibitemShut {NoStop}%
\bibitem [{\citenamefont {{Shakura}}\ and\ \citenamefont {{Sunyaev}}(1973)}]{SS_1973}%
  \BibitemOpen
  \bibfield  {author} {\bibinfo {author} {\bibfnamefont {N.~I.}\ \bibnamefont {{Shakura}}}\ and\ \bibinfo {author} {\bibfnamefont {R.~A.}\ \bibnamefont {{Sunyaev}}},\ }\href@noop {} {\bibfield  {journal} {\bibinfo  {journal} {\aap}\ }\textbf {\bibinfo {volume} {24}},\ \bibinfo {pages} {337} (\bibinfo {year} {1973})}\BibitemShut {NoStop}%
\bibitem [{\citenamefont {{Kaaz}}\ \emph {et~al.}(2023)\citenamefont {{Kaaz}}, \citenamefont {{Murguia-Berthier}}, \citenamefont {{Chatterjee}}, \citenamefont {{Liska}},\ and\ \citenamefont {{Tchekhovskoy}}}]{Kaaz_2023}%
  \BibitemOpen
  \bibfield  {author} {\bibinfo {author} {\bibfnamefont {N.}~\bibnamefont {{Kaaz}}}, \bibinfo {author} {\bibfnamefont {A.}~\bibnamefont {{Murguia-Berthier}}}, \bibinfo {author} {\bibfnamefont {K.}~\bibnamefont {{Chatterjee}}}, \bibinfo {author} {\bibfnamefont {M.~T.~P.}\ \bibnamefont {{Liska}}},\ and\ \bibinfo {author} {\bibfnamefont {A.}~\bibnamefont {{Tchekhovskoy}}},\ }\href {https://doi.org/10.3847/1538-4357/acc7a1} {\bibfield  {journal} {\bibinfo  {journal} {\apj}\ }\textbf {\bibinfo {volume} {950}},\ \bibinfo {eid} {31} (\bibinfo {year} {2023})},\ \Eprint {https://arxiv.org/abs/2201.11753} {arXiv:2201.11753 [astro-ph.HE]} \BibitemShut {NoStop}%
\bibitem [{\citenamefont {{Blandford}}\ and\ \citenamefont {{Znajek}}(1977)}]{Blandford_1977}%
  \BibitemOpen
  \bibfield  {author} {\bibinfo {author} {\bibfnamefont {R.~D.}\ \bibnamefont {{Blandford}}}\ and\ \bibinfo {author} {\bibfnamefont {R.~L.}\ \bibnamefont {{Znajek}}},\ }\href {https://doi.org/10.1093/mnras/179.3.433} {\bibfield  {journal} {\bibinfo  {journal} {\mnras}\ }\textbf {\bibinfo {volume} {179}},\ \bibinfo {pages} {433} (\bibinfo {year} {1977})}\BibitemShut {NoStop}%
\bibitem [{\citenamefont {{Gruzinov}}\ \emph {et~al.}(2020)\citenamefont {{Gruzinov}}, \citenamefont {{Levin}},\ and\ \citenamefont {{Matzner}}}]{Gruzinov_2020}%
  \BibitemOpen
  \bibfield  {author} {\bibinfo {author} {\bibfnamefont {A.}~\bibnamefont {{Gruzinov}}}, \bibinfo {author} {\bibfnamefont {Y.}~\bibnamefont {{Levin}}},\ and\ \bibinfo {author} {\bibfnamefont {C.~D.}\ \bibnamefont {{Matzner}}},\ }\href {https://doi.org/10.1093/mnras/staa013} {\bibfield  {journal} {\bibinfo  {journal} {\mnras}\ }\textbf {\bibinfo {volume} {492}},\ \bibinfo {pages} {2755} (\bibinfo {year} {2020})},\ \Eprint {https://arxiv.org/abs/1906.01186} {arXiv:1906.01186 [astro-ph.HE]} \BibitemShut {NoStop}%
\bibitem [{\citenamefont {{Li}}\ \emph {et~al.}(2020)\citenamefont {{Li}}, \citenamefont {{Chang}}, \citenamefont {{Levin}}, \citenamefont {{Matzner}},\ and\ \citenamefont {{Armitage}}}]{Li_2020}%
  \BibitemOpen
  \bibfield  {author} {\bibinfo {author} {\bibfnamefont {X.}~\bibnamefont {{Li}}}, \bibinfo {author} {\bibfnamefont {P.}~\bibnamefont {{Chang}}}, \bibinfo {author} {\bibfnamefont {Y.}~\bibnamefont {{Levin}}}, \bibinfo {author} {\bibfnamefont {C.~D.}\ \bibnamefont {{Matzner}}},\ and\ \bibinfo {author} {\bibfnamefont {P.~J.}\ \bibnamefont {{Armitage}}},\ }\href {https://doi.org/10.1093/mnras/staa900} {\bibfield  {journal} {\bibinfo  {journal} {\mnras}\ }\textbf {\bibinfo {volume} {494}},\ \bibinfo {pages} {2327} (\bibinfo {year} {2020})},\ \Eprint {https://arxiv.org/abs/1912.06864} {arXiv:1912.06864 [astro-ph.HE]} \BibitemShut {NoStop}%
\bibitem [{\citenamefont {{Fabrika}}\ \emph {et~al.}(2021)\citenamefont {{Fabrika}}, \citenamefont {{Atapin}}, \citenamefont {{Vinokurov}},\ and\ \citenamefont {{Sholukhova}}}]{Fabrika_2021}%
  \BibitemOpen
  \bibfield  {author} {\bibinfo {author} {\bibfnamefont {S.~N.}\ \bibnamefont {{Fabrika}}}, \bibinfo {author} {\bibfnamefont {K.~E.}\ \bibnamefont {{Atapin}}}, \bibinfo {author} {\bibfnamefont {A.~S.}\ \bibnamefont {{Vinokurov}}},\ and\ \bibinfo {author} {\bibfnamefont {O.~N.}\ \bibnamefont {{Sholukhova}}},\ }\href {https://doi.org/10.1134/S1990341321010077} {\bibfield  {journal} {\bibinfo  {journal} {Astrophysical Bulletin}\ }\textbf {\bibinfo {volume} {76}},\ \bibinfo {pages} {6} (\bibinfo {year} {2021})},\ \Eprint {https://arxiv.org/abs/2105.10537} {arXiv:2105.10537 [astro-ph.GA]} \BibitemShut {NoStop}%
\bibitem [{\citenamefont {{Abaroa}}\ \emph {et~al.}(2023)\citenamefont {{Abaroa}}, \citenamefont {{Romero}},\ and\ \citenamefont {{Sotomayor}}}]{Abaroa_2023}%
  \BibitemOpen
  \bibfield  {author} {\bibinfo {author} {\bibfnamefont {L.}~\bibnamefont {{Abaroa}}}, \bibinfo {author} {\bibfnamefont {G.~E.}\ \bibnamefont {{Romero}}},\ and\ \bibinfo {author} {\bibfnamefont {P.}~\bibnamefont {{Sotomayor}}},\ }\href {https://doi.org/10.1051/0004-6361/202245285} {\bibfield  {journal} {\bibinfo  {journal} {\aap}\ }\textbf {\bibinfo {volume} {671}},\ \bibinfo {eid} {A9} (\bibinfo {year} {2023})},\ \Eprint {https://arxiv.org/abs/2301.08635} {arXiv:2301.08635 [astro-ph.HE]} \BibitemShut {NoStop}%
\bibitem [{\citenamefont {{Sirko}}\ and\ \citenamefont {{Goodman}}(2003)}]{SG_2003}%
  \BibitemOpen
  \bibfield  {author} {\bibinfo {author} {\bibfnamefont {E.}~\bibnamefont {{Sirko}}}\ and\ \bibinfo {author} {\bibfnamefont {J.}~\bibnamefont {{Goodman}}},\ }\href {https://doi.org/10.1046/j.1365-8711.2003.06431.x} {\bibfield  {journal} {\bibinfo  {journal} {\mnras}\ }\textbf {\bibinfo {volume} {341}},\ \bibinfo {pages} {501} (\bibinfo {year} {2003})},\ \Eprint {https://arxiv.org/abs/astro-ph/0209469} {arXiv:astro-ph/0209469 [astro-ph]} \BibitemShut {NoStop}%
\bibitem [{\citenamefont {{Thompson}}\ \emph {et~al.}(2005)\citenamefont {{Thompson}}, \citenamefont {{Quataert}},\ and\ \citenamefont {{Murray}}}]{TQM_2005}%
  \BibitemOpen
  \bibfield  {author} {\bibinfo {author} {\bibfnamefont {T.~A.}\ \bibnamefont {{Thompson}}}, \bibinfo {author} {\bibfnamefont {E.}~\bibnamefont {{Quataert}}},\ and\ \bibinfo {author} {\bibfnamefont {N.}~\bibnamefont {{Murray}}},\ }\href {https://doi.org/10.1086/431923} {\bibfield  {journal} {\bibinfo  {journal} {\apj}\ }\textbf {\bibinfo {volume} {630}},\ \bibinfo {pages} {167} (\bibinfo {year} {2005})},\ \Eprint {https://arxiv.org/abs/astro-ph/0503027} {arXiv:astro-ph/0503027 [astro-ph]} \BibitemShut {NoStop}%
\bibitem [{\citenamefont {{Gangardt}}\ \emph {et~al.}(2024)\citenamefont {{Gangardt}}, \citenamefont {{Trani}}, \citenamefont {{Bonnerot}},\ and\ \citenamefont {{Gerosa}}}]{Gangardt_2024}%
  \BibitemOpen
  \bibfield  {author} {\bibinfo {author} {\bibfnamefont {D.}~\bibnamefont {{Gangardt}}}, \bibinfo {author} {\bibfnamefont {A.~A.}\ \bibnamefont {{Trani}}}, \bibinfo {author} {\bibfnamefont {C.}~\bibnamefont {{Bonnerot}}},\ and\ \bibinfo {author} {\bibfnamefont {D.}~\bibnamefont {{Gerosa}}},\ }\href {https://doi.org/10.1093/mnras/stae1117} {\bibfield  {journal} {\bibinfo  {journal} {\mnras}\ }\textbf {\bibinfo {volume} {530}},\ \bibinfo {pages} {3689} (\bibinfo {year} {2024})},\ \Eprint {https://arxiv.org/abs/2403.00060} {arXiv:2403.00060 [astro-ph.HE]} \BibitemShut {NoStop}%
\bibitem [{\citenamefont {{Pihajoki}}(2016)}]{Pihajoki_2016}%
  \BibitemOpen
  \bibfield  {author} {\bibinfo {author} {\bibfnamefont {P.}~\bibnamefont {{Pihajoki}}},\ }\href {https://doi.org/10.1093/mnras/stv3023} {\bibfield  {journal} {\bibinfo  {journal} {\mnras}\ }\textbf {\bibinfo {volume} {457}},\ \bibinfo {pages} {1145} (\bibinfo {year} {2016})},\ \Eprint {https://arxiv.org/abs/1510.07642} {arXiv:1510.07642 [astro-ph.HE]} \BibitemShut {NoStop}%
\bibitem [{\citenamefont {{Rodr{\'\i}guez-Ram{\'\i}rez}}\ \emph {et~al.}(2020)\citenamefont {{Rodr{\'\i}guez-Ram{\'\i}rez}}, \citenamefont {{Kushwaha}}, \citenamefont {{de Gouveia Dal Pino}},\ and\ \citenamefont {{Santos-Lima}}}]{Rodriguez-Ramirez_2020}%
  \BibitemOpen
  \bibfield  {author} {\bibinfo {author} {\bibfnamefont {J.~C.}\ \bibnamefont {{Rodr{\'\i}guez-Ram{\'\i}rez}}}, \bibinfo {author} {\bibfnamefont {P.}~\bibnamefont {{Kushwaha}}}, \bibinfo {author} {\bibfnamefont {E.~M.}\ \bibnamefont {{de Gouveia Dal Pino}}},\ and\ \bibinfo {author} {\bibfnamefont {R.}~\bibnamefont {{Santos-Lima}}},\ }\href {https://doi.org/10.1093/mnras/staa2664} {\bibfield  {journal} {\bibinfo  {journal} {\mnras}\ }\textbf {\bibinfo {volume} {498}},\ \bibinfo {pages} {5424} (\bibinfo {year} {2020})},\ \Eprint {https://arxiv.org/abs/2005.01276} {arXiv:2005.01276 [astro-ph.HE]} \BibitemShut {NoStop}%
\bibitem [{\citenamefont {{Fukue}}(2004)}]{Fukue_2004}%
  \BibitemOpen
  \bibfield  {author} {\bibinfo {author} {\bibfnamefont {J.}~\bibnamefont {{Fukue}}},\ }\href {https://doi.org/10.1093/pasj/56.3.569} {\bibfield  {journal} {\bibinfo  {journal} {\pasj}\ }\textbf {\bibinfo {volume} {56}},\ \bibinfo {pages} {569} (\bibinfo {year} {2004})}\BibitemShut {NoStop}%
\bibitem [{\citenamefont {{Ivanov}}\ \emph {et~al.}(1998)\citenamefont {{Ivanov}}, \citenamefont {{Igumenshchev}},\ and\ \citenamefont {{Novikov}}}]{Ivanov_1998}%
  \BibitemOpen
  \bibfield  {author} {\bibinfo {author} {\bibfnamefont {P.~B.}\ \bibnamefont {{Ivanov}}}, \bibinfo {author} {\bibfnamefont {I.~V.}\ \bibnamefont {{Igumenshchev}}},\ and\ \bibinfo {author} {\bibfnamefont {I.~D.}\ \bibnamefont {{Novikov}}},\ }\href {https://doi.org/10.1086/306324} {\bibfield  {journal} {\bibinfo  {journal} {\apj}\ }\textbf {\bibinfo {volume} {507}},\ \bibinfo {pages} {131} (\bibinfo {year} {1998})}\BibitemShut {NoStop}%
\bibitem [{\citenamefont {{Arnett}}(1980)}]{Arnett_1980}%
  \BibitemOpen
  \bibfield  {author} {\bibinfo {author} {\bibfnamefont {W.~D.}\ \bibnamefont {{Arnett}}},\ }\href {https://doi.org/10.1086/157898} {\bibfield  {journal} {\bibinfo  {journal} {\apj}\ }\textbf {\bibinfo {volume} {237}},\ \bibinfo {pages} {541} (\bibinfo {year} {1980})}\BibitemShut {NoStop}%
\bibitem [{\citenamefont {{Arnett}}(1996)}]{Arnett_1996}%
  \BibitemOpen
  \bibfield  {author} {\bibinfo {author} {\bibfnamefont {D.}~\bibnamefont {{Arnett}}},\ }\href@noop {} {\emph {\bibinfo {title} {{Supernovae and Nucleosynthesis: An Investigation of the History of Matter from the Big Bang to the Present}}}}\ (\bibinfo {year} {1996})\BibitemShut {NoStop}%
\bibitem [{\citenamefont {{Chatzopoulos}}\ \emph {et~al.}(2012)\citenamefont {{Chatzopoulos}}, \citenamefont {{Wheeler}},\ and\ \citenamefont {{Vinko}}}]{Chatzopoulos_2012}%
  \BibitemOpen
  \bibfield  {author} {\bibinfo {author} {\bibfnamefont {E.}~\bibnamefont {{Chatzopoulos}}}, \bibinfo {author} {\bibfnamefont {J.~C.}\ \bibnamefont {{Wheeler}}},\ and\ \bibinfo {author} {\bibfnamefont {J.}~\bibnamefont {{Vinko}}},\ }\href {https://doi.org/10.1088/0004-637X/746/2/121} {\bibfield  {journal} {\bibinfo  {journal} {\apj}\ }\textbf {\bibinfo {volume} {746}},\ \bibinfo {eid} {121} (\bibinfo {year} {2012})}\BibitemShut {NoStop}%
\bibitem [{\citenamefont {{Weaver}}\ \emph {et~al.}(1977)\citenamefont {{Weaver}}, \citenamefont {{McCray}}, \citenamefont {{Castor}}, \citenamefont {{Shapiro}},\ and\ \citenamefont {{Moore}}}]{Weaver_1977}%
  \BibitemOpen
  \bibfield  {author} {\bibinfo {author} {\bibfnamefont {R.}~\bibnamefont {{Weaver}}}, \bibinfo {author} {\bibfnamefont {R.}~\bibnamefont {{McCray}}}, \bibinfo {author} {\bibfnamefont {J.}~\bibnamefont {{Castor}}}, \bibinfo {author} {\bibfnamefont {P.}~\bibnamefont {{Shapiro}}},\ and\ \bibinfo {author} {\bibfnamefont {R.}~\bibnamefont {{Moore}}},\ }\href {https://doi.org/10.1086/155692} {\bibfield  {journal} {\bibinfo  {journal} {\apj}\ }\textbf {\bibinfo {volume} {218}},\ \bibinfo {pages} {377} (\bibinfo {year} {1977})}\BibitemShut {NoStop}%
\bibitem [{\citenamefont {{Mandel}}\ and\ \citenamefont {{Farmer}}(2022)}]{Mandel_2022}%
  \BibitemOpen
  \bibfield  {author} {\bibinfo {author} {\bibfnamefont {I.}~\bibnamefont {{Mandel}}}\ and\ \bibinfo {author} {\bibfnamefont {A.}~\bibnamefont {{Farmer}}},\ }\href {https://doi.org/10.1016/j.physrep.2022.01.003} {\bibfield  {journal} {\bibinfo  {journal} {\physrep}\ }\textbf {\bibinfo {volume} {955}},\ \bibinfo {pages} {1} (\bibinfo {year} {2022})},\ \Eprint {https://arxiv.org/abs/1806.05820} {arXiv:1806.05820 [astro-ph.HE]} \BibitemShut {NoStop}%
\bibitem [{\citenamefont {{Greene}}\ and\ \citenamefont {{Ho}}(2007)}]{Greene_2007}%
  \BibitemOpen
  \bibfield  {author} {\bibinfo {author} {\bibfnamefont {J.~E.}\ \bibnamefont {{Greene}}}\ and\ \bibinfo {author} {\bibfnamefont {L.~C.}\ \bibnamefont {{Ho}}},\ }\href {https://doi.org/10.1086/520497} {\bibfield  {journal} {\bibinfo  {journal} {\apj}\ }\textbf {\bibinfo {volume} {667}},\ \bibinfo {pages} {131} (\bibinfo {year} {2007})},\ \Eprint {https://arxiv.org/abs/0705.0020} {arXiv:0705.0020 [astro-ph]} \BibitemShut {NoStop}%
\bibitem [{\citenamefont {{Li}}\ \emph {et~al.}(2011)\citenamefont {{Li}}, \citenamefont {{Ho}},\ and\ \citenamefont {{Wang}}}]{Li_2011}%
  \BibitemOpen
  \bibfield  {author} {\bibinfo {author} {\bibfnamefont {Y.-R.}\ \bibnamefont {{Li}}}, \bibinfo {author} {\bibfnamefont {L.~C.}\ \bibnamefont {{Ho}}},\ and\ \bibinfo {author} {\bibfnamefont {J.-M.}\ \bibnamefont {{Wang}}},\ }\href {https://doi.org/10.1088/0004-637X/742/1/33} {\bibfield  {journal} {\bibinfo  {journal} {\apj}\ }\textbf {\bibinfo {volume} {742}},\ \bibinfo {eid} {33} (\bibinfo {year} {2011})},\ \Eprint {https://arxiv.org/abs/1109.0089} {arXiv:1109.0089 [astro-ph.CO]} \BibitemShut {NoStop}%
\bibitem [{\citenamefont {{Narayan}}\ and\ \citenamefont {{McClintock}}(2008)}]{Narayan_2008}%
  \BibitemOpen
  \bibfield  {author} {\bibinfo {author} {\bibfnamefont {R.}~\bibnamefont {{Narayan}}}\ and\ \bibinfo {author} {\bibfnamefont {J.~E.}\ \bibnamefont {{McClintock}}},\ }\href {https://doi.org/10.1016/j.newar.2008.03.002} {\bibfield  {journal} {\bibinfo  {journal} {\nar}\ }\textbf {\bibinfo {volume} {51}},\ \bibinfo {pages} {733} (\bibinfo {year} {2008})},\ \Eprint {https://arxiv.org/abs/0803.0322} {arXiv:0803.0322 [astro-ph]} \BibitemShut {NoStop}%
\bibitem [{\citenamefont {{Grishin}}\ \emph {et~al.}(2024)\citenamefont {{Grishin}}, \citenamefont {{Gilbaum}},\ and\ \citenamefont {{Stone}}}]{Grishin_2024}%
  \BibitemOpen
  \bibfield  {author} {\bibinfo {author} {\bibfnamefont {E.}~\bibnamefont {{Grishin}}}, \bibinfo {author} {\bibfnamefont {S.}~\bibnamefont {{Gilbaum}}},\ and\ \bibinfo {author} {\bibfnamefont {N.~C.}\ \bibnamefont {{Stone}}},\ }\href {https://doi.org/10.1093/mnras/stae828} {\bibfield  {journal} {\bibinfo  {journal} {\mnras}\ }\textbf {\bibinfo {volume} {530}},\ \bibinfo {pages} {2114} (\bibinfo {year} {2024})},\ \Eprint {https://arxiv.org/abs/2307.07546} {arXiv:2307.07546 [astro-ph.HE]} \BibitemShut {NoStop}%
\bibitem [{\citenamefont {{Gonz{\'a}lez}}\ \emph {et~al.}(2007)\citenamefont {{Gonz{\'a}lez}}, \citenamefont {{Sperhake}}, \citenamefont {{Br{\"u}gmann}}, \citenamefont {{Hannam}},\ and\ \citenamefont {{Husa}}}]{Gonzalez_2007}%
  \BibitemOpen
  \bibfield  {author} {\bibinfo {author} {\bibfnamefont {J.~A.}\ \bibnamefont {{Gonz{\'a}lez}}}, \bibinfo {author} {\bibfnamefont {U.}~\bibnamefont {{Sperhake}}}, \bibinfo {author} {\bibfnamefont {B.}~\bibnamefont {{Br{\"u}gmann}}}, \bibinfo {author} {\bibfnamefont {M.}~\bibnamefont {{Hannam}}},\ and\ \bibinfo {author} {\bibfnamefont {S.}~\bibnamefont {{Husa}}},\ }\href {https://doi.org/10.1103/PhysRevLett.98.091101} {\bibfield  {journal} {\bibinfo  {journal} {\prl}\ }\textbf {\bibinfo {volume} {98}},\ \bibinfo {eid} {091101} (\bibinfo {year} {2007})},\ \Eprint {https://arxiv.org/abs/gr-qc/0610154} {arXiv:gr-qc/0610154 [gr-qc]} \BibitemShut {NoStop}%
\bibitem [{\citenamefont {{Ho}}(2008)}]{Ho_2008}%
  \BibitemOpen
  \bibfield  {author} {\bibinfo {author} {\bibfnamefont {L.~C.}\ \bibnamefont {{Ho}}},\ }\href {https://doi.org/10.1146/annurev.astro.45.051806.110546} {\bibfield  {journal} {\bibinfo  {journal} {\araa}\ }\textbf {\bibinfo {volume} {46}},\ \bibinfo {pages} {475} (\bibinfo {year} {2008})},\ \Eprint {https://arxiv.org/abs/0803.2268} {arXiv:0803.2268 [astro-ph]} \BibitemShut {NoStop}%
\bibitem [{\citenamefont {Aghanim}\ \emph {et~al.}(2020)\citenamefont {Aghanim} \emph {et~al.}}]{PlanckCollab_2020}%
  \BibitemOpen
  \bibfield  {author} {\bibinfo {author} {\bibfnamefont {N.}~\bibnamefont {Aghanim}} \emph {et~al.} (\bibinfo {collaboration} {Planck}),\ }\href {https://doi.org/10.1051/0004-6361/201833910} {\bibfield  {journal} {\bibinfo  {journal} {Astron. Astrophys.}\ }\textbf {\bibinfo {volume} {641}},\ \bibinfo {pages} {A6} (\bibinfo {year} {2020})},\ \bibinfo {note} {[Erratum: Astron.Astrophys. 652, C4 (2021)]},\ \Eprint {https://arxiv.org/abs/1807.06209} {arXiv:1807.06209 [astro-ph.CO]} \BibitemShut {NoStop}%
\bibitem [{Note2()}]{Note2}%
  \BibitemOpen
  \bibinfo {note} {\protect \href {https://speclite.readthedocs.io/en/latest/index.html}{https://speclite.readthedocs.io/en/latest/index.html}}\BibitemShut {NoStop}%
\bibitem [{\citenamefont {Blanton}\ \emph {et~al.}(2017)\citenamefont {Blanton}, \citenamefont {Bershady}, \citenamefont {Abolfathi}, \citenamefont {Albareti}, \citenamefont {Prieto}, \citenamefont {Almeida}, \citenamefont {Alonso-Garc{\'\i}a}, \citenamefont {Anders}, \citenamefont {Anderson}, \citenamefont {Andrews} \emph {et~al.}}]{blanton2017sloan}%
  \BibitemOpen
  \bibfield  {author} {\bibinfo {author} {\bibfnamefont {M.~R.}\ \bibnamefont {Blanton}}, \bibinfo {author} {\bibfnamefont {M.~A.}\ \bibnamefont {Bershady}}, \bibinfo {author} {\bibfnamefont {B.}~\bibnamefont {Abolfathi}}, \bibinfo {author} {\bibfnamefont {F.~D.}\ \bibnamefont {Albareti}}, \bibinfo {author} {\bibfnamefont {C.~A.}\ \bibnamefont {Prieto}}, \bibinfo {author} {\bibfnamefont {A.}~\bibnamefont {Almeida}}, \bibinfo {author} {\bibfnamefont {J.}~\bibnamefont {Alonso-Garc{\'\i}a}}, \bibinfo {author} {\bibfnamefont {F.}~\bibnamefont {Anders}}, \bibinfo {author} {\bibfnamefont {S.~F.}\ \bibnamefont {Anderson}}, \bibinfo {author} {\bibfnamefont {B.}~\bibnamefont {Andrews}}, \emph {et~al.},\ }\href@noop {} {\bibfield  {journal} {\bibinfo  {journal} {The Astronomical Journal}\ }\textbf {\bibinfo {volume} {154}},\ \bibinfo {pages} {28} (\bibinfo {year} {2017})}\BibitemShut {NoStop}%
\bibitem [{\citenamefont {{Dichiara}}\ \emph {et~al.}(2021)\citenamefont {{Dichiara}}, \citenamefont {{Becerra}}, \citenamefont {{Chase}}, \citenamefont {{Troja}}, \citenamefont {{Lee}}, \citenamefont {{Watson}}, \citenamefont {{Butler}}, \citenamefont {{O'Connor}}, \citenamefont {{Pereyra}}, \citenamefont {{L{\'o}pez}}, \citenamefont {{Lien}}, \citenamefont {{Gottlieb}},\ and\ \citenamefont {{Kutyrev}}}]{Dichiara_2021}%
  \BibitemOpen
  \bibfield  {author} {\bibinfo {author} {\bibfnamefont {S.}~\bibnamefont {{Dichiara}}}, \bibinfo {author} {\bibfnamefont {R.~L.}\ \bibnamefont {{Becerra}}}, \bibinfo {author} {\bibfnamefont {E.~A.}\ \bibnamefont {{Chase}}}, \bibinfo {author} {\bibfnamefont {E.}~\bibnamefont {{Troja}}}, \bibinfo {author} {\bibfnamefont {W.~H.}\ \bibnamefont {{Lee}}}, \bibinfo {author} {\bibfnamefont {A.~M.}\ \bibnamefont {{Watson}}}, \bibinfo {author} {\bibfnamefont {N.~R.}\ \bibnamefont {{Butler}}}, \bibinfo {author} {\bibfnamefont {B.}~\bibnamefont {{O'Connor}}}, \bibinfo {author} {\bibfnamefont {M.}~\bibnamefont {{Pereyra}}}, \bibinfo {author} {\bibfnamefont {K.~O.~C.}\ \bibnamefont {{L{\'o}pez}}}, \bibinfo {author} {\bibfnamefont {A.~Y.}\ \bibnamefont {{Lien}}}, \bibinfo {author} {\bibfnamefont {A.}~\bibnamefont {{Gottlieb}}},\ and\ \bibinfo {author} {\bibfnamefont {A.~S.}\ \bibnamefont {{Kutyrev}}},\ }\href {https://doi.org/10.3847/2041-8213/ac4259} {\bibfield  {journal} {\bibinfo  {journal} {\apjl}\ }\textbf {\bibinfo
  {volume} {923}},\ \bibinfo {eid} {L32} (\bibinfo {year} {2021})},\ \Eprint {https://arxiv.org/abs/2110.12047} {arXiv:2110.12047 [astro-ph.HE]} \BibitemShut {NoStop}%
\bibitem [{\citenamefont {Mendes~de Oliveira}\ \emph {et~al.}(2019)\citenamefont {Mendes~de Oliveira} \emph {et~al.}}]{MendesdeOliveira_2019}%
  \BibitemOpen
  \bibfield  {author} {\bibinfo {author} {\bibfnamefont {C.}~\bibnamefont {Mendes~de Oliveira}} \emph {et~al.},\ }\href {https://doi.org/10.1093/mnras/stz1985} {\bibfield  {journal} {\bibinfo  {journal} {Mon. Not. Roy. Astron. Soc.}\ }\textbf {\bibinfo {volume} {489}},\ \bibinfo {pages} {241} (\bibinfo {year} {2019})},\ \Eprint {https://arxiv.org/abs/1907.01567} {arXiv:1907.01567 [astro-ph.GA]} \BibitemShut {NoStop}%
\bibitem [{\citenamefont {{Almeida-Fernandes}}\ \emph {et~al.}(2022)\citenamefont {{Almeida-Fernandes}}, \citenamefont {{SamPedro}}, \citenamefont {{Herpich}}, \citenamefont {{Molino}}, \citenamefont {{Barbosa}}, \citenamefont {{Buzzo}}, \citenamefont {{Overzier}}, \citenamefont {{de Lima}}, \citenamefont {{Nakazono}}, \citenamefont {{Oliveira Schwarz}}, \citenamefont {{Perottoni}}, \citenamefont {{Bolutavicius}}, \citenamefont {{Guti{\'e}rrez-Soto}}, \citenamefont {{Santos-Silva}}, \citenamefont {{Vitorelli}}, \citenamefont {{Werle}}, \citenamefont {{Whitten}}, \citenamefont {{Costa Duarte}}, \citenamefont {{Bom}}, \citenamefont {{Coelho}}, \citenamefont {{Sodr{\'e}}}, \citenamefont {{Placco}}, \citenamefont {{Teixeira}}, \citenamefont {{Alonso-Garc{\'\i}a}}, \citenamefont {{Barbosa}}, \citenamefont {{Beers}}, \citenamefont {{Bonatto}}, \citenamefont {{Chies-Santos}}, \citenamefont {{Hartmann}}, \citenamefont {{Lopes de Oliveira}}, \citenamefont {{Navarete}}, \citenamefont {{Kanaan}}, \citenamefont
  {{Ribeiro}}, \citenamefont {{Schoenell}},\ and\ \citenamefont {{Mendes de Oliveira}}}]{Almeida-Fernandes_2022}%
  \BibitemOpen
  \bibfield  {author} {\bibinfo {author} {\bibfnamefont {F.}~\bibnamefont {{Almeida-Fernandes}}}, \bibinfo {author} {\bibfnamefont {L.}~\bibnamefont {{SamPedro}}}, \bibinfo {author} {\bibfnamefont {F.~R.}\ \bibnamefont {{Herpich}}}, \bibinfo {author} {\bibfnamefont {A.}~\bibnamefont {{Molino}}}, \bibinfo {author} {\bibfnamefont {C.~E.}\ \bibnamefont {{Barbosa}}}, \bibinfo {author} {\bibfnamefont {M.~L.}\ \bibnamefont {{Buzzo}}}, \bibinfo {author} {\bibfnamefont {R.~A.}\ \bibnamefont {{Overzier}}}, \bibinfo {author} {\bibfnamefont {E.~V.~R.}\ \bibnamefont {{de Lima}}}, \bibinfo {author} {\bibfnamefont {L.~M.~I.}\ \bibnamefont {{Nakazono}}}, \bibinfo {author} {\bibfnamefont {G.~B.}\ \bibnamefont {{Oliveira Schwarz}}}, \bibinfo {author} {\bibfnamefont {H.~D.}\ \bibnamefont {{Perottoni}}}, \bibinfo {author} {\bibfnamefont {G.~F.}\ \bibnamefont {{Bolutavicius}}}, \bibinfo {author} {\bibfnamefont {L.~A.}\ \bibnamefont {{Guti{\'e}rrez-Soto}}}, \bibinfo {author} {\bibfnamefont {T.}~\bibnamefont {{Santos-Silva}}},
  \bibinfo {author} {\bibfnamefont {A.~Z.}\ \bibnamefont {{Vitorelli}}}, \bibinfo {author} {\bibfnamefont {A.}~\bibnamefont {{Werle}}}, \bibinfo {author} {\bibfnamefont {D.~D.}\ \bibnamefont {{Whitten}}}, \bibinfo {author} {\bibfnamefont {M.~V.}\ \bibnamefont {{Costa Duarte}}}, \bibinfo {author} {\bibfnamefont {C.~R.}\ \bibnamefont {{Bom}}}, \bibinfo {author} {\bibfnamefont {P.}~\bibnamefont {{Coelho}}}, \bibinfo {author} {\bibfnamefont {L.}~\bibnamefont {{Sodr{\'e}}}}, \bibinfo {author} {\bibfnamefont {V.~M.}\ \bibnamefont {{Placco}}}, \bibinfo {author} {\bibfnamefont {G.~S.~M.}\ \bibnamefont {{Teixeira}}}, \bibinfo {author} {\bibfnamefont {J.}~\bibnamefont {{Alonso-Garc{\'\i}a}}}, \bibinfo {author} {\bibfnamefont {C.~L.}\ \bibnamefont {{Barbosa}}}, \bibinfo {author} {\bibfnamefont {T.~C.}\ \bibnamefont {{Beers}}}, \bibinfo {author} {\bibfnamefont {C.~J.}\ \bibnamefont {{Bonatto}}}, \bibinfo {author} {\bibfnamefont {A.~L.}\ \bibnamefont {{Chies-Santos}}}, \bibinfo {author} {\bibfnamefont {E.~A.}\
  \bibnamefont {{Hartmann}}}, \bibinfo {author} {\bibfnamefont {R.}~\bibnamefont {{Lopes de Oliveira}}}, \bibinfo {author} {\bibfnamefont {F.}~\bibnamefont {{Navarete}}}, \bibinfo {author} {\bibfnamefont {A.}~\bibnamefont {{Kanaan}}}, \bibinfo {author} {\bibfnamefont {T.}~\bibnamefont {{Ribeiro}}}, \bibinfo {author} {\bibfnamefont {W.}~\bibnamefont {{Schoenell}}},\ and\ \bibinfo {author} {\bibfnamefont {C.}~\bibnamefont {{Mendes de Oliveira}}},\ }\href {https://doi.org/10.1093/mnras/stac284} {\bibfield  {journal} {\bibinfo  {journal} {\mnras}\ }\textbf {\bibinfo {volume} {511}},\ \bibinfo {pages} {4590} (\bibinfo {year} {2022})},\ \Eprint {https://arxiv.org/abs/2104.00020} {arXiv:2104.00020 [astro-ph.IM]} \BibitemShut {NoStop}%
\bibitem [{\citenamefont {Ivezi\'c}\ \emph {et~al.}(2019)\citenamefont {Ivezi\'c} \emph {et~al.}}]{LSST_2019}%
  \BibitemOpen
  \bibfield  {author} {\bibinfo {author} {\bibfnamefont {v.}~\bibnamefont {Ivezi\'c}} \emph {et~al.} (\bibinfo {collaboration} {LSST}),\ }\href {https://doi.org/10.3847/1538-4357/ab042c} {\bibfield  {journal} {\bibinfo  {journal} {Astrophys. J.}\ }\textbf {\bibinfo {volume} {873}},\ \bibinfo {pages} {111} (\bibinfo {year} {2019})},\ \Eprint {https://arxiv.org/abs/0805.2366} {arXiv:0805.2366 [astro-ph]} \BibitemShut {NoStop}%
\bibitem [{\citenamefont {{Fragione}}\ and\ \citenamefont {{Loeb}}(2021)}]{Fragione_2021}%
  \BibitemOpen
  \bibfield  {author} {\bibinfo {author} {\bibfnamefont {G.}~\bibnamefont {{Fragione}}}\ and\ \bibinfo {author} {\bibfnamefont {A.}~\bibnamefont {{Loeb}}},\ }\href {https://doi.org/10.1093/mnras/stab247} {\bibfield  {journal} {\bibinfo  {journal} {\mnras}\ }\textbf {\bibinfo {volume} {502}},\ \bibinfo {pages} {3879} (\bibinfo {year} {2021})},\ \Eprint {https://arxiv.org/abs/2011.08935} {arXiv:2011.08935 [astro-ph.HE]} \BibitemShut {NoStop}%
\bibitem [{\citenamefont {{Abbott}}\ \emph {et~al.}(2023)\citenamefont {{Abbott}}, \citenamefont {{Abbott}}, \citenamefont {{Acernese}}, \citenamefont {{Ackley}}, \citenamefont {{Adams}}, \citenamefont {{Adhikari}}, \citenamefont {{Adhikari}}, \citenamefont {{Adya}}, \citenamefont {{Affeldt}}, \citenamefont {{Agarwal}}, \citenamefont {{Agathos}}, \citenamefont {{Agatsuma}}, \citenamefont {{Aggarwal}}, \citenamefont {{Aguiar}}, \citenamefont {{Aiello}}, \citenamefont {{Ain}}, \citenamefont {{Ajith}}, \citenamefont {{Akutsu}}, \citenamefont {{de Alarc{\'o}n}}, \citenamefont {{Akcay}}, \citenamefont {{Albanesi}}, \citenamefont {{Allocca}}, \citenamefont {{Altin}}, \citenamefont {{Amato}}, \citenamefont {{Anand}}, \citenamefont {{Anand}}, \citenamefont {{Ananyeva}}, \citenamefont {{Anderson}}, \citenamefont {{Anderson}}, \citenamefont {{Ando}}, \citenamefont {{Andrade}}, \citenamefont {{Andres}}, \citenamefont {{Andri{\'c}}}, \citenamefont {{Angelova}}, \citenamefont {{Ansoldi}}, \citenamefont {{Antelis}},
  \citenamefont {{Antier}}, \citenamefont {{Antonini}}, \citenamefont {{Appert}}, \citenamefont {{Arai}}, \citenamefont {{Arai}}, \citenamefont {{Arai}}, \citenamefont {{Araki}}, \citenamefont {{Araya}}, \citenamefont {{Araya}}, \citenamefont {{Areeda}}, \citenamefont {{Ar{\`e}ne}}, \citenamefont {{Aritomi}}, \citenamefont {{Arnaud}}, \citenamefont {{Arogeti}}, \citenamefont {{Aronson}}, \citenamefont {{Arun}}, \citenamefont {{Asada}}, \citenamefont {{Asali}}, \citenamefont {{Ashton}}, \citenamefont {{Aso}}, \citenamefont {{Assiduo}}, \citenamefont {{Aston}}, \citenamefont {{Astone}}, \citenamefont {{Aubin}}, \citenamefont {{Austin}}, \citenamefont {{Babak}}, \citenamefont {{Badaracco}}, \citenamefont {{Bader}}, \citenamefont {{Badger}}, \citenamefont {{Bae}}, \citenamefont {{Bae}}, \citenamefont {{Baer}}, \citenamefont {{Bagnasco}}, \citenamefont {{Bai}}, \citenamefont {{Baiotti}}, \citenamefont {{Baird}}, \citenamefont {{Bajpai}}, \citenamefont {{Ball}}, \citenamefont {{Ballardin}}, \citenamefont
  {{Ballmer}}, \citenamefont {{Balsamo}}, \citenamefont {{Baltus}}, \citenamefont {{Banagiri}}, \citenamefont {{Bankar}}, \citenamefont {{Barayoga}}, \citenamefont {{Barbieri}}, \citenamefont {{Barish}}, \citenamefont {{Barker}}, \citenamefont {{Barneo}}, \citenamefont {{Barone}}, \citenamefont {{Barr}}, \citenamefont {{Barsotti}}, \citenamefont {{Barsuglia}}, \citenamefont {{Barta}}, \citenamefont {{Bartlett}}, \citenamefont {{Barton}}, \citenamefont {{Bartos}}, \citenamefont {{Bassiri}}, \citenamefont {{Basti}}, \citenamefont {{Bawaj}}, \citenamefont {{Bayley}}, \citenamefont {{Baylor}}, \citenamefont {{Bazzan}}, \citenamefont {{B{\'e}csy}}, \citenamefont {{Bedakihale}}, \citenamefont {{Bejger}}, \citenamefont {{Belahcene}}, \citenamefont {{Benedetto}}, \citenamefont {{Beniwal}}, \citenamefont {{Bennett}}, \citenamefont {{Bentley}}, \citenamefont {{Benyaala}}, \citenamefont {{Bergamin}}, \citenamefont {{Berger}}, \citenamefont {{Bernuzzi}}, \citenamefont {{Berry}}, \citenamefont {{Bersanetti}},
  \citenamefont {{Bertolini}}, \citenamefont {{Betzwieser}}, \citenamefont {{Beveridge}}, \citenamefont {{Bhandare}}, \citenamefont {{Bhardwaj}}, \citenamefont {{Bhattacharjee}}, \citenamefont {{Bhaumik}}, \citenamefont {{Bilenko}}, \citenamefont {{Billingsley}}, \citenamefont {{Bini}}, \citenamefont {{Birney}}, \citenamefont {{Birnholtz}}, \citenamefont {{Biscans}}, \citenamefont {{Bischi}}, \citenamefont {{Biscoveanu}}, \citenamefont {{Bisht}}, \citenamefont {{Biswas}}, \citenamefont {{Bitossi}}, \citenamefont {{Bizouard}}, \citenamefont {{Blackburn}}, \citenamefont {{Blair}}, \citenamefont {{Blair}}, \citenamefont {{Blair}}, \citenamefont {{Bobba}}, \citenamefont {{Bode}}, \citenamefont {{Boer}}, \citenamefont {{Bogaert}}, \citenamefont {{Boldrini}}, \citenamefont {{Bonavena}}, \citenamefont {{Bondu}}, \citenamefont {{Bonilla}}, \citenamefont {{Bonnand}}, \citenamefont {{Booker}}, \citenamefont {{Boom}}, \citenamefont {{Bork}}, \citenamefont {{Boschi}}, \citenamefont {{Bose}}, \citenamefont {{Bose}},
  \citenamefont {{Bossilkov}}, \citenamefont {{Boudart}}, \citenamefont {{Bouffanais}}, \citenamefont {{Bozzi}}, \citenamefont {{Bradaschia}}, \citenamefont {{Brady}}, \citenamefont {{Bramley}}, \citenamefont {{Branch}}, \citenamefont {{Branchesi}}, \citenamefont {{Brandt}}, \citenamefont {{Brau}}, \citenamefont {{Breschi}}, \citenamefont {{Briant}}, \citenamefont {{Briggs}}, \citenamefont {{Brillet}}, \citenamefont {{Brinkmann}}, \citenamefont {{Brockill}}, \citenamefont {{Brooks}}, \citenamefont {{Brooks}}, \citenamefont {{Brown}}, \citenamefont {{Brunett}}, \citenamefont {{Bruno}}, \citenamefont {{Bruntz}}, \citenamefont {{Bryant}}, \citenamefont {{Bulik}}, \citenamefont {{Bulten}}, \citenamefont {{Buonanno}}, \citenamefont {{Buscicchio}}, \citenamefont {{Buskulic}}, \citenamefont {{Buy}}, \citenamefont {{Byer}}, \citenamefont {{Cadonati}}, \citenamefont {{Cagnoli}}, \citenamefont {{Cahillane}}, \citenamefont {{Bustillo}}, \citenamefont {{Callaghan}}, \citenamefont {{Callister}}, \citenamefont {{Calloni}},
  \citenamefont {{Cameron}}, \citenamefont {{Camp}}, \citenamefont {{Canepa}}, \citenamefont {{Canevarolo}}, \citenamefont {{Cannavacciuolo}}, \citenamefont {{Cannon}}, \citenamefont {{Cao}}, \citenamefont {{Cao}}, \citenamefont {{Capocasa}}, \citenamefont {{Capote}},\ and\ \citenamefont {{Carapella}}}]{Abbott_2023}%
  \BibitemOpen
  \bibfield  {author} {\bibinfo {author} {\bibfnamefont {R.}~\bibnamefont {{Abbott}}}, \bibinfo {author} {\bibfnamefont {T.~D.}\ \bibnamefont {{Abbott}}}, \bibinfo {author} {\bibfnamefont {F.}~\bibnamefont {{Acernese}}}, \bibinfo {author} {\bibfnamefont {K.}~\bibnamefont {{Ackley}}}, \bibinfo {author} {\bibfnamefont {C.}~\bibnamefont {{Adams}}}, \bibinfo {author} {\bibfnamefont {N.}~\bibnamefont {{Adhikari}}}, \bibinfo {author} {\bibfnamefont {R.~X.}\ \bibnamefont {{Adhikari}}}, \bibinfo {author} {\bibfnamefont {V.~B.}\ \bibnamefont {{Adya}}}, \bibinfo {author} {\bibfnamefont {C.}~\bibnamefont {{Affeldt}}}, \bibinfo {author} {\bibfnamefont {D.}~\bibnamefont {{Agarwal}}}, \bibinfo {author} {\bibfnamefont {M.}~\bibnamefont {{Agathos}}}, \bibinfo {author} {\bibfnamefont {K.}~\bibnamefont {{Agatsuma}}}, \bibinfo {author} {\bibfnamefont {N.}~\bibnamefont {{Aggarwal}}}, \bibinfo {author} {\bibfnamefont {O.~D.}\ \bibnamefont {{Aguiar}}}, \bibinfo {author} {\bibfnamefont {L.}~\bibnamefont {{Aiello}}}, \bibinfo
  {author} {\bibfnamefont {A.}~\bibnamefont {{Ain}}}, \bibinfo {author} {\bibfnamefont {P.}~\bibnamefont {{Ajith}}}, \bibinfo {author} {\bibfnamefont {T.}~\bibnamefont {{Akutsu}}}, \bibinfo {author} {\bibfnamefont {P.~F.}\ \bibnamefont {{de Alarc{\'o}n}}}, \bibinfo {author} {\bibfnamefont {S.}~\bibnamefont {{Akcay}}}, \bibinfo {author} {\bibfnamefont {S.}~\bibnamefont {{Albanesi}}}, \bibinfo {author} {\bibfnamefont {A.}~\bibnamefont {{Allocca}}}, \bibinfo {author} {\bibfnamefont {P.~A.}\ \bibnamefont {{Altin}}}, \bibinfo {author} {\bibfnamefont {A.}~\bibnamefont {{Amato}}}, \bibinfo {author} {\bibfnamefont {C.}~\bibnamefont {{Anand}}}, \bibinfo {author} {\bibfnamefont {S.}~\bibnamefont {{Anand}}}, \bibinfo {author} {\bibfnamefont {A.}~\bibnamefont {{Ananyeva}}}, \bibinfo {author} {\bibfnamefont {S.~B.}\ \bibnamefont {{Anderson}}}, \bibinfo {author} {\bibfnamefont {W.~G.}\ \bibnamefont {{Anderson}}}, \bibinfo {author} {\bibfnamefont {M.}~\bibnamefont {{Ando}}}, \bibinfo {author} {\bibfnamefont
  {T.}~\bibnamefont {{Andrade}}}, \bibinfo {author} {\bibfnamefont {N.}~\bibnamefont {{Andres}}}, \bibinfo {author} {\bibfnamefont {T.}~\bibnamefont {{Andri{\'c}}}}, \bibinfo {author} {\bibfnamefont {S.~V.}\ \bibnamefont {{Angelova}}}, \bibinfo {author} {\bibfnamefont {S.}~\bibnamefont {{Ansoldi}}}, \bibinfo {author} {\bibfnamefont {J.~M.}\ \bibnamefont {{Antelis}}}, \bibinfo {author} {\bibfnamefont {S.}~\bibnamefont {{Antier}}}, \bibinfo {author} {\bibfnamefont {F.}~\bibnamefont {{Antonini}}}, \bibinfo {author} {\bibfnamefont {S.}~\bibnamefont {{Appert}}}, \bibinfo {author} {\bibfnamefont {K.}~\bibnamefont {{Arai}}}, \bibinfo {author} {\bibfnamefont {K.}~\bibnamefont {{Arai}}}, \bibinfo {author} {\bibfnamefont {Y.}~\bibnamefont {{Arai}}}, \bibinfo {author} {\bibfnamefont {S.}~\bibnamefont {{Araki}}}, \bibinfo {author} {\bibfnamefont {A.}~\bibnamefont {{Araya}}}, \bibinfo {author} {\bibfnamefont {M.~C.}\ \bibnamefont {{Araya}}}, \bibinfo {author} {\bibfnamefont {J.~S.}\ \bibnamefont {{Areeda}}}, \bibinfo
  {author} {\bibfnamefont {M.}~\bibnamefont {{Ar{\`e}ne}}}, \bibinfo {author} {\bibfnamefont {N.}~\bibnamefont {{Aritomi}}}, \bibinfo {author} {\bibfnamefont {N.}~\bibnamefont {{Arnaud}}}, \bibinfo {author} {\bibfnamefont {M.}~\bibnamefont {{Arogeti}}}, \bibinfo {author} {\bibfnamefont {S.~M.}\ \bibnamefont {{Aronson}}}, \bibinfo {author} {\bibfnamefont {K.~G.}\ \bibnamefont {{Arun}}}, \bibinfo {author} {\bibfnamefont {H.}~\bibnamefont {{Asada}}}, \bibinfo {author} {\bibfnamefont {Y.}~\bibnamefont {{Asali}}}, \bibinfo {author} {\bibfnamefont {G.}~\bibnamefont {{Ashton}}}, \bibinfo {author} {\bibfnamefont {Y.}~\bibnamefont {{Aso}}}, \bibinfo {author} {\bibfnamefont {M.}~\bibnamefont {{Assiduo}}}, \bibinfo {author} {\bibfnamefont {S.~M.}\ \bibnamefont {{Aston}}}, \bibinfo {author} {\bibfnamefont {P.}~\bibnamefont {{Astone}}}, \bibinfo {author} {\bibfnamefont {F.}~\bibnamefont {{Aubin}}}, \bibinfo {author} {\bibfnamefont {C.}~\bibnamefont {{Austin}}}, \bibinfo {author} {\bibfnamefont {S.}~\bibnamefont
  {{Babak}}}, \bibinfo {author} {\bibfnamefont {F.}~\bibnamefont {{Badaracco}}}, \bibinfo {author} {\bibfnamefont {M.~K.~M.}\ \bibnamefont {{Bader}}}, \bibinfo {author} {\bibfnamefont {C.}~\bibnamefont {{Badger}}}, \bibinfo {author} {\bibfnamefont {S.}~\bibnamefont {{Bae}}}, \bibinfo {author} {\bibfnamefont {Y.}~\bibnamefont {{Bae}}}, \bibinfo {author} {\bibfnamefont {A.~M.}\ \bibnamefont {{Baer}}}, \bibinfo {author} {\bibfnamefont {S.}~\bibnamefont {{Bagnasco}}}, \bibinfo {author} {\bibfnamefont {Y.}~\bibnamefont {{Bai}}}, \bibinfo {author} {\bibfnamefont {L.}~\bibnamefont {{Baiotti}}}, \bibinfo {author} {\bibfnamefont {J.}~\bibnamefont {{Baird}}}, \bibinfo {author} {\bibfnamefont {R.}~\bibnamefont {{Bajpai}}}, \bibinfo {author} {\bibfnamefont {M.}~\bibnamefont {{Ball}}}, \bibinfo {author} {\bibfnamefont {G.}~\bibnamefont {{Ballardin}}}, \bibinfo {author} {\bibfnamefont {S.~W.}\ \bibnamefont {{Ballmer}}}, \bibinfo {author} {\bibfnamefont {A.}~\bibnamefont {{Balsamo}}}, \bibinfo {author} {\bibfnamefont
  {G.}~\bibnamefont {{Baltus}}}, \bibinfo {author} {\bibfnamefont {S.}~\bibnamefont {{Banagiri}}}, \bibinfo {author} {\bibfnamefont {D.}~\bibnamefont {{Bankar}}}, \bibinfo {author} {\bibfnamefont {J.~C.}\ \bibnamefont {{Barayoga}}}, \bibinfo {author} {\bibfnamefont {C.}~\bibnamefont {{Barbieri}}}, \bibinfo {author} {\bibfnamefont {B.~C.}\ \bibnamefont {{Barish}}}, \bibinfo {author} {\bibfnamefont {D.}~\bibnamefont {{Barker}}}, \bibinfo {author} {\bibfnamefont {P.}~\bibnamefont {{Barneo}}}, \bibinfo {author} {\bibfnamefont {F.}~\bibnamefont {{Barone}}}, \bibinfo {author} {\bibfnamefont {B.}~\bibnamefont {{Barr}}}, \bibinfo {author} {\bibfnamefont {L.}~\bibnamefont {{Barsotti}}}, \bibinfo {author} {\bibfnamefont {M.}~\bibnamefont {{Barsuglia}}}, \bibinfo {author} {\bibfnamefont {D.}~\bibnamefont {{Barta}}}, \bibinfo {author} {\bibfnamefont {J.}~\bibnamefont {{Bartlett}}}, \bibinfo {author} {\bibfnamefont {M.~A.}\ \bibnamefont {{Barton}}}, \bibinfo {author} {\bibfnamefont {I.}~\bibnamefont {{Bartos}}}, \bibinfo
  {author} {\bibfnamefont {R.}~\bibnamefont {{Bassiri}}}, \bibinfo {author} {\bibfnamefont {A.}~\bibnamefont {{Basti}}}, \bibinfo {author} {\bibfnamefont {M.}~\bibnamefont {{Bawaj}}}, \bibinfo {author} {\bibfnamefont {J.~C.}\ \bibnamefont {{Bayley}}}, \bibinfo {author} {\bibfnamefont {A.~C.}\ \bibnamefont {{Baylor}}}, \bibinfo {author} {\bibfnamefont {M.}~\bibnamefont {{Bazzan}}}, \bibinfo {author} {\bibfnamefont {B.}~\bibnamefont {{B{\'e}csy}}}, \bibinfo {author} {\bibfnamefont {V.~M.}\ \bibnamefont {{Bedakihale}}}, \bibinfo {author} {\bibfnamefont {M.}~\bibnamefont {{Bejger}}}, \bibinfo {author} {\bibfnamefont {I.}~\bibnamefont {{Belahcene}}}, \bibinfo {author} {\bibfnamefont {V.}~\bibnamefont {{Benedetto}}}, \bibinfo {author} {\bibfnamefont {D.}~\bibnamefont {{Beniwal}}}, \bibinfo {author} {\bibfnamefont {T.~F.}\ \bibnamefont {{Bennett}}}, \bibinfo {author} {\bibfnamefont {J.~D.}\ \bibnamefont {{Bentley}}}, \bibinfo {author} {\bibfnamefont {M.}~\bibnamefont {{Benyaala}}}, \bibinfo {author} {\bibfnamefont
  {F.}~\bibnamefont {{Bergamin}}}, \bibinfo {author} {\bibfnamefont {B.~K.}\ \bibnamefont {{Berger}}}, \bibinfo {author} {\bibfnamefont {S.}~\bibnamefont {{Bernuzzi}}}, \bibinfo {author} {\bibfnamefont {C.~P.~L.}\ \bibnamefont {{Berry}}}, \bibinfo {author} {\bibfnamefont {D.}~\bibnamefont {{Bersanetti}}}, \bibinfo {author} {\bibfnamefont {A.}~\bibnamefont {{Bertolini}}}, \bibinfo {author} {\bibfnamefont {J.}~\bibnamefont {{Betzwieser}}}, \bibinfo {author} {\bibfnamefont {D.}~\bibnamefont {{Beveridge}}}, \bibinfo {author} {\bibfnamefont {R.}~\bibnamefont {{Bhandare}}}, \bibinfo {author} {\bibfnamefont {U.}~\bibnamefont {{Bhardwaj}}}, \bibinfo {author} {\bibfnamefont {D.}~\bibnamefont {{Bhattacharjee}}}, \bibinfo {author} {\bibfnamefont {S.}~\bibnamefont {{Bhaumik}}}, \bibinfo {author} {\bibfnamefont {I.~A.}\ \bibnamefont {{Bilenko}}}, \bibinfo {author} {\bibfnamefont {G.}~\bibnamefont {{Billingsley}}}, \bibinfo {author} {\bibfnamefont {S.}~\bibnamefont {{Bini}}}, \bibinfo {author} {\bibfnamefont
  {R.}~\bibnamefont {{Birney}}}, \bibinfo {author} {\bibfnamefont {O.}~\bibnamefont {{Birnholtz}}}, \bibinfo {author} {\bibfnamefont {S.}~\bibnamefont {{Biscans}}}, \bibinfo {author} {\bibfnamefont {M.}~\bibnamefont {{Bischi}}}, \bibinfo {author} {\bibfnamefont {S.}~\bibnamefont {{Biscoveanu}}}, \bibinfo {author} {\bibfnamefont {A.}~\bibnamefont {{Bisht}}}, \bibinfo {author} {\bibfnamefont {B.}~\bibnamefont {{Biswas}}}, \bibinfo {author} {\bibfnamefont {M.}~\bibnamefont {{Bitossi}}}, \bibinfo {author} {\bibfnamefont {M.~A.}\ \bibnamefont {{Bizouard}}}, \bibinfo {author} {\bibfnamefont {J.~K.}\ \bibnamefont {{Blackburn}}}, \bibinfo {author} {\bibfnamefont {C.~D.}\ \bibnamefont {{Blair}}}, \bibinfo {author} {\bibfnamefont {D.~G.}\ \bibnamefont {{Blair}}}, \bibinfo {author} {\bibfnamefont {R.~M.}\ \bibnamefont {{Blair}}}, \bibinfo {author} {\bibfnamefont {F.}~\bibnamefont {{Bobba}}}, \bibinfo {author} {\bibfnamefont {N.}~\bibnamefont {{Bode}}}, \bibinfo {author} {\bibfnamefont {M.}~\bibnamefont {{Boer}}},
  \bibinfo {author} {\bibfnamefont {G.}~\bibnamefont {{Bogaert}}}, \bibinfo {author} {\bibfnamefont {M.}~\bibnamefont {{Boldrini}}}, \bibinfo {author} {\bibfnamefont {L.~D.}\ \bibnamefont {{Bonavena}}}, \bibinfo {author} {\bibfnamefont {F.}~\bibnamefont {{Bondu}}}, \bibinfo {author} {\bibfnamefont {E.}~\bibnamefont {{Bonilla}}}, \bibinfo {author} {\bibfnamefont {R.}~\bibnamefont {{Bonnand}}}, \bibinfo {author} {\bibfnamefont {P.}~\bibnamefont {{Booker}}}, \bibinfo {author} {\bibfnamefont {B.~A.}\ \bibnamefont {{Boom}}}, \bibinfo {author} {\bibfnamefont {R.}~\bibnamefont {{Bork}}}, \bibinfo {author} {\bibfnamefont {V.}~\bibnamefont {{Boschi}}}, \bibinfo {author} {\bibfnamefont {N.}~\bibnamefont {{Bose}}}, \bibinfo {author} {\bibfnamefont {S.}~\bibnamefont {{Bose}}}, \bibinfo {author} {\bibfnamefont {V.}~\bibnamefont {{Bossilkov}}}, \bibinfo {author} {\bibfnamefont {V.}~\bibnamefont {{Boudart}}}, \bibinfo {author} {\bibfnamefont {Y.}~\bibnamefont {{Bouffanais}}}, \bibinfo {author} {\bibfnamefont
  {A.}~\bibnamefont {{Bozzi}}}, \bibinfo {author} {\bibfnamefont {C.}~\bibnamefont {{Bradaschia}}}, \bibinfo {author} {\bibfnamefont {P.~R.}\ \bibnamefont {{Brady}}}, \bibinfo {author} {\bibfnamefont {A.}~\bibnamefont {{Bramley}}}, \bibinfo {author} {\bibfnamefont {A.}~\bibnamefont {{Branch}}}, \bibinfo {author} {\bibfnamefont {M.}~\bibnamefont {{Branchesi}}}, \bibinfo {author} {\bibfnamefont {J.}~\bibnamefont {{Brandt}}}, \bibinfo {author} {\bibfnamefont {J.~E.}\ \bibnamefont {{Brau}}}, \bibinfo {author} {\bibfnamefont {M.}~\bibnamefont {{Breschi}}}, \bibinfo {author} {\bibfnamefont {T.}~\bibnamefont {{Briant}}}, \bibinfo {author} {\bibfnamefont {J.~H.}\ \bibnamefont {{Briggs}}}, \bibinfo {author} {\bibfnamefont {A.}~\bibnamefont {{Brillet}}}, \bibinfo {author} {\bibfnamefont {M.}~\bibnamefont {{Brinkmann}}}, \bibinfo {author} {\bibfnamefont {P.}~\bibnamefont {{Brockill}}}, \bibinfo {author} {\bibfnamefont {A.~F.}\ \bibnamefont {{Brooks}}}, \bibinfo {author} {\bibfnamefont {J.}~\bibnamefont {{Brooks}}},
  \bibinfo {author} {\bibfnamefont {D.~D.}\ \bibnamefont {{Brown}}}, \bibinfo {author} {\bibfnamefont {S.}~\bibnamefont {{Brunett}}}, \bibinfo {author} {\bibfnamefont {G.}~\bibnamefont {{Bruno}}}, \bibinfo {author} {\bibfnamefont {R.}~\bibnamefont {{Bruntz}}}, \bibinfo {author} {\bibfnamefont {J.}~\bibnamefont {{Bryant}}}, \bibinfo {author} {\bibfnamefont {T.}~\bibnamefont {{Bulik}}}, \bibinfo {author} {\bibfnamefont {H.~J.}\ \bibnamefont {{Bulten}}}, \bibinfo {author} {\bibfnamefont {A.}~\bibnamefont {{Buonanno}}}, \bibinfo {author} {\bibfnamefont {R.}~\bibnamefont {{Buscicchio}}}, \bibinfo {author} {\bibfnamefont {D.}~\bibnamefont {{Buskulic}}}, \bibinfo {author} {\bibfnamefont {C.}~\bibnamefont {{Buy}}}, \bibinfo {author} {\bibfnamefont {R.~L.}\ \bibnamefont {{Byer}}}, \bibinfo {author} {\bibfnamefont {L.}~\bibnamefont {{Cadonati}}}, \bibinfo {author} {\bibfnamefont {G.}~\bibnamefont {{Cagnoli}}}, \bibinfo {author} {\bibfnamefont {C.}~\bibnamefont {{Cahillane}}}, \bibinfo {author} {\bibfnamefont {J.~C.}\
  \bibnamefont {{Bustillo}}}, \bibinfo {author} {\bibfnamefont {J.~D.}\ \bibnamefont {{Callaghan}}}, \bibinfo {author} {\bibfnamefont {T.~A.}\ \bibnamefont {{Callister}}}, \bibinfo {author} {\bibfnamefont {E.}~\bibnamefont {{Calloni}}}, \bibinfo {author} {\bibfnamefont {J.}~\bibnamefont {{Cameron}}}, \bibinfo {author} {\bibfnamefont {J.~B.}\ \bibnamefont {{Camp}}}, \bibinfo {author} {\bibfnamefont {M.}~\bibnamefont {{Canepa}}}, \bibinfo {author} {\bibfnamefont {S.}~\bibnamefont {{Canevarolo}}}, \bibinfo {author} {\bibfnamefont {M.}~\bibnamefont {{Cannavacciuolo}}}, \bibinfo {author} {\bibfnamefont {K.~C.}\ \bibnamefont {{Cannon}}}, \bibinfo {author} {\bibfnamefont {H.}~\bibnamefont {{Cao}}}, \bibinfo {author} {\bibfnamefont {Z.}~\bibnamefont {{Cao}}}, \bibinfo {author} {\bibfnamefont {E.}~\bibnamefont {{Capocasa}}}, \bibinfo {author} {\bibfnamefont {E.}~\bibnamefont {{Capote}}},\ and\ \bibinfo {author} {\bibfnamefont {G.}~\bibnamefont {{Carapella}}},\ }\href {https://doi.org/10.1103/PhysRevX.13.011048}
  {\bibfield  {journal} {\bibinfo  {journal} {Physical Review X}\ }\textbf {\bibinfo {volume} {13}},\ \bibinfo {eid} {011048} (\bibinfo {year} {2023})},\ \Eprint {https://arxiv.org/abs/2111.03634} {arXiv:2111.03634 [astro-ph.HE]} \BibitemShut {NoStop}%
\bibitem [{\citenamefont {{Adamcewicz}}\ and\ \citenamefont {{Thrane}}(2022)}]{Adamcewicz_2022}%
  \BibitemOpen
  \bibfield  {author} {\bibinfo {author} {\bibfnamefont {C.}~\bibnamefont {{Adamcewicz}}}\ and\ \bibinfo {author} {\bibfnamefont {E.}~\bibnamefont {{Thrane}}},\ }\href {https://doi.org/10.1093/mnras/stac2961} {\bibfield  {journal} {\bibinfo  {journal} {\mnras}\ }\textbf {\bibinfo {volume} {517}},\ \bibinfo {pages} {3928} (\bibinfo {year} {2022})},\ \Eprint {https://arxiv.org/abs/2208.03405} {arXiv:2208.03405 [astro-ph.HE]} \BibitemShut {NoStop}%
\bibitem [{\citenamefont {{Shvartzvald}}\ \emph {et~al.}(2024)\citenamefont {{Shvartzvald}}, \citenamefont {{Waxman}}, \citenamefont {{Gal-Yam}}, \citenamefont {{Ofek}}, \citenamefont {{Ben-Ami}}, \citenamefont {{Berge}}, \citenamefont {{Kowalski}}, \citenamefont {{B{\"u}hler}}, \citenamefont {{Worm}}, \citenamefont {{Rhoads}}, \citenamefont {{Arcavi}}, \citenamefont {{Maoz}}, \citenamefont {{Polishook}}, \citenamefont {{Stone}}, \citenamefont {{Trakhtenbrot}}, \citenamefont {{Ackermann}}, \citenamefont {{Aharonson}}, \citenamefont {{Birnholtz}}, \citenamefont {{Chelouche}}, \citenamefont {{Guetta}}, \citenamefont {{Hallakoun}}, \citenamefont {{Horesh}}, \citenamefont {{Kushnir}}, \citenamefont {{Mazeh}}, \citenamefont {{Nordin}}, \citenamefont {{Ofir}}, \citenamefont {{Ohm}}, \citenamefont {{Parsons}}, \citenamefont {{Pe'er}}, \citenamefont {{Perets}}, \citenamefont {{Perdelwitz}}, \citenamefont {{Poznanski}}, \citenamefont {{Sadeh}}, \citenamefont {{Sagiv}}, \citenamefont {{Shahaf}}, \citenamefont
  {{Soumagnac}}, \citenamefont {{Tal-Or}}, \citenamefont {{Santen}}, \citenamefont {{Zackay}}, \citenamefont {{Guttman}}, \citenamefont {{Rekhi}}, \citenamefont {{Townsend}}, \citenamefont {{Weinstein}},\ and\ \citenamefont {{Wold}}}]{Shvartzvald_2024}%
  \BibitemOpen
  \bibfield  {author} {\bibinfo {author} {\bibfnamefont {Y.}~\bibnamefont {{Shvartzvald}}}, \bibinfo {author} {\bibfnamefont {E.}~\bibnamefont {{Waxman}}}, \bibinfo {author} {\bibfnamefont {A.}~\bibnamefont {{Gal-Yam}}}, \bibinfo {author} {\bibfnamefont {E.~O.}\ \bibnamefont {{Ofek}}}, \bibinfo {author} {\bibfnamefont {S.}~\bibnamefont {{Ben-Ami}}}, \bibinfo {author} {\bibfnamefont {D.}~\bibnamefont {{Berge}}}, \bibinfo {author} {\bibfnamefont {M.}~\bibnamefont {{Kowalski}}}, \bibinfo {author} {\bibfnamefont {R.}~\bibnamefont {{B{\"u}hler}}}, \bibinfo {author} {\bibfnamefont {S.}~\bibnamefont {{Worm}}}, \bibinfo {author} {\bibfnamefont {J.~E.}\ \bibnamefont {{Rhoads}}}, \bibinfo {author} {\bibfnamefont {I.}~\bibnamefont {{Arcavi}}}, \bibinfo {author} {\bibfnamefont {D.}~\bibnamefont {{Maoz}}}, \bibinfo {author} {\bibfnamefont {D.}~\bibnamefont {{Polishook}}}, \bibinfo {author} {\bibfnamefont {N.}~\bibnamefont {{Stone}}}, \bibinfo {author} {\bibfnamefont {B.}~\bibnamefont {{Trakhtenbrot}}}, \bibinfo {author}
  {\bibfnamefont {M.}~\bibnamefont {{Ackermann}}}, \bibinfo {author} {\bibfnamefont {O.}~\bibnamefont {{Aharonson}}}, \bibinfo {author} {\bibfnamefont {O.}~\bibnamefont {{Birnholtz}}}, \bibinfo {author} {\bibfnamefont {D.}~\bibnamefont {{Chelouche}}}, \bibinfo {author} {\bibfnamefont {D.}~\bibnamefont {{Guetta}}}, \bibinfo {author} {\bibfnamefont {N.}~\bibnamefont {{Hallakoun}}}, \bibinfo {author} {\bibfnamefont {A.}~\bibnamefont {{Horesh}}}, \bibinfo {author} {\bibfnamefont {D.}~\bibnamefont {{Kushnir}}}, \bibinfo {author} {\bibfnamefont {T.}~\bibnamefont {{Mazeh}}}, \bibinfo {author} {\bibfnamefont {J.}~\bibnamefont {{Nordin}}}, \bibinfo {author} {\bibfnamefont {A.}~\bibnamefont {{Ofir}}}, \bibinfo {author} {\bibfnamefont {S.}~\bibnamefont {{Ohm}}}, \bibinfo {author} {\bibfnamefont {D.}~\bibnamefont {{Parsons}}}, \bibinfo {author} {\bibfnamefont {A.}~\bibnamefont {{Pe'er}}}, \bibinfo {author} {\bibfnamefont {H.~B.}\ \bibnamefont {{Perets}}}, \bibinfo {author} {\bibfnamefont {V.}~\bibnamefont
  {{Perdelwitz}}}, \bibinfo {author} {\bibfnamefont {D.}~\bibnamefont {{Poznanski}}}, \bibinfo {author} {\bibfnamefont {I.}~\bibnamefont {{Sadeh}}}, \bibinfo {author} {\bibfnamefont {I.}~\bibnamefont {{Sagiv}}}, \bibinfo {author} {\bibfnamefont {S.}~\bibnamefont {{Shahaf}}}, \bibinfo {author} {\bibfnamefont {M.}~\bibnamefont {{Soumagnac}}}, \bibinfo {author} {\bibfnamefont {L.}~\bibnamefont {{Tal-Or}}}, \bibinfo {author} {\bibfnamefont {J.~V.}\ \bibnamefont {{Santen}}}, \bibinfo {author} {\bibfnamefont {B.}~\bibnamefont {{Zackay}}}, \bibinfo {author} {\bibfnamefont {O.}~\bibnamefont {{Guttman}}}, \bibinfo {author} {\bibfnamefont {P.}~\bibnamefont {{Rekhi}}}, \bibinfo {author} {\bibfnamefont {A.}~\bibnamefont {{Townsend}}}, \bibinfo {author} {\bibfnamefont {A.}~\bibnamefont {{Weinstein}}},\ and\ \bibinfo {author} {\bibfnamefont {I.}~\bibnamefont {{Wold}}},\ }\href {https://doi.org/10.3847/1538-4357/ad2704} {\bibfield  {journal} {\bibinfo  {journal} {\apj}\ }\textbf {\bibinfo {volume} {964}},\ \bibinfo {eid}
  {74} (\bibinfo {year} {2024})},\ \Eprint {https://arxiv.org/abs/2304.14482} {arXiv:2304.14482 [astro-ph.IM]} \BibitemShut {NoStop}%
\bibitem [{\citenamefont {{Kulkarni}}\ \emph {et~al.}(2021)\citenamefont {{Kulkarni}}, \citenamefont {{Harrison}}, \citenamefont {{Grefenstette}}, \citenamefont {{Earnshaw}}, \citenamefont {{Andreoni}}, \citenamefont {{Berg}}, \citenamefont {{Bloom}}, \citenamefont {{Cenko}}, \citenamefont {{Chornock}}, \citenamefont {{Christiansen}}, \citenamefont {{Coughlin}}, \citenamefont {{Wuollet Criswell}}, \citenamefont {{Darvish}}, \citenamefont {{Das}}, \citenamefont {{De}}, \citenamefont {{Dessart}}, \citenamefont {{Dixon}}, \citenamefont {{Dorsman}}, \citenamefont {{El-Badry}}, \citenamefont {{Evans}}, \citenamefont {{Ford}}, \citenamefont {{Fremling}}, \citenamefont {{Gansicke}}, \citenamefont {{Gezari}}, \citenamefont {{Goetberg}}, \citenamefont {{Green}}, \citenamefont {{Graham}}, \citenamefont {{Heida}}, \citenamefont {{Ho}}, \citenamefont {{Jaodand}}, \citenamefont {{Johns-Krull}}, \citenamefont {{Kasliwal}}, \citenamefont {{Lazzarini}}, \citenamefont {{Lu}}, \citenamefont {{Margutti}}, \citenamefont
  {{Martin}}, \citenamefont {{Masters}}, \citenamefont {{McKernan}}, \citenamefont {{Naze}}, \citenamefont {{Nissanke}}, \citenamefont {{Parazin}}, \citenamefont {{Perley}}, \citenamefont {{Phinney}}, \citenamefont {{Piro}}, \citenamefont {{Raaijmakers}}, \citenamefont {{Rauw}}, \citenamefont {{Rodriguez}}, \citenamefont {{Sana}}, \citenamefont {{Senchyna}}, \citenamefont {{Singer}}, \citenamefont {{Spake}}, \citenamefont {{Stassun}}, \citenamefont {{Stern}}, \citenamefont {{Teplitz}}, \citenamefont {{Weisz}},\ and\ \citenamefont {{Yao}}}]{Kulkarni_2021}%
  \BibitemOpen
  \bibfield  {author} {\bibinfo {author} {\bibfnamefont {S.~R.}\ \bibnamefont {{Kulkarni}}}, \bibinfo {author} {\bibfnamefont {F.~A.}\ \bibnamefont {{Harrison}}}, \bibinfo {author} {\bibfnamefont {B.~W.}\ \bibnamefont {{Grefenstette}}}, \bibinfo {author} {\bibfnamefont {H.~P.}\ \bibnamefont {{Earnshaw}}}, \bibinfo {author} {\bibfnamefont {I.}~\bibnamefont {{Andreoni}}}, \bibinfo {author} {\bibfnamefont {D.~A.}\ \bibnamefont {{Berg}}}, \bibinfo {author} {\bibfnamefont {J.~S.}\ \bibnamefont {{Bloom}}}, \bibinfo {author} {\bibfnamefont {S.~B.}\ \bibnamefont {{Cenko}}}, \bibinfo {author} {\bibfnamefont {R.}~\bibnamefont {{Chornock}}}, \bibinfo {author} {\bibfnamefont {J.~L.}\ \bibnamefont {{Christiansen}}}, \bibinfo {author} {\bibfnamefont {M.~W.}\ \bibnamefont {{Coughlin}}}, \bibinfo {author} {\bibfnamefont {A.}~\bibnamefont {{Wuollet Criswell}}}, \bibinfo {author} {\bibfnamefont {B.}~\bibnamefont {{Darvish}}}, \bibinfo {author} {\bibfnamefont {K.~K.}\ \bibnamefont {{Das}}}, \bibinfo {author} {\bibfnamefont
  {K.}~\bibnamefont {{De}}}, \bibinfo {author} {\bibfnamefont {L.}~\bibnamefont {{Dessart}}}, \bibinfo {author} {\bibfnamefont {D.}~\bibnamefont {{Dixon}}}, \bibinfo {author} {\bibfnamefont {B.}~\bibnamefont {{Dorsman}}}, \bibinfo {author} {\bibfnamefont {K.}~\bibnamefont {{El-Badry}}}, \bibinfo {author} {\bibfnamefont {C.}~\bibnamefont {{Evans}}}, \bibinfo {author} {\bibfnamefont {K.~E.~S.}\ \bibnamefont {{Ford}}}, \bibinfo {author} {\bibfnamefont {C.}~\bibnamefont {{Fremling}}}, \bibinfo {author} {\bibfnamefont {B.~T.}\ \bibnamefont {{Gansicke}}}, \bibinfo {author} {\bibfnamefont {S.}~\bibnamefont {{Gezari}}}, \bibinfo {author} {\bibfnamefont {Y.}~\bibnamefont {{Goetberg}}}, \bibinfo {author} {\bibfnamefont {G.~M.}\ \bibnamefont {{Green}}}, \bibinfo {author} {\bibfnamefont {M.~J.}\ \bibnamefont {{Graham}}}, \bibinfo {author} {\bibfnamefont {M.}~\bibnamefont {{Heida}}}, \bibinfo {author} {\bibfnamefont {A.~Y.~Q.}\ \bibnamefont {{Ho}}}, \bibinfo {author} {\bibfnamefont {A.~D.}\ \bibnamefont {{Jaodand}}},
  \bibinfo {author} {\bibfnamefont {C.~M.}\ \bibnamefont {{Johns-Krull}}}, \bibinfo {author} {\bibfnamefont {M.~M.}\ \bibnamefont {{Kasliwal}}}, \bibinfo {author} {\bibfnamefont {M.}~\bibnamefont {{Lazzarini}}}, \bibinfo {author} {\bibfnamefont {W.}~\bibnamefont {{Lu}}}, \bibinfo {author} {\bibfnamefont {R.}~\bibnamefont {{Margutti}}}, \bibinfo {author} {\bibfnamefont {D.~C.}\ \bibnamefont {{Martin}}}, \bibinfo {author} {\bibfnamefont {D.~C.}\ \bibnamefont {{Masters}}}, \bibinfo {author} {\bibfnamefont {B.}~\bibnamefont {{McKernan}}}, \bibinfo {author} {\bibfnamefont {Y.}~\bibnamefont {{Naze}}}, \bibinfo {author} {\bibfnamefont {S.~M.}\ \bibnamefont {{Nissanke}}}, \bibinfo {author} {\bibfnamefont {B.}~\bibnamefont {{Parazin}}}, \bibinfo {author} {\bibfnamefont {D.~A.}\ \bibnamefont {{Perley}}}, \bibinfo {author} {\bibfnamefont {E.~S.}\ \bibnamefont {{Phinney}}}, \bibinfo {author} {\bibfnamefont {A.~L.}\ \bibnamefont {{Piro}}}, \bibinfo {author} {\bibfnamefont {G.}~\bibnamefont {{Raaijmakers}}}, \bibinfo
  {author} {\bibfnamefont {G.}~\bibnamefont {{Rauw}}}, \bibinfo {author} {\bibfnamefont {A.~C.}\ \bibnamefont {{Rodriguez}}}, \bibinfo {author} {\bibfnamefont {H.}~\bibnamefont {{Sana}}}, \bibinfo {author} {\bibfnamefont {P.}~\bibnamefont {{Senchyna}}}, \bibinfo {author} {\bibfnamefont {L.~P.}\ \bibnamefont {{Singer}}}, \bibinfo {author} {\bibfnamefont {J.~J.}\ \bibnamefont {{Spake}}}, \bibinfo {author} {\bibfnamefont {K.~G.}\ \bibnamefont {{Stassun}}}, \bibinfo {author} {\bibfnamefont {D.}~\bibnamefont {{Stern}}}, \bibinfo {author} {\bibfnamefont {H.~I.}\ \bibnamefont {{Teplitz}}}, \bibinfo {author} {\bibfnamefont {D.~R.}\ \bibnamefont {{Weisz}}},\ and\ \bibinfo {author} {\bibfnamefont {Y.}~\bibnamefont {{Yao}}},\ }\href {https://doi.org/10.48550/arXiv.2111.15608} {\bibfield  {journal} {\bibinfo  {journal} {arXiv e-prints}\ ,\ \bibinfo {eid} {arXiv:2111.15608}} (\bibinfo {year} {2021})},\ \Eprint {https://arxiv.org/abs/2111.15608} {arXiv:2111.15608 [astro-ph.GA]} \BibitemShut {NoStop}%
\bibitem [{\citenamefont {{Kato}}\ \emph {et~al.}(2008)\citenamefont {{Kato}}, \citenamefont {{Fukue}},\ and\ \citenamefont {{Mineshige}}}]{Kato_2008}%
  \BibitemOpen
  \bibfield  {author} {\bibinfo {author} {\bibfnamefont {S.}~\bibnamefont {{Kato}}}, \bibinfo {author} {\bibfnamefont {J.}~\bibnamefont {{Fukue}}},\ and\ \bibinfo {author} {\bibfnamefont {S.}~\bibnamefont {{Mineshige}}},\ }\href@noop {} {\emph {\bibinfo {title} {{Black-Hole Accretion Disks --- Towards a New Paradigm ---}}}}\ (\bibinfo {year} {2008})\BibitemShut {NoStop}%
\bibitem [{Note3()}]{Note3}%
  \BibitemOpen
  \bibinfo {note} {\protect \href {https://github.com/DariaGangardt/pAGN}{https://github.com/DariaGangardt/pAGN}}\BibitemShut {NoStop}%
\bibitem [{\citenamefont {{Kormendy}}\ and\ \citenamefont {{Ho}}(2013)}]{Kormendy_2013}%
  \BibitemOpen
  \bibfield  {author} {\bibinfo {author} {\bibfnamefont {J.}~\bibnamefont {{Kormendy}}}\ and\ \bibinfo {author} {\bibfnamefont {L.~C.}\ \bibnamefont {{Ho}}},\ }\href {https://doi.org/10.1146/annurev-astro-082708-101811} {\bibfield  {journal} {\bibinfo  {journal} {\araa}\ }\textbf {\bibinfo {volume} {51}},\ \bibinfo {pages} {511} (\bibinfo {year} {2013})},\ \Eprint {https://arxiv.org/abs/1304.7762} {arXiv:1304.7762 [astro-ph.CO]} \BibitemShut {NoStop}%
\end{thebibliography}%

\end{document}